\documentclass[twocolumn]{aastex63}

\usepackage{graphicx}
\usepackage{amssymb,amsmath}
\usepackage{epstopdf}
\usepackage{textcomp,txfonts} 

\usepackage{rotating,multirow}

\usepackage{natbib}
\setcitestyle{aysep={}}

\usepackage{lipsum}

\shortauthors{Pineda et al.}
\shorttitle{FUMES}

\begin{document}

\title{The Far Ultraviolet M-dwarf Evolution Survey. I. The Rotational Evolution of High-Energy Emissions \footnote{This research is based on observations made with the NASA/ESA Hubble Space Telescope obtained from the Space Telescope Science Institute, which is operated by the Association of Universities for Research in Astronomy, Inc., under NASA contract NAS 5–26555.}}

\author{J.~Sebastian Pineda}
\altaffiliation{\href{mailto:sebastian.pineda@lasp.colorado.edu}{sebastian.pineda@lasp.colorado.edu}}
\affiliation{University of Colorado Boulder, Laboratoy for Atmospheric and Space Physics, 3665 Discovery Drive, Boulder CO, 80303, USA}
\affiliation{Visiting astronomer, Cerro Tololo Inter-American Observatory, National Optical Astronomy Observatory, which is operated by the Association of Universities for Research in Astronomy (AURA) under a cooperative agreement with the National Science Foundation.}

\author{Allison Youngblood}
\affiliation{University of Colorado Boulder, Laboratoy for Atmospheric and Space Physics, 3665 Discovery Drive, Boulder CO, 80303, USA}
\affiliation{NASA Goddard Space Flight Center, 8800 Greenbelt Rd, Greenbelt, MD 20771, USA}

\author{Kevin France}
\affiliation{University of Colorado Boulder, Laboratoy for Atmospheric and Space Physics, 3665 Discovery Drive, Boulder CO, 80303, USA}

\begin{abstract}
	
	M-dwarf stars are prime targets for exoplanet searches because of their close proximity and favorable properties for both planet detection and characterization. However, the potential habitability and atmospheric characterization of these exoplanetary systems depends critically on the history of high-energy stellar radiation from X-rays to NUV, which drive atmospheric mass loss and photochemistry in the planetary atmospheres. With the Far Ultraviolet M-dwarf Evolution Survey (FUMES) we have assessed the evolution of the FUV radiation, specifically 8 prominent emission lines, including Ly$\alpha$, of M-dwarf stars with stellar rotation period and age. We demonstrate tight power-law correlations between the spectroscopic FUV features, and measure the intrinsic scatter of the quiescent FUV emissions. The luminosity evolution with rotation of these spectroscopic features is well described by a broken power-law, saturated for fast rotators, and decaying with increasing Rossby number, with a typical power-law slope of $-2$, although likely shallower for Ly$\alpha$. Our regression fits enable FUV emission line luminosity estimates relative to bolometric from known rotation periods to within $\sim$0.3 dex, across 8 distinct UV emission lines, with possible trends in the fit parameters as a function of source layer in the stellar atmosphere. Our detailed analysis of the UV luminosity evolution with age further shows that habitable zone planets orbiting lower-mass stars experience much greater high-energy radiative exposure relative the same planets orbiting more massive hosts. Around early-to-mid M-dwarfs these exoplanets, at field ages, accumulate up to 10-20$\times$ more EUV energy relative to modern Earth. Moreover, the bulk of this UV exposure likely takes place within the first Gyr of the stellar lifetime.
	
\end{abstract}

\section{Introduction}

The presence of self-sustained magnetic fields in low-mass stars has important consequences for their upper atmospheric structure and their high-energy radiative environments. Due to non-thermal magnetic heating processes \citep[see within][]{Linsky1980, Hall2008}, thought to be either wave dissipation \citep[e.g.,][]{Narain1996} or Joule heating from magnetic reconnection \citep[e.g.,][]{Klimchuk2006}, these low-mass stars exhibit significant temperature inversions in their outer atmospheres. These portions of the stellar atmosphere, the chromosphere, transition region and corona, are largely responsible for the entire high-energy spectrum from the near ultraviolet (NUV) to the X-rays in low-mass stars \citep[see][for recent review]{Linsky2017}. While much effort has been devoted to studying stellar chromospheres/coronae, the nature of these structures, the underlying processes that generate them, and their evolution remain poorly understood, especially for M-dwarf stars.

Addressing these questions has attained renewed urgency given the prevalence of terrestrial exoplanets orbiting M-dwarfs, with at least $\sim$20\% of these stars hosting an Earth-sized planet within their habitable zones \citep[HZ; e.g.,][]{Dressing2015, Vanderburg2020}. Moreover, the best systems for detailed atmospheric characterization with the \textit{James Webb Space Telescope} will be around nearby low-mass M-dwarfs \citep[e.g.,][]{Morley2017}. Understanding the high-energy emissions of these systems is crucial because of the role they play in planetary atmospheric mass-loss and photochemistry \citep[e.g.,][]{Scalo2007, Owen2012,Tian2015, Luger2015}. For example, for a Neptune/Earth-sized planet in the HZ of an M-dwarf, strong radiation shortward of $\lesssim$911 \AA\ (X-rays + extreme ultraviolet, XUV) can dictate the ultimate water content of the planet through atmospheric evaporation \citep{Owen2012}. Furthermore, the balance of NUV (1700-3200 \AA) to FUV (912-1700 \AA) emissions can determine equilibrium levels of abiotically produced O$_{2}$, complicating the search for biosignatures \citep[see within][]{Meadows2018}.

The completion of the MUSCLES Treasury Survey provided the first comprehensive constraints of M-dwarf X-ray and UV luminosities from panchromatic observations of exoplanet hosting M-dwarfs \citep{France2016,Youngblood2016,Loyd2016, Youngblood2017}. Their work further showed that the entire XUV and FUV broadband fluxes could be estimated based on a couple of FUV or NUV spectroscopic features \citep{France2016,Youngblood2017}. These measurements have since become important inputs into models of planetary atmospheres \citep[e.g.,][]{Gao2015,Ranjan2017}. However, interpreting future atmospheric observations, potential biosignature detections and the ability of such atmospheres to develop life, also depends on the evolution of the planetary atmosphere and hence the evolution of the incident radiation field. The MUSCLES survey focused on older field objects with confirmed planet detections, however, at early ages these M-dwarf hosts likely exhibited much stronger high-energy emissions \citep[e.g.,][]{Shkolnik2014a}, capable of desiccating terrestrial worlds early in their lifetimes \citep{Tian2015}. 

Understanding this evolution has been challenging because of the difficulty in determining stellar ages in the M-dwarf regime \citep{Guinan2016}. Although the ages of some objects can be determined through membership in clusters or young moving groups \citep[see within][]{Zuckerman2004}, the vast majority of M-dwarfs lack precise age determinations. Instead, rotation can be used as a proxy for stellar age, as in gyrochronology \citep[e.g.,][]{Skumanich1972,Barnes2003,Meibom2015,vanSaders2016}, although understanding the angular momentum evolution of M-dwarfs remains a topic of continued work \citep[][]{Barnes2010,Reiners2012b,Garraffo2015,Guinan2016,Garraffo2018}. Nevertheless, there is a fundamental physical interplay between stellar age, rotation and magnetic activity in low-mass stars \citep[e.g.,][]{Skumanich1972,Noyes1984,Vidotto2014b}, a consequence of the feedback between magnetic field generation in the internal dynamo and angular-momentum loss over time through coronally driven stellar winds.

Indeed, rotation-activity correlations have been used extensively as probes of these magnetic processes and their evolution in M-dwarfs, confirming the strong rotational dependence of the magnetic emissions, even across the fully convective boundary toward late M-dwarfs, and a saturation of the activity at fast rotation rates and young ages \citep[e.g.,][]{Pizzolato2003,Stelzer2013,Wright2016,Newton2017,Houdebine2017,AstudilloDefru2017, Shulyak2017, Wright2018}. These studies have focused predominantly on X-rays or optical emission lines like H$\alpha$ to trace the magnetic activity. The quiescent UV spectra of active M-dwarfs had been largely unexplored except for a few well known flare stars \citep{Rutten1989,Hawley1991,Ayres2003,Hawley2003b, Hawley2007}, limiting our ability to probe the rotational evolution of these features. Understanding these UV emissions is not only important for the incident radiation field impacting planetary atmospheres, but the various emission lines spanning a range of formation temperatures in the FUV spectra serve as unique probes throughout the different layers of the transition region ($T_{\mathrm{f}}\sim 10^{4.0-5.2}$), where the temperature in the outer atmosphere is rising rapidly from the chromosphere to the corona \citep[see within][]{Linsky2017}. 

These developments motivate the Far Ultraviolet M-dwarf Evolution Survey (FUMES) with the \textit{Hubble Space Telescope} (\textit{HST}) to examine the rotational evolution of the FUV spectral features in early-to-mid M-dwarfs, provide important benchmarks for their high energy emission over time, and provide constraints to the chromospheric/coronal structure of active low-mass stars. In this paper, the first of several, we focus on the quiescent emissions of our FUMES sample as a function of rotation/age. In Section~\ref{sec:sample}, we introduce the FUMES sample and assess their stellar properties. In Section~\ref{sec:uvdata}, we discuss our \textit{HST} observations and spectral measurements. In Section~\ref{sec:rotact}, we examine the rotation-activity correlations of UV emission in low-mass stars incorporating literature data. In Section~\ref{sec:age_evo}, we discuss the implications of our measurements for the temporal evolution of high-energy emissions around low-mass stars. Lastly, in Section~\ref{sec:conc} we provide our conclusions and summarize our findings in Section~\ref{sec:summary}.

\section{Sample and Stellar Properties}\label{sec:sample}

In contrast to previous samples of M-dwarf stars selected for UV observations, either exoplanet hosts (e.g., MUSCLES, \citealt{France2013}) or known flare stars \citep[e.g.,][]{Hawley2003b,Hawley2007}, the FUMES target list of 10 objects was chosen to span a range of rotation periods from $\sim$1-55 d, to trace the rotational evolution of low-mass stars and fill the gap in rotation parameter space between the active flare stars and the slowly rotating exoplanet hosts. The rotation period measurements were typically determined from photometric monitoring \citep[e.g.,][]{Messina2010,Newton2016} or long-term variability of optical emission lines \citep[e.g.,][]{SuarezMascareno2015}. We also included many objects with known ages from likely membership in a young moving group to facilitate age comparisons and provide multiple benchmarks for high energy emissions at early ages. We further focused on early-to-mid M-dwarf systems because thanks to the success of previous studies like MUSCLES \citep{France2016}, there already exists multiple benchmarks in UV emission for slowly rotating ($\gtrsim$70 d) objects in this spectral type range, allowing us to focus on providing the comparison with more active targets.

\begin{deluxetable*}{l c c c c c c c c c}
	\tablecaption{ IR Data Observing Log
		\label{tab:IR_obslog} }
	\tablehead{
		\colhead{Name} & \colhead{UT Date} & \colhead{Instrument} & Airmass & \colhead{A0 Calibrator} & \colhead{Seeing} & \colhead{Exp. Time (s)} & \colhead{SNR\tablenotemark{a}} & \colhead{Weather}
	}
	\startdata
	GJ 4334 & 2017-11-22 & TSPEC & 1.08 &  HD 240290 &  2'' & 20 & 160 & Windy, Clear \\
	GJ 49 & 2017-11-22 & TSPEC &  1.15 &  HD 5031 & 2''  & 8  & 380  & Windy, Clear \\
	HIP 112312 & 2017-11-05 & ARCoIRIS  & 1.17 &  HD 213044 &  1.8'' & 8 &  280 & Clear \\
	LP 247-13 &  2017-11-22 & TSPEC & 1.02 & HD 21038 & 2'' &  20  & 275 & Windy, Clear \\
	HIP 17695 & 2017-11-05 &   ARCoIRIS & 1.24  & HD 24003 & 1.8'' & 8 & 360 & Partly Cloudy \\
	HIP 23309 & 2017-11-06 & ARCoIRIS & 1.38 &  HD 32507 &  1.3'' & 10 &  500 & Clear \\
	CD-35 2722 & 2017-11-06 & ARCoIRIS & 1.38 & HD 42681& 1.3'' & 8  & 390 & Partly Cloudy\\
	GJ 410 & 2019-04-30 &  TSPEC &  1.04 & HD101060 &  1.2'' & 20-25  & 170 & Cloudy \\
	LP 55-41 & 2017-11-22  &  TSPEC &  1.28  &  HD 32781 & 2'' & 25  & 120 & Windy, Clear \\
	G 249-11 & 2017-11-22  &  TSPEC &  1.3  &  HD 32781 & 2'' & 25  & 180 & Windy, Clear  \\
	\enddata
	\tablenotetext{a}{This column reflects the typical signal-to-noise ratio in the H-band of each spectrum.}
\end{deluxetable*}

Before delving into the analysis of the UV emissions (Section~\ref{sec:uvdata}), it is important to estimate the physical properties of our targets. To these ends, we obtained infrared spectra of all 10 stars, which enabled us to consistently determine spectral types, measure metallicity indicators and compare estimates for effective temperature and bolometric luminosity \citep[e.g.,][]{Newton2015, Terrien2015}. These data are discussed in Section~\ref{sec:irdata}. The IR data were most valuable for estimating spectral types and ruling out the possible influence of cool unknown companions in the photometry. Ultimately, we rely on the results of Pineda et al. (\textit{submitted}) for the physical properties of the stars used in this work. Those methods are summarized below in Section~\ref{sec:starprop}.

\subsection{IR Data}\label{sec:irdata}

To measure the NIR spectra of our sample targets we used the TripleSpec (TSPEC) instrument \citep{Wilson2004} on the ARC 3.5 m telescope at the Apache Point Observatory and the similarly designed ARCoIRIS instrument\footnote{Instrument info can be found here: \href{http://www.ctio.noao.edu/noao/content/Arcoiris}{ARCoIRIS}} on the Blanco 4 m telescope at NOAO's Cerro Tololo Inter-American Observatory. A summary of these observations can be found in Table~\ref{tab:IR_obslog}. Both instruments, with a 1.1'' slit, provide $R$$\sim$3500 spectra continuously across NIR wavelengths in 5-6 echelle orders, with TSPEC spanning 0.95-2.46 $\mu$m and the updated ARCoIRIS design covering 0.80-2.47  $\mu$m. 

For all of our data we took the same observing approach, using an ABBA slit nod sequence with short exposures ($<$30 s), mitigating sky emission line variability, to provide clean sky subtraction from subsequent frames, and remaining on target over several tens of minutes to obtain a high signal-to-noise observation for each target (see Table~\ref{tab:IR_obslog}). We also observed a nearby A0 star close in time and at a similar airmass to provide a reference for flux and telluric calibration \citep{Vacca2003}.

\begin{figure}[tbp]
	\centering
	\includegraphics[width=0.5\textwidth]{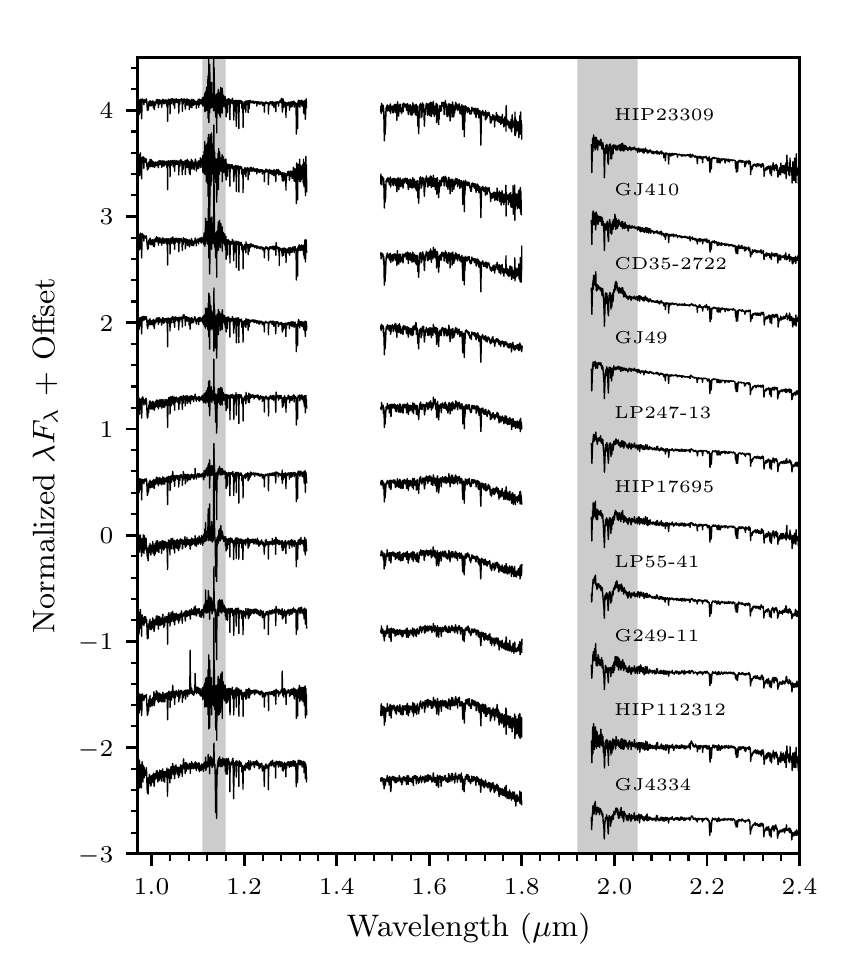} 
	\caption{The NIR spectra, shown here, allow us to provide spectral types for our sample in the NIR and assess physical properties using spectroscopic calibrations (see Section~\ref{sec:sample}). The spectra have been normalized in the $H$-band at the median flux between 1.55 and 1.75 $\mu$m. Regions of heavy atmospheric absorption have been removed, with gray shaded areas denoting wavelengths still influenced by some telluric water bands. }
	\label{fig:irspec}
\end{figure}

We reduced the data using modified versions of \texttt{Spextool} \citep{Cushing2004}, one for TSPEC and a separate one for ARCoIRIS.\footnote{For TSPEC see \href{https://www.apo.nmsu.edu/arc35m/Instruments/TRIPLESPEC/TspecTool/index.html}{TriplespecTool} and for the ARCoIRIS version, developed by Dr. Allers, see \href{http://www.ctio.noao.edu/noao/content/TS4-Data-Reduction}{TS4 Reduction}.} We summarize the data reduction procedure as follows. We first created the master flatfield for each observing night by median combining several dome lamp exposures, and subtracting off the median thermal contribution from dome exposures taken with the lamps off. The thermal contribution is most significant in the $K$-band. The wavelength calibration for each target was then determined from the median sky spectrum, created from each target's science observations.\footnote{Given an AB nod pair the sky contribution can be estimated as $S = 0.5[(A+B) - |(A-B)|]$. For short exposures, one needs to accumulate several frames to produce sufficient signal in sky emission.} Initial sky subtraction was performed from differencing AB nod pairs, from which we determined the object trace in each order. We extracted the spectrum in windows centered along the trace, applying the normalized flatfield and wavelength calibration. When variable cloud cover was evident, we also applied additional sky subtraction, removing a linear fit to the residual background. The spectra from individual frames were averaged together to increase the signal-to-noise ratio and then we used the similarly extracted A0 calibrator spectra to correct for telluric absorption and provide a flux calibration using \texttt{xtellcorr} \citep{Vacca2003}. We then merged the different echelle orders, averaging the spectra in overlapping wavelength regions to create the final spectrum of each target. The IR spectra of the FUMES sample is shown in Figure~\ref{fig:irspec}.

\subsection{Spectral Types}\label{sec:spt}

We determine the NIR spectral type classifications for the FUMES sample using the composite spectral standards compiled from multiple stars by \cite{Newton2014}.\footnote{\url{https://www.dartmouth.edu/~ernewton/nirsurvey.html}} These NIR spectra, taken with the NASA Infrared Telescope Facility (IRTF)/SpeX, are classified on the KHM system originally defined by \cite{Kirkpatrick1991,Kirkpatrick1995,Kirkpatrick1999} at red optical wavelengths. We first computed the H$_{2}$O-K2 index defined by \cite{RojasAyala2012} as a $K$-band feature sensitive to spectral type, and used the calibration from \cite{Newton2014} (their Equation~2) to convert this measurement to an initial type estimate.\footnote{We prefer the conversion from H$_{2}$O-K2 index to NIR spectral type from \cite{Newton2014} rather than \cite{RojasAyala2012} because the former is based on classifications using the entire NIR spectra.} We then used this classification to select a range of nearby spectral standards, spanning one type earlier and later, to compare against our observations to classify the spectra by eye using features across the entire infrared spectrum. This holistic approach helped mitigate potential feature mismatches introduced by metallicity differences between the targets and the standards by not relying on any single features in the spectra. We determined a single best type for each of the $YJHK$ bands and took the median as our best classification. 

Our NIR classifications for the FUMES sample are shown in Table~\ref{tab:fumes}, and we estimate spectral type uncertainties of half a subtype. We also include the literature optical spectral types for comparison, typically from the Palomar/Michigan State University (PMSU) survey \citep{Reid1995,Hawley1996} for the field objects, but from additional sources for the younger stars. As discussed in \cite{RojasAyala2012} and \cite{Newton2014}, the literature optical spectral types are often dependent on metallicity across the M1-M4 range, and we consider the NIR classifications to be the more consistent metrics. We similarly report the NIR/optical spectral types for the literature sample in Table~\ref{tab:fumes}. These classifications will be important in Section~\ref{sec:age_evo}.
 
A couple of objects deserve further attention. Since the \cite{Newton2014} standards do not include spectra for types earlier than M1, we also used the M0 and K7 standards within the IRTF spectral library \citep{Rayner2009} when classifying GJ410 and HIP23309. The M1 star, CD-35~2722, has an L4 companion that is 5 magnitudes fainter across the $JHK$-bands \citep{Wahhaj2011} at a separation of $\sim$70 AU. This companion is unresolved in our spectroscopic observations, but the $\sim$1\% contribution to the integrated NIR fluxes did not meaningfully distort the observed spectra. However, there was evidence for a slightly deeper 0.99 $\mu$m FeH feature than would be expected for an M1 dwarf, perhaps due to the strength of this feature in L-dwarf spectra \citep[e.g.,][]{Kirkpatrick2005}. The young star HIP~112312 has an optical classification as a subgiant from \cite{Torres2006}. Without subgiant standards with which to compare in the NIR, we cannot confirm this classification with our observations. For this star we also compared the M-giant IRTF NIR standards \citep{Rayner2009}, finding that the HIP~112312 NIR spectrum largely agrees with the dwarf sequence except for deeper CO lines in the $K$-band, which are very prominent in the M-giant spectra, reflecting the relatively low gravity of this young object. The other young FUMES stars (see Table~\ref{tab:fumes}) showed spectra consistent with the dwarf standards.

\begin{deluxetable*}{l c c c c c c c c}
	\tablecaption{ UV Sample\tablenotemark{a}
		\label{tab:fumes} }
	\tablehead{
		\colhead{Name}  & \colhead{SpT} & \colhead{$L_{\mathrm{bol}}$}  & \colhead{Mass} & \colhead{Radius} & \colhead{$T_{\mathrm{eff}}$}&   \colhead{Rot. Period\tablenotemark{b}}  & \colhead{References\tablenotemark{c}} \\
		& Opt/NIR  & \colhead{(10$^{31}$ erg s$^{-1}$)} & \colhead{($M_{\odot}$)} & \colhead{($R_{\odot}$)} & \colhead{(K)}&   \colhead{(d)}  & 	
	}
	\startdata
	G 249-11  &  M4\tablenotemark{d}/M4 & $2.59 \pm0.07$ & $0.237 \pm 0.006 $ &$ 0.255 \pm 0.010$  & $3277 \pm_{68}^{71}$ & 52.76 & 14, 14, 12 \\
	HIP 112312\tablenotemark{e} & M4IV/M4.5 &  $ 16.6 \pm 0.5$& $ 0.247 \pm_{ 0.015}^{ 0.018 }$ & $ 0.688 \pm_{ 0.016}^{ 0.015 }$ & $ 3173 \pm_{ 22}^{ 25}$  & $2.355 \pm 0.005$& 19, 14, 9 \\ 
	GJ 4334  & M4.5/M5 &  $3.68 \pm0.09$ & $0.294 \pm 0.007 $ &$ 0.307 \pm 0.012 $  & $3260 \pm _{66}^{70} $  & 23.54 & 15, 14, 12\\
	LP 55-41  & M3.5\tablenotemark{d}/M3 & $8.10 \pm0.22$ & $0.41 \pm 0.01 $ &$ 0.416 \pm 0.017$  & $3412 \pm _{73}^{ 75} $ & 53.44 & 14, 14, 12\\
	HIP 17695\tablenotemark{f}  & M3/M4 & $11.5 \pm0.2$ & $0.435 \pm_{ 0.014}^{ 0.011}$ &$0.502 \pm_{ 0.007}^{ 0.008}$ & $ 3393 \pm_{ 21}^{ 20}$ & $3.87 \pm 0.01$& 1, 14, 9 \\ 
	LP 247-13  & M2.7/M3.5 & $12.68 \pm 0.35$ & $0.495 \pm0.013 $ &$ 0.49 \pm0.02$  & $3511 \pm _{76}^{ 79}$ & 1.289& 16, 14, 5 \\
	GJ 49  & M1.5/M1 & $18.7\pm0.4$ & $0.541 \pm 0.015 $ &$ 0.534 \pm 0.023$  & $3713  \pm_{81}^{86} $& $18.6 \pm 0.3$&  15, 14, 4 \\
	GJ 410 &  M0/M0.5 & $21.3 \pm 0.5$ & $0.557 \pm 0.015 $ &$ 0.549 \pm 0.024$  & $3786 \pm_{83}^{89} $ & 14.0 & 15, 14, 4 \\  
	CD-35 2722 \tablenotemark{fg} & M1/M1 & $ 21.0 \pm0.3 $ & $ 0.572 \pm 0.002$ & $ 0.561 \pm0.003 $ &  $3727 \pm_{ 6}^{8 }$ & $1.717 \pm 0.004$ & 19, 14, 9 \\ 
	HIP 23309\tablenotemark{e}  & M0/M0 & $68.2 \pm _{1.1}^{1.2}$& $0.785 \pm_{0.010}^{0.009}$ & $0.932 \pm 0.014 $ &  $3886 \pm_{27}^{28}$ & $8.6 \pm 0.07$ & 19, 14, 9 \\  
	\hline
	Prox. Cen. *  & M5.5/- & $0.5997 \pm_{0.0075}^{0.0076}$ & $0.123 \pm 0.003 $ &$ 0.147 \pm 0.005$  & $2992 \pm_{47}^{49} $ & 88.977 & 6, - , 13 \\ 
	GJ 1061   &  M5.5/- & $0.628  \pm 0.014$ & $0.125 \pm 0.003 $ &$ 0.152 \pm 0.007$  & $2976 \pm _{72}^{ 69} $  & 133 & 6, - , 2 \\   
	GJ 1214  & M4.5/M4 &  $1.343 \pm_{0.034}^{0.036}$ & $0.181 \pm0.005$ &$ 0.204 \pm_{0.0084}^{0.0085}$  & $3111 \pm_{66}^{69} $ & $125 \pm 5$ & 15, 11, 8 \\  
	GJ 1132   & M3.5/- & $1.67 \pm 0.05$ & $0.194 \pm 0.005 $ &$ 0.215 \pm 0.009$  & $3196  \pm_{69}^{72} $  & 129.15 & 6, - ,13 \\  
	Gl 213  & M4/M4 & $2.41\pm 0.04$ & $0.218 \pm 0.005 $ &$ 0.238 \pm 0.009$  & $3334 \pm _{63}^{67} $ & 170 & 15, 11, 2\\
	GJ 628 * & M3.5/M3 & $4.21 \pm 0.04$ & $0.304 \pm 0.007 $ &$ 0.319 \pm 0.007$  & $3307 \pm_{36}^{38} $  & $119.3 \pm 0.5$&  15, 11, 18 \\  
	Gl 581 * & M3/M2 &   $4.56 \pm 0.04$ & $0.307 \pm 0.007 $ &$ 0.310 \pm 0.008$  & $3424 \pm_{42}^{43} $  & $132.5\pm6.3$ & 15, 11, 17 \\ 
	YZ CMi &  M4.5/M5 & $4.35 \pm_{0.14}^{0.15}$ & $0.316 \pm 0.008 $ &$ 0.328 \pm0.013$  & $3293 \pm_{71}^{ 74}$  & $2.7758 \pm 0.0002$& 15, 11, 10 \\
	EV Lac &  M3.5/M3 &  $4.88 \pm0.15$ & $0.320 \pm 0.008 $ &$ 0.331\pm 0.013$  & $3370 \pm _{70} ^{75} $  & $4.3715 \pm 0.0002$ & 15, 11, 10\\
	GJ 667C  & M1.5/- & $5.51 \pm 0.13$ & $0.327 \pm 0.008$ &$ 0.337 \pm 0.014$  & $3443 \pm_{71}^{75} $ & $103.9\pm0.7$ & 6, - , 17 \\
	GJ 876 *    & M4/M3 & $5.01 \pm0.04$ & $0.346 \pm 0.007 $ & $ 0.372 \pm 0.004 $  & $3201 \pm _{19}^{20} $ & $87.3\pm5.7$ & 15, 11, 17 \\  
	Gl 821 & M1/- &  $6.98 \pm 0.09$ & $0.355 \pm0.009$ &$ 0.363 \pm_{0.014}^{0.015}$  & $3512 \pm _{69}^{73}$ & 107 & 15, - , 2 \\
	Gl 436 * & M2.5/M3 & $9.42 \pm 0.11$ & $0.425 \pm 0.009 $ &$ 0.432 \pm 0.011$  & $3477 \pm_{44}^{46}$ & $44.09\pm 0.08$ & 15, 11, 3 \\
	AD Leo  & M3/M3 &  $8.7 \pm0.1 $ & $0.426 \pm0.010 $ &$ 0.43 \pm 0.02 $  & $3425 \pm_{68}^{69}$ & $2.2399 \pm 0.0002$ & 15, 11, 10 \\ 
	GJ 832  & M1.5/- &   $10.6 \pm 0.3$ & $0.441 \pm 0.011  $ &$ 0.442 \pm 0.018 $  & $3539 \pm_{74}^{79} $ & $45.7\pm9.3$ & 6, - , 17 \\ 
	Gl 887 *  &  M0.5/& $14.08 \pm 0.22 $ & $0.48 \pm 0.01$ &$ 0.474 \pm 0.008 $  & $3672 \pm_{34}^{36} $ & 33 & 6, - , 2 \\
	GJ 176 *  & M2/M2 & $13.46 \pm 0.12$ & $0.485 \pm 0.012$ &$ 0.474 \pm 0.015$  & $3632 \pm_{56}^{58}$ & $39.3 \pm 0.1$ & 15, 11, 17\\
	AU Mic\tablenotemark{e}  & M0/- &  $ 37.7 \pm_{0.7}^{0.8}$ & $ 0.667 \pm_{0.008}^{0.006} $ & $0.798 \pm_{0.013}^{0.014}$& $ 3619 \pm_{25}^{23}$ & 4.86& 6, - , 7
	\enddata
	\tablenotetext{a}{Stellar properties are from Pineda et al.\ (\textit{in prep}), quoting medians and the central 68\% confidence interval, see Section~\ref{sec:starprop}. The horizontal division separates the new FUMES (\textit{top}) targets from the literature stars (\textit{bottom}).}
	\tablenotetext{b}{Rotation periods are taken from the literature, typically from either photometric variations or long-term monitoring of periodic emission lines. Uncertainties are as reported in the literature if available.}
	\tablenotetext{c}{References in order denote source for optical spectral type, infrared spectral type and rotation period.}
	\tablenotetext{d}{Optical spectral types were not available in the literature, so we used our NIR classification and Eqn.\ 3 from \cite{Newton2014} with the metallicity determined from their calibration of the equivalent widths of the Na doublet at 2.2 $\mu$m (their Equation 10).}
	\tablenotetext{e}{HIP 23309, HIP112312, and AU Mic are members of $\beta$ Pic which has a mean age of $\sim$24 Myr \citep{Bell2015}.}
	\tablenotetext{f}{CD-35 2722, and HIP17695 are members of AB dor which has a mean age of $\sim$150 Myr \citep{Bell2015}.}
	\tablenotetext{g}{CD-35 2722 is a binary with a substellar L4 companion with separation of $\sim$70 AU \citep{Wahhaj2011}.}	
	\tablenotetext{*}{Star names listed with an asterisk have measured angular diameters from interferometric measurements \citep{vonBraun2011,Boyajian2012,vonBraun2014,Kane2017}, incorporated in the radius estiamtes, see Pineda et al.\ (\textit{in prep}).}
	\tablenotetext{}{\textbf{References.} -- (1) \cite{AlonsoFloriano2015}, (2) \cite{AstudilloDefru2017}, (3) \cite{Bourrier2018}, (4) \cite{Donati2008}, (5) \cite{Hartman2011}, (6) \cite{Hawley1996}, (7) \cite{Kuker2019}, (8) \cite{Mallonn2018}, (9) \cite{Messina2010}, (10) \cite{Morin2008}, (11) \cite{Newton2014}, (12) \cite{Newton2016}, (13) \cite{Newton2018}, (14) This Work, (15) \cite{Reid1995}, (16) \cite{Shkolnik2009}, (17) \cite{SuarezMascareno2015}, (18) \cite{SuarezMascareno2016}, (19) \cite{Torres2006}. }
\end{deluxetable*}

\subsection{Stellar Properties}\label{sec:starprop}

Our aim in this paper is to analyze the relation between FUV emissions and the physical properties of low-mass stars. To these ends, we desired self-consistent properties, mass, radius, bolometric luminosity, and effective temperature for each FUMES target, and any suitable additional targets found in the literature. Consistently determined properties are crucial to mitigate potential systematic effects introduced by relying on an assortment of eclectic literature determinations for these stellar properties. Although our IR spectra allowed us to utilize spectroscopic property calibrations \citep[e.g.,][]{Mann2015,Newton2015,Terrien2015}, such data were not uniformly available for both the FUMES and literature targets, and we thus rely instead on the largely photometric results from our companion paper summarized below (Pineda et al.\ \textit{in prep}). The corresponding stellar properties are shown in Table~\ref{tab:fumes}.

\subsubsection{Field Stars}

To determine the properties of low-mass field stars Pineda et al.\ (\textit{in prep}) use a Bayesian framework to combine multiple empirical calibrations, largely photometry based, to jointly constrain mass, radius, bolometric luminosity, and derive the stellar effective temperature. Their methods fully incorporate measurement uncertainties, and intrinsic scatter within the utilized calibrations, and produce well defined joint posterior distributions for the full set of physical properties. Pineda et al.\ (\textit{in prep}) jointly uses the mass-luminosity relation of \cite{Mann2019}, the bolometric correction calibration of \cite{Mann2015}, and a new mass-radius relation valid across $0.1$-$0.7$ $M_{\odot}$ with 3.1\% uncertainties at fixed mass specifically developed in that work. 

The Bayesian framework further allowed them to freely incorporate additional measurements whenever available, such as bolometric fluxes, or angular diameters from interferometry. Many of the individual objects in the sample (see Table~\ref{tab:fumes}) had these additional measurements which largely improved the precision of the physical properties of a given sample object. Full details on these methods are available in Pineda et al.\ (\textit{in prep}).

As compared to literature estimates of the M-dwarf ensemble analyzed in Pineda et al.\ (\textit{in prep}), their methods yielded a stellar sequence with less scatter, consistent property estimates for objects with interferometric angular diameter measurements, and stellar densities consistent with independent data inferred from exoplanetary transits of low eccentricity planets.

\subsubsection{Young Stars}

Of the objects shown in Table~\ref{tab:fumes}, five are high probability members of known young moving groups: HIP~112312, HIP~17695, CD-35~2722, HIP~23309, and AU Mic (Pineda et al.\ \textit{in prep}). Because the available empirical calibrations are only applicable to field age stars, these five young objects required a different approach for estimating their stellar properties. For these stars, Table~\ref{tab:fumes} also quotes the stellar model based results from Pineda et al.\ (\textit{in prep}). We summarize their methods as follows. 

Within a Bayesian framework, Pineda et al.\ (\textit{in prep}) couple stellar evolutionary models with spectral energy distribution fitting of model spectra using blue optical to far infrared photometry. Using Monte Carlo sampling, a given mass and age within the evolutionary model defines the corresponding bolometric luminosity and radius, and thus the effective temperature and gravity of a sample point. These properties are then used for interpolation of a model atmosphere grid and generation of synthetic photometry for comparison with the observed data points. The best-fit stellar properties are those that best reproduce the photometry consistently within the evolutionary models. This approach self-consistently produces parameter estimates for all of the properties with well defined posterior distributions for each. For the five young objects, Table~\ref{tab:fumes} reproduces the parameter results using magnetic stellar evolutionary models \citep{Feiden2013,Feiden2014,Feiden2016}. Full details of these methods and analysis of likely systematic effects in the model choices are explained in Pineda et al.\ (\textit{in prep}). 

Those properties most directly constrained by the SED fitting, namely $L_{\mathrm{bol}}$, $T_{\mathrm{eff}}$, and $R$ are consistent within errors to literature values (Pineda et al.\ \textit{in prep}). The young stars used in this work are all active M-dwarfs, and we discuss further how their modeling choices and thus the inferred mass impact our rotation activity analysis in Section~\ref{sec:rotact_fitting}.

\section{Far Ultraviolet Emissions}\label{sec:uvdata}

\begin{figure*}[tbp]
	\centering
	\includegraphics{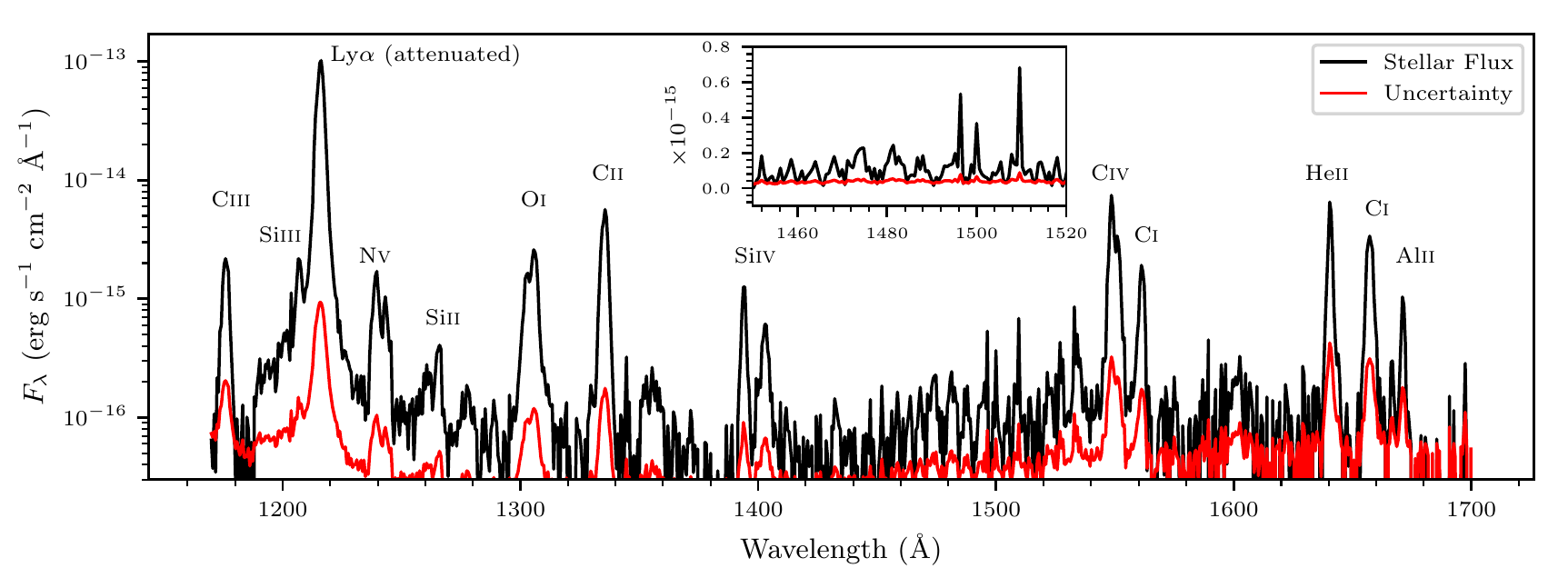} 
	\caption{The FUV spectra of M-dwarfs show several discrete emission features spanning lines that probe different temperatures in the transition region from C\textsc{ii} to N\textsc{v}. This example illustrates a typical spectrum in the FUMES data set (GJ49, T$_{\mathrm{exp}} = 5585$ s) spanning 1170-1700 \AA\ using \textit{HST}-STIS G140L. The inset figure shows the low-level continuum emission from 1450-1520 \AA. }
	\label{fig:quiesUV_G140L}
\end{figure*}

\begin{figure*}[tbp]
	\centering
	\includegraphics[height=0.61\textheight]{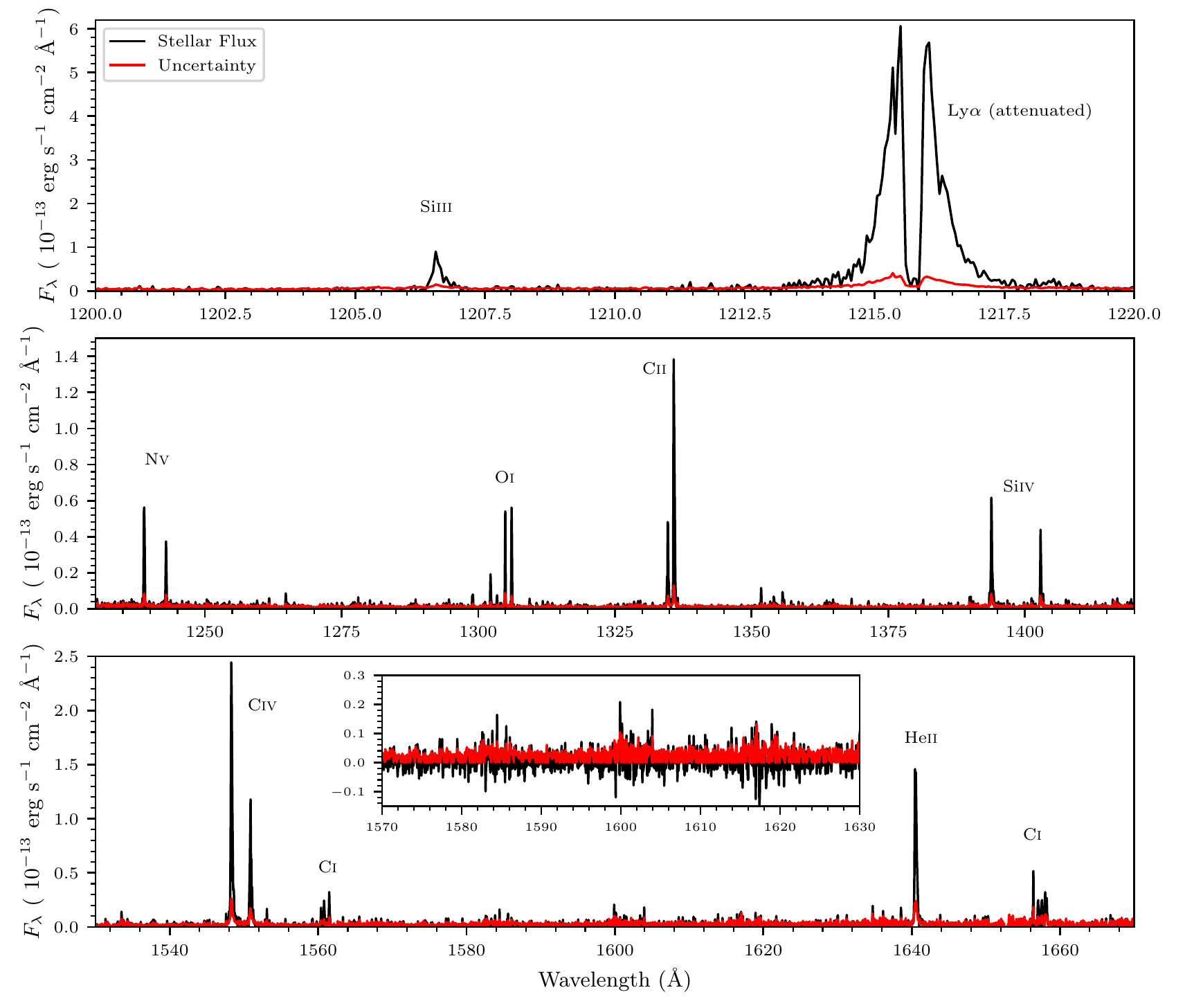} 
	\caption{This \textit{HST}-STIS E140M data for HIP~17695 (T$_{\mathrm{exp}} = 1974$ s) shows the discrete FUV emission features of M-dwarfs prominently, with no continuum FUV flux detected. Important emission features are labeled, with the multiplet nature of some of the lines more clearly resolved. The inset figure shows an example continuum emission region from 1570-1630 \AA.}
	\label{fig:quiesUV_E140M}
\end{figure*}

\subsection{HST Data}

We used the Space Telescope Imaging Spectrograph (STIS) on \textit{HST} to measure the FUV spectra of our FUMES sample through program HST GO-14640 (PI - Pineda). We show a summary of our observations in Table~\ref{tab:UV_obslog}. Typically data were taken using the G140L grating with the FUV-MAMA detector, providing a typical resolving power of $\sim$1000 across 1150-1730 \AA. For the brighter targets we used the echelle grating E140M instead, with similar wavelength coverage and a resolving power of $\sim$45,000. Although the lower resolution of the G140L grating does not permit measurement of the typical FUV line-widths, and doublet lines are often blended, our focus in this study is simply the total flux in the strongest FUV lines, which we can measure using G140L for fainter targets than is possible at high resolution using STIS. All observations were taken in photon counting TIME-TAG mode, providing high time resolution to our program. We focus on the quiescent emissions of the FUMES sample here, integrated across all exposures and orbits, after removing flares, with the time resolved analysis to be presented in a follow-up paper. 

To extract the spectra from the FUV-MAMA photon counting detector we used the \texttt{spectralPhoton} routines from R.\ O.\ P.\ Loyd previously used in the \textit{HST}-STIS and \textit{HST}-COS analyses of M-dwarf UV spectra \citep[e.g.,][]{Loyd2014,Loyd2018}.\footnote{The codes are available on GitHub: \url{https://github.com/parkus/spectralPhoton}. } To summarize, the reduction procedure sums the photons in a narrow ribbon identified as the stellar trace subtracting off a background count rate determined from the signal in offset regions with the same wavelengths. The reduction then uses the flux calibration from the full exposures to convert the photon counts to a calibrated spectrum. The advantage of \texttt{spectralPhoton} over the standard output from the STScI pipelines is that it allows for the definition of custom wavelength extraction regions, trace locations, and integrated time intervals. This was important because we found for the faint targets LP~55-41 and G~249-11 that the standard pipeline products did not correctly identify the stellar trace. Additionally, some of the targets flared during our observations, which we removed by manually identifying when the flares took place, and defining custom time intervals during the exposures for extraction of photons corresponding to the quiescent emission spectra.

\begin{deluxetable}{l c c c c}
	\tablecaption{ UV Data Summary
		\label{tab:UV_obslog} }
	\tablehead{
		\colhead{Name} & \colhead{UT Date} & \colhead{Grating} & Aperture\tablenotemark{a} & \colhead{Orbits} 
	}
	\startdata
	GJ 4334 & 2017-09-20 & G140L & 0.2" &  2 \\
	GJ 49 & 2017-09-20 & G140L &  0.2" &  2 \\
	HIP 112312 & 2017-08-22 & E140M  & 0.2" &  1 \\
	LP 247-13 &  2017-09-13 & G140L & 0.2" & 1 \\
	HIP 17695 & 2017-12-27 &   E140M  & 0.2"  & 1 \\
	HIP 23309 & 2017-11-24 & G140L & 0.1" &  1\\
	CD-35 2722 & 2017-09-26 & G140L & 0.2" & 1\\
	GJ 410 & 2017-12-18 &  G140L &  0.1" & 1 \\
	LP 55-41 & 2017-09-13  &  G140L &  0.2"  &  3 \\
	G 249-11 & 2017-09-10  &  G140L &  0.2" &  3 
	\enddata
	\tablenotetext{a}{Aperature denotes the width of the STIS slit.}
\end{deluxetable}

\begin{deluxetable}{l c c}
	\tablecaption{Model Line Fitting \label{tab:linefits} }
	\tablehead{
		\colhead{Line ID} & \colhead{G140L} & \colhead{E140M}
	}
	\startdata
	Ly$\alpha$ \tablenotemark{a} & Voigt + ISM & Voigt + ISM \\
	He\textsc{ii} & Gaussian + Continuum & Gaussian \\
	C\textsc{ii}\tablenotemark{b}& Gaussian + Continuum & Voigt \\
	C\textsc{iii} & Gaussian+ Continuum & 6x Gaussian \\
	C\textsc{iv} & 2x Voigt + Continuum & 2x Voigt \\
	N\textsc{v} & 2x Voigt + Continuum & 2x Voigt \\
	Si\textsc{iii} \tablenotemark{a} &  Gaussian + Ly$\alpha$ & Gaussian \\
	Si\textsc{iv} &  2x Voigt + Continuum & 2x Voigt \\
	\enddata
	\tablenotetext{a}{For G140L data the Ly$\alpha$ line is jointly fit, accounting for ISM absorption, with Si\textsc{iii} (Youngblood et al.\ \textit{accepted}), but for E140M data the Si\textsc{iii} line is fit independently. }
	\tablenotetext{b}{For the E140M data we fit only the single redward line in the C\textsc{ii} doublet, $\lambda$ 1335.71 \AA, since the $\lambda$ 1334.54 \AA\ line is affected by the ISM; for G140L data, these lines are blended, see Table~\ref{tab:fluxes}.}
\end{deluxetable}

The majority of our targets were observed in the wider 0.2" STIS slit mitigating potential slit-loss effects, affecting the flux calibration, with poor acquisition, target centering or guiding. For one target, HIP~23309, observed with the narrower 0.1" slit for bright object protection considerations, the time series analysis showed a long-term trend in the photon counts over the course of the orbit (this effect was not seen in the data for GJ~410, the other program observation employing the 0.1" slit). To correct for this we fit a third-order polynomial to the trend (in the 10 s binned light curve) to divide it out and scaled to the average peak count rate. The quiescent spectrum was subsequently extracted from the appropriately scaled spectrum. For each target we extracted spectra from each exposure (usually 1 per orbit), and then co-added them together rebinning onto a constant linear wavelength grid. For the echelle spectra we also merged the wavelength regions in which the echelle orders overlap, averaging flux points falling within the same 0.05 \AA\ bins, and preserving the integrated flux of the spectral bin. We show an example G140L spectrum of one of our targets in Figure~\ref{fig:quiesUV_G140L}, and an echelle E140M spectrum in Figure~\ref{fig:quiesUV_E140M}.

By using STIS, we were able to observe Ly$\alpha$ in all of our objects, as well as all of the significant FUV emission lines He\textsc{ii} $\lambda1640.4$ \AA, C\textsc{ii} $\lambda \lambda1334.54, 1335.71$ \AA, C\textsc{iii} $\lambda1175.7$ \AA, C\textsc{iv} $\lambda\lambda 1548.19,1550.78$ \AA, Si\textsc{iii} $\lambda1206.5$ \AA, Si\textsc{iv} $\lambda\lambda 1393.76,1402.77$ \AA, and N\textsc{v} $\lambda \lambda 1238.82, 1242.806$ \AA. These lines span mean formation temperatures $\log T$ (K) $= $ 4.5-5.2, probing the transition region of the stellar coronal atmosphere.\footnote{The mean formation temperatures listed in Table~\ref{tab:UVcorr} depend on the differential emission measure and may differ (significantly) from the peak formation temperature.} We focused on these lines as both the most prominent in the data, being well measured both in the FUMES targets and the literature sample.

\subsection{FUV Line Fitting}\label{sec:linefits}

To measure the target emission line fluxes, we used PyMC3 to fit the observed line shapes, typically with either a Voigt or Gaussian profile convolved with the instrument line spread function.\footnote{ The LSFs have spectral resolutions of 1.7-1.5 pixels at FWHM for STIS-G140L and 1.4-1.3 pixels at FWHM for E140M. We used the LSFs obtained from the STSci STIS instrument documentation. } Although we must assume a particular profile shape, these shapes are well motivated physically, and this approach has several advantages to simply summing the flux in the appropriate region for each line. At high resolution, we can measure the line widths, compare the emission core to the line wings, and examine centroid offsets, indicative of the stellar radial velocity. We can also simultaneously fit a continuum level below each line and incorporate the uncertainty in that estimate to our reported emission line fluxes. This effect was especially important in the G140L data, where there appeared to be some continuum level below many of the lines. For example in the region around N\textsc{v}, some of the low-level extended Ly$\alpha$ wing emission could be seen. In the E140M data, no continua were evident. We quantify continuum levels in Section~\ref{sec:UVcont}. Additionally, the line fitting is robust to potentially badly characterized data points because it allows us to include a fit scatter term to the data to account for underestimated errors.

\begin{figure*}[htbp]
	\centering
	\includegraphics[]{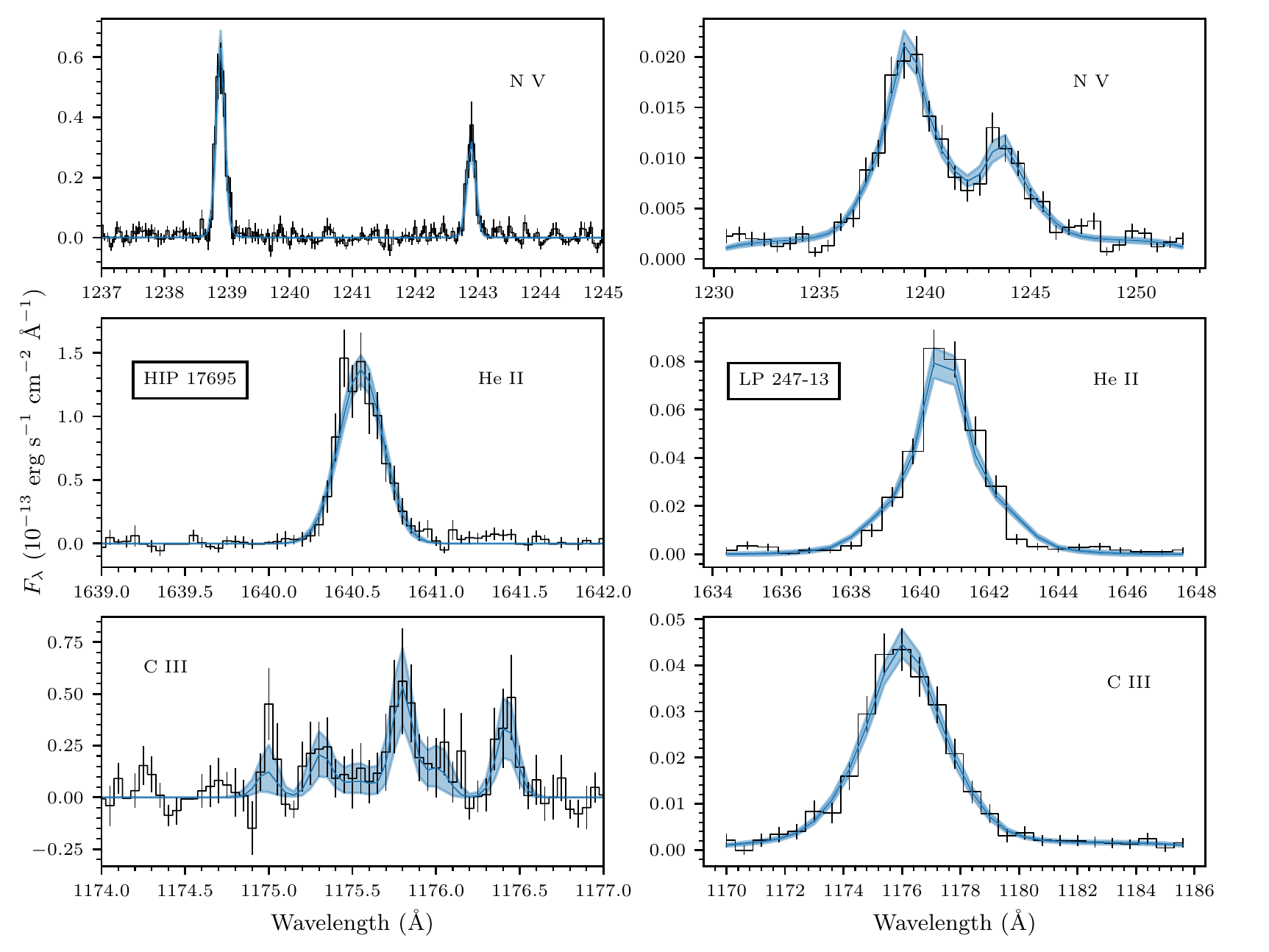} 
	\caption{We show example results displaying our typical line fits for N\textsc{v} (\textit{top}), He\textsc{ii} (\textit{middle}), and C\textsc{iii} (\textit{bottom}), with both E140M high resolution data (\textit{left}), and G140L low resolution data (\textit{right}). Data (black) are from HIP~17695 (\textit{left}) and LP247-13 (\textit{right}), with the shaded region (blue) denoting the 1$\sigma$ confidence interval for the model about its median. Emission lines that are separable at high resolution can become blended in the lower resolution data.}
	\label{fig:linefits}
\end{figure*}

\begin{deluxetable*}{l c c c c c c c c}
	\tablecaption{FUV Line Fluxes\tablenotemark{a} \label{tab:fluxes}  }
	\tablehead{
		\colhead{Name}  & \colhead{Ly$\alpha$\tablenotemark{b}} & \colhead{He\textsc{ii}}  & \colhead{C\textsc{ii}\tablenotemark{c}} & \colhead{C\textsc{iii}} & \colhead{C\textsc{iv}}&   \colhead{N\textsc{v}} & \colhead{Si\textsc{iii}\tablenotemark{d}}  & \colhead{Si\textsc{iv}} 
	}
	\startdata
	G 249-11  &  --- & $<0.012$ & $<0.012$ & $<0.014$  & $0.013 \pm 0.003 $  & $0.0065 \pm 0.0012$ & --- & $<0.01$ \\
	HIP 112312 & $214 \pm_{6}^{7}$ & $ 9.14 \pm 0.48  $ & $ 3.97 \pm 0.18  $  & $ 3.01\pm_{0.47}^{0.53}  $  & $ 11.87 \pm_{0.53}^{0.52}  $  & $ 1.90 \pm_{0.13}^{0.14}  $ & $ 1.67 \pm_{0.28}^{0.29}  $  & $ 2.20 \pm_ {0.14}^{0.13}  $  \\ 
	GJ 4334  & $7.03\pm^{3.93}_{1.16}$&  $ 0.280 \pm_{0.035}^{0.036}  $  &  $ 0.68 \pm 0.03  $  & $ 0.15 \pm 0.01 $ & $ 1.62 \pm_{0.06}^{0.07}  $  & $ 0.337 \pm 0.024 $   & $0.211\pm^{0.017}_{0.018}$& $ 0.36 \pm 0.02 $\\
	LP 55-41  & --- &  $<0.018$& $0.008 \pm   0.002$& $<0.019$& $<0.034 $ &  $0.014 \pm 0.004$  & --- & $<0.003$ \\
	HIP 17695  & $110 \pm_{3}^{2} $ & $ 4.50 \pm_{0.30}^{0.29}  $  & $ 2.98 \pm 0.14 $  & $ 2.44 \pm_{0.38}^{0.41}  $ & $ 8.88 \pm_ {0.39}^{0.40}  $ & $ 1.74 \pm 0.11  $ & $ 1.82 \pm_ {0.18}^{0.19}  $  & $ 2.14 \pm_  {0.14}^{0.15}  $  \\ 
	LP 247-13  & $442\pm^{374}_{103}$ & $ 2.04 \pm 0.13 $ & $ 1.77 \pm 0.07  $ & $ 1.48 \pm 0.081  $ & $ 3.57 \pm 0.11  $  & $ 0.918 \pm_{0.052}^{0.056}  $ & $0.731\pm^{0.041}_{0.040}$& $ 0.60 \pm 0.04  $  \\
	GJ 49  & $249\pm^{58}_{93}$ & $ 1.38 \pm_{0.07}^{0.08}  $  & $ 1.84\pm 0.09 $& $ 0.686 \pm 0.035 $  & $ 2.28 \pm0.07 $ & $ 0.674 \pm_{0.031}^{0.032}  $ &  $0.550\pm^{0.021}_{0.013}$& $ 0.47 \pm 0.02  $\\
	GJ 410 &  $158 \pm_{53}^{163} $&  $ 2.54 \pm 0.17  $ & $ 2.68 \pm 0.09 $ &  $ 0.920 \pm 0.099  $ & $ 2.99 \pm_{0.16}^{0.17}  $   & $ 0.609 \pm 0.051 $  & $0.876 \pm^{0.072}_{0.074} $& $ 1.06 \pm 0.06  $  \\  
	CD-35 2722  & $25.0 \pm^{2.2}_{1.3}$  & $ 3.85 \pm 0.18  $ & $ 2.90 \pm_{0.11}^{0.12}  $ & $ 1.74 \pm 0.09 $  & $ 4.49 \pm 0.12 $  & $ 0.90 \pm 0.05 $  &$1.32 \pm 0.06$& $ 1.29 \pm 0.05  $ \\ 
	HIP 23309  & $54.0\pm^{2.0}_{1.9} $& $ 6.60 \pm_{0.28}^{0.29}  $ & $ 3.82 \pm0.09$ & $ 2.23 \pm_{0.12}^{0.13}  $ &  $ 6.87 \pm 0.19  $  &  $ 1.54 \pm 0.06  $ &$ 2.01 \pm^{0.09}_{0.08} $ & $ 2.42 \pm 0.08  $\\  
	\enddata	
	\tablenotetext{a}{All fluxes are as observed at Earth, in units of $10^{-14}$ erg s$^{-1}$ cm$^{-2}$, and the flux from multiplet lines are added together, unless indicated otherwise. Uncertainties reflect the central 68\% confidence interval about the median.}			
	\tablenotetext{b}{Ly$\alpha$ fluxes as reconstructed in FUMES paper II (Youngblood et al.\ \textit{accepted}). For G249-11 and LP 55-41, the weak emission precluded a confident fit.}			
	\tablenotetext{c}{For HIP112312 and HIP17695 with E140M data we report the flux of just the $\lambda$ 1335 \AA\ line. For the rest of the sample with G140L data, we report the total flux of the blended doublet including ISM absorption. We estimate that these measurements should be corrected upward by $\sim$7\% to account for the influence of the ISM (see Section~\ref{sec:linefits}).}			
	\tablenotetext{d}{Si\textsc{iii} fluxes are from FUMES paper II (Youngblood et al.\ \textit{accepted}), jointly fit with Ly$\alpha$ in G140L STIS data.}			
	
\end{deluxetable*}

We summarize the model profiles fit to the data in Table~\ref{tab:linefits}, where the number in front of the profile shape indicates that multiple lines were fit simultaneously, and the continuum component is described by a line with an unknown slope and constant offset. When multiple lines are fit simultaneously, as in the C\textsc{iv} doublet, the widths of the Gaussian and Lorentzian components in the Voigt profile shape are constrained to be identical in each line. This reduces the number of free parameters in the fit and is justified because each line comes from the same species, forming under similar temperature conditions with only a small difference in energy levels between the transitions. As priors in our Bayesian regression analysis, we used uniform distributions for the line flux, center, and when necessary the continuum offset and slope. For the width parameters in the Gaussian and Lorentzian components we used a weakly informative Half-Cauchy distribution, with a characteristic width of 100 km s$^{-1}$.\footnote{For the non-negative Half-Cauchy distribution, half of the probability mass density is contained at parameter values less than the characteristic width with the distribution tail decaying in a Lorentzian fashion.} This choice had a negligible affect on the parameter posterior distribution, but greatly improved MCMC sampling efficiency and convergence relative to a log uniform prior within PyMC3. 

We show examples of our fit line profiles in Figure~\ref{fig:linefits}, with the results for the total line fluxes shown in Table~\ref{tab:fluxes}. The Ly$\alpha$ emission is the focus of paper II (Youngblood et al.\ \textit{accepted}), and we reproduce those results here. For the two targets with E140M data, we fit a single Gaussian line for Si\textsc{iii}, to be consistent with the approach for the G140L data, that required simultaneous fitting with Ly$\alpha$ (Youngblood et al.\ \textit{accepted}). At the higher resolution, the low flux C\textsc{iii} lines are somewhat blended but distinguishable, so we fit six components to the multiplet, but only a single Gaussian feature at low-resolution. The He\textsc{ii} multiplet, by contrast, remains unresolved in all the datasets and we fit a single Gaussian for the combined emission accordingly. In general for the He\textsc{ii} line fits, the best model scatter term was often larger than the fits for other lines of the same star possibly indicating that the shape choice was not ideally suited to the data. For this unresolved multiplet, however, our model choice remains the simplest way to capture the line flux.

For the E140M data the C\textsc{ii} doublet is resolved, but at low-resolution the two lines are blended, which impacts the flux measurements since one of the lines is impacted by ISM absorption. We estimated the typical ISM attenuation of the C\textsc{ii} 1334 \AA\ line by assuming typical ISM parameters from \cite{Redfield2004} and a stellar line approximated as a Voigt profile with a width of 30 km s$^{-1}$ (0.05 \AA) for the Gaussian component and 20 km s$^{-1}$ (0.04 \AA) for the Lorentzian component based on our E140M spectra of HIP 17695's unattenuated C\textsc{ii} 1335 \AA~line. The range of C\textsc{ii} ISM parameters identified by \cite{Redfield2004} for stars inside 40 pc is 13.1-15.2 for the log column density, 2.9-6.6 km/s for the Doppler $b$ value, and a radial velocity of -30 km s$^{-1}$ to 30 km s$^{-1}$. We created 10,000 realizations of the attenuated 1334 \AA~flux where column density, $b$ value, and radial velocity of the absorbers relative to the star's rest frame were drawn from uniform distributions with the ranges described previously. We found that the median attenuation for the 1334 \AA~line is 14\%, with a 68\% confidence interval range of 4-45\%. Assuming the 1334 and 1335 \AA~C\textsc{ii} lines are equally populated (a reasonable assumption for thermal equilibrium), total C\textsc{ii} fluxes derived from the low-resolution spectra where the two lines are blended should be corrected upward $\sim$7\%, with a 68\% confidence interval range of 2-23\%.

The STIS data for G249-11 and LP~55-41 were much fainter than the rest of the FUMES targets, and many of the typical emission lines were barely visible. We determined the 3$\sigma$ upper limits for these lines by simply summing the flux in the region around the expected line center, and computing the associated uncertainty in that flux. For the few lines that were measurable, C\textsc{ii}, C\textsc{iv}, and N\textsc{v}, we accommodated the lower signal-to-noise ratio by only fitting single Gaussian line profiles in each case, one component for the blended C\textsc{ii}, and two components each for the C\textsc{iv} and N\textsc{v} doublets. These results are included in Table~\ref{tab:fluxes}.

\begin{deluxetable}{l c c}
	\tablecaption{ FUV Continuum Measurements
		\label{tab:UVcont} }
	\tablehead{
		\colhead{Name} & \colhead{Obs Flux}  & \colhead{S/N}	\\ 
		& (erg s$^{-1}$ cm$^{-2}$ \AA$^{-1}$) & }
	\startdata
	GJ 4334 & $5.5\pm0.3 \times 10^{-17}$ &  18 \\
	GJ 49 & $7.4 \pm 0.3 \times 10^{-17}$ &  25 \\
	HIP 112312 \tablenotemark{a} & $5.8 \pm 0.8 \times 10^{-16}$ & 7.25  \\
	LP 247-13 &  $1.7 \pm 0.05 \times 10^{-16}$ &  34 \\
	HIP 17695 \tablenotemark{a} & $3.3 \pm 0.4 \times 10^{-16}$ & 8.25   \\
	HIP 23309 & $3.10 \pm 0.09 \times 10^{-16}$ & 34 \\
	CD-35 2722 & $1.91 \pm 0.05 \times 10^{-16}$ & 38 \\
	GJ 410 & $1.3\pm 0.1 \times 10^{-16}$ & 13  \\
	LP 55-41 & $0.06 \pm 1 \times 10^{-18} $  & --- \\
	G 249-11 & $5 \pm 6 \times 10^{-19}$  & ---
	\enddata
	\tablenotetext{a}{The E140M spectra show very little continuum if any, leading to likely poorly characterized uncertainties, and thus these values should be taken with caution.}
\end{deluxetable}

\subsection{FUV Continuum Emission}\label{sec:UVcont} 

Although the FUV spectra of low-mass stars is dominated by discrete emission line features, there is a weak underlying FUV continuum that is likely defined by the recombination edges of species like Si \citep{Loyd2016,Peacock2019b}. As seen in Figures~\ref{fig:quiesUV_G140L} and~\ref{fig:quiesUV_E140M}, continuum emission between the strong line features was evident in the low-resolution G140L spectra, but unapparent in the high-resolution E140M spectra. This is a consequence of the higher sensitivity and lower resolution of the G140L grating, making it easier to detect weak continuum levels. Compared to \textit{HST}-COS medium-resolution gratings (G130M, G160M), the \textit{HST}-STIS low-resolution grating (G140L) is slightly more sensitive at detecting faint continuum emission. \footnote{Using the STScI exposure time calculator (\url{http://etc.stsci.edu}), detecting a low-level flat continuum ($F_{\lambda} = 10^{-16}$ erg s$^{-1}$ cm$^{-2}$ \AA$^{-1}$ at 1350 \AA) at a signal-to-noise threshold of 3 would take twice as long with COS G130M as opposed to STIS G140L (6.1 ks vs.\ 2.9 ks) when accounting for the factor of $\sim$15 difference in resolution.} For the full FUMES sample we quantify the FUV continuum levels apparent in our spectra following the work of \cite{Loyd2016}.

As part of the MUSCLES program they defined an ensemble of narrow bands (0.7-1 \AA\ in width) interspersed between the prominent FUV emission lines to assess the FUV continuum flux across 1300-1700 \AA. We used these same bands (obtained through private communication, Loyd, R.\ O.\ P.) transformed to the radial velocity frame of our targets to integrate the FUV continuum.\footnote{All the FUMES targets had measured RVs, except G249-11 and LP~55-41, for which we used a 0 km s$^{-1}$ frame. The shift is small in general and has no impact on the result, since these two targets were too faint to detect any continuum emission.} We were able to use the bands unaltered for the E140M data sets; however, for the lower resolution G140L data, we visually verified that the bands did not overlap with any of the strong emission lines. This removed 6 narrow bands ($\sim$6 \AA) that encompassed parts of emission line wings.

The integrated fluxes in these narrow bands are shown in Table~\ref{tab:UVcont}, averaged over the $\sim$150 \AA\ cumulative width of the passbands. For 8 of the 10 targets, including the two observed with the E140M grating, we measure a non-zero FUV continuum level, all except for G149-11 and LP~55-41. Although in each narrow band the integrated flux is usually insignificant, revealing no clear continuum shape, when adding up the emissions across all of the narrow passbands of the spectra the overall continuum emission is detectable. For the quoted uncertainties of Table~\ref{tab:UVcont}, we propagated the data uncertainty through the continuum flux summation.

Our results are broadly consistent with those of \cite{Loyd2016} accounting for the distance to each target; however, with this broader sample of stars we see evidence for a wide array of continuum levels. Higher signal-to-noise spectra are needed to verify the detection of these average continua and determine what defines their flux levels and spectral composition.

\begin{figure*}[tbp]
	\centering
	\includegraphics[width=\textwidth]{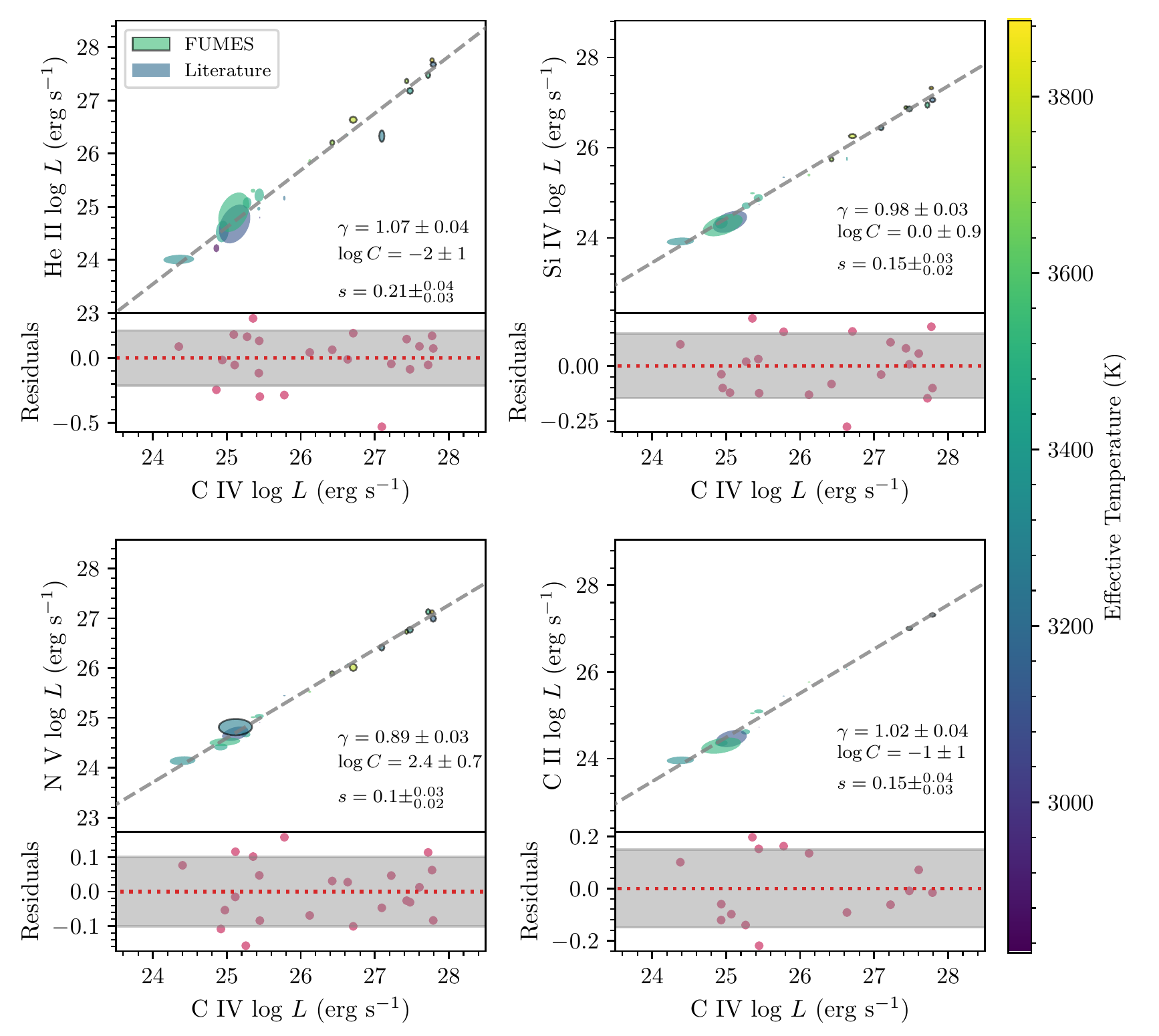} 
	\caption{Our FUV emission line correlations for the combined FUMES (black outline) and literature samples show tight power-law fits, $L_{y} = C L_{\mathrm{CIV}}^{\gamma}$, with scatter $s$. Each star is shown as a representative error ellipse corresponding to a 2$\sigma$ confidence level. Dashed lines show the individual best fit relations for each line-pair. Any FUV line can be used to predict any other emission line in quiescence. The lower panel of each plot shows the central residuals from the power-law fit with the intrinsic scatter $s$ shaded in gray.}
	\label{fig:lineline_a}
\end{figure*}

\begin{figure*}[tbp]
	\centering
	\includegraphics[width=\textwidth]{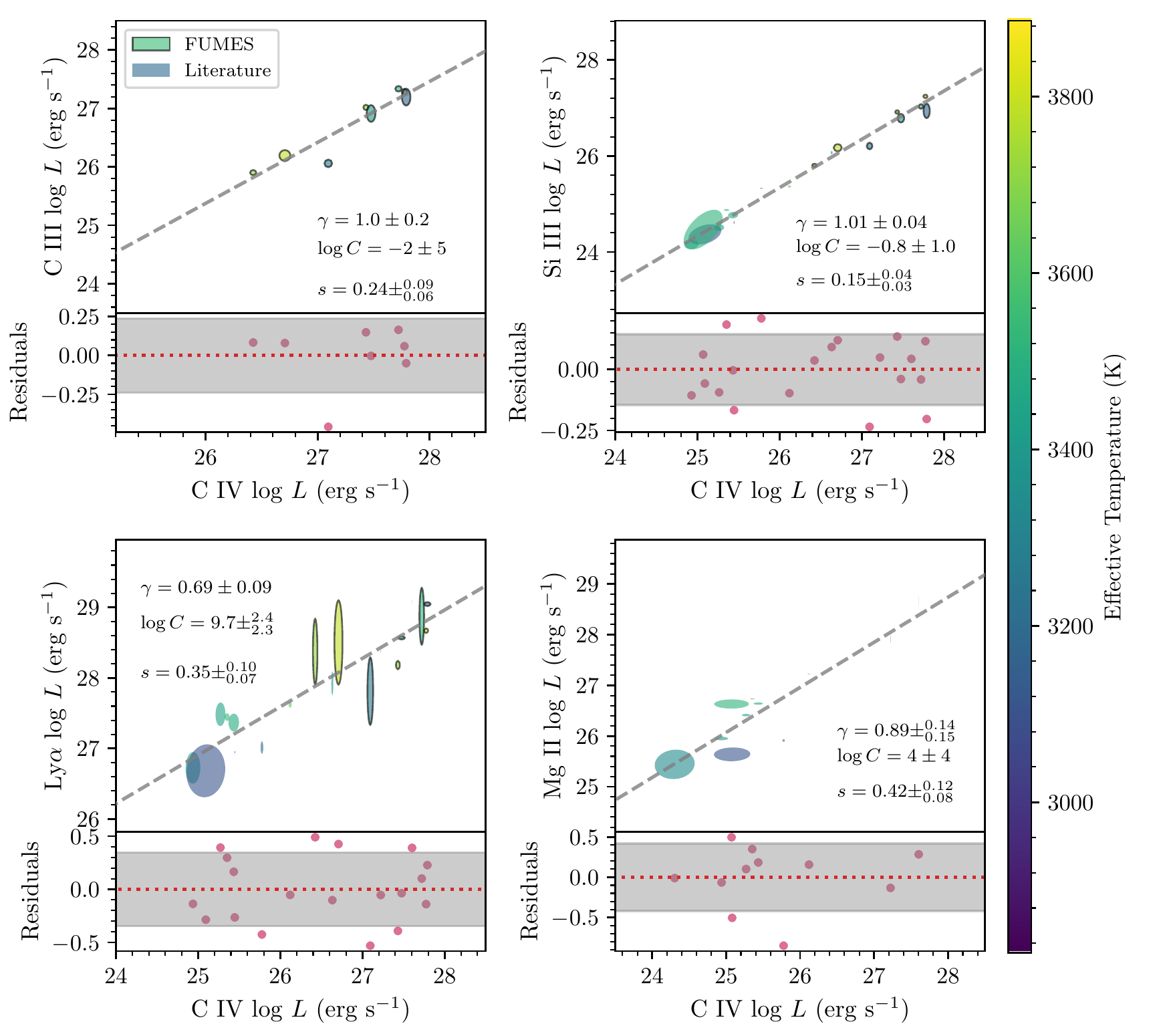} 
	\caption{Continued from Figure~\ref{fig:lineline_a}. The correlation for the NUV Mg\textsc{ii} line shows greater scatter than any of the FUV relations.}
	\label{fig:lineline_b}
\end{figure*}

\subsection{Line-Line Correlations}\label{sec:lineline}

Our FUV emission line measurements are diagnostic of the stellar upper atmospheres and how they change with magnetic heating. Because they form in similar regions, the line strengths are strongly correlated, as has been well illustrated in the literature \citep[e..g,][]{Youngblood2017}. We build on these results by expanding the available UV samples using our new measurements with the FUMES targets. In addition to the FUMES data we also analyzed the Mg\textsc{ii} NUV data for the literature sample, as representative of an additional atmospheric layer and spectral region of interest. 

Since C\textsc{iv} is a bright and readily accessible emission feature, we show in Figures~\ref{fig:lineline_a}-\ref{fig:lineline_b} the correlation of each of the emission lines of Table~\ref{tab:fluxes} against C\textsc{iv}. We plot the luminosities of each feature as two-dimensional error ellipses (2$\sigma$ contours) representative of the bivariate error distributions for the luminosities of each line pair. We further discuss the difference in undertaking this analysis in luminosity as compared to surface flux in Appendix~\ref{sec:ap_surfaceF}. Since each luminosity depends on the known parallax and its uncertainty, the luminosity determinations have correlated uncertainties, revealed by diagonally oriented ellipses. This correlation is more prominent when the parallax uncertainty dominates the luminosity measurements. Figures~\ref{fig:lineline_a} and~\ref{fig:lineline_b} shows these data with the filled shading of each shape indicating the effective temperature of the stars. The emission fluxes are not primarily determined by the stellar effective temperature, as both warmer and cooler objects in the FUMES and literature samples show high and low FUV luminosities, but instead by each object's activity regime, as will be discussed in Section~\ref{sec:rotact}. 

The correlations shown in Figures~\ref{fig:lineline_a}-\ref{fig:lineline_b} further demonstrate that the C\textsc{iv} strength can be used to predict the quiescent emission of any of the other prominent transition region emission lines across four orders of magnitude in luminosity. To these data, we fit a power-law relation using C\textsc{iv} as the predictor variable, within a Bayesian framework \citep{Kelly2007}, accounting for uncertainties in both dimensions and incorporating an intrinsic scatter at fixed C\textsc{iv} luminosity. We defined the regression model for the line luminosities, $L_{y}$, as 

\begin{equation}
\log L_{y,i} \, | \, \gamma, C, s \sim\mathcal{N}_{i}(\gamma \log L_{x,i} + \log C, s) \; , 
\label{eq:linepow}
\end{equation}

\noindent where $x$ corresponds to C\textsc{iv} and $y$ any of the other lines (e.g., N\textsc{v}, Si\textsc{iv} etc.), $\mathcal{N}(\mu, \sigma)$ denotes a Normal distribution of mean $\mu$ and standard deviation $\sigma$, $i$ denotes each star in the dataset, and $s$ is the intrinsic scatter of the power-law relation in $\log$-$\log$ space.\footnote{The notation of Equation~\ref{eq:linepow}, e.g., $z \, | \, u , w \sim \mathcal{N}(u,w)$, indicates that the random variable $z$ conditioned on $u$ and $w$ follows the probability density function defined by $\mathcal{N}$. }, The $s$ parameter is fit along with the power-law model, with its marginalized posterior distribution describing the set of intrinsic scatters consistent with the uncertainty of the data, and the posterior distributions on $\gamma$ and $C$. The peak of that distribution is our best estimate of the representative intrinsic scatter in the line-line correlations. As an example, a scatter value, $s = 0.1$, would suggest an individual line luminosity could be predicted to within $\sim$25\%, corresponding to the central 68\% confidence interval, with a known C\textsc{iv} luminosity.

\begin{deluxetable}{l c c c c c}
	\tablecaption{ Line Correlations with C\textsc{iv}\tablenotemark{a}
		\label{tab:UVcorr} }
	\tablehead{
		\colhead{Line} & \colhead{$\log T_{\textrm{f}}$}& \colhead{$N_{\mathrm{samp}}$} & \colhead{$\gamma$} &  \colhead{$\log C$} & \colhead{$s$} 
	}
	\startdata
	Mg\textsc{ii} &  4.5 & 11 & $0.89 \pm ^{0.14}_{0.15} $& $4 \pm 4 $& $0.42 \pm ^{0.12}_{0.08}$\\		
	Ly$\alpha$ &  4.5 & 11 & $0.69 \pm 0.09 $& $9.7 \pm ^{2.4}_{2.3} $& $0.35 \pm ^{0.10}_{0.07}$ \\
	C\textsc{ii} &  4.5 & 15 & $1.02 \pm 0.04$& $-1 \pm 1 $& $0.15 \pm ^{0.04}_{0.03}$ \\
	Si\textsc{iii} &  4.7 &20& $1.01 \pm 0.04 $& $-0.8 \pm 1.0 $& $0.15 \pm ^{0.04}_{0.03}$ \\
	C\textsc{iii} &  4.8 & 8 &  $1.0 \pm 0.2 $& $-2 \pm 5 $& $0.24 \pm ^{0.09}_{0.06}$ \\
	He\textsc{ii} & 4.9 & 23 & $1.07 \pm 0.04 $& $-2 \pm 1 $& $0.21 \pm ^{0.04}_{0.03}$ \\
	Si\textsc{iv} &  4.9 &21& $0.98 \pm 0.03 $& $0.0 \pm 0.9 $& $0.15 \pm ^{0.03}_{0.02}$ \\
	N\textsc{v}	 &  5.2 &22 & $0.89 \pm 0.03 $& $2.4 \pm 0.7$& $0.10 \pm ^{0.03}_{0.02}$ \\
	\enddata
	\tablenotetext{a}{C\textsc{iv} has $\log T_{\textrm{f}} = 5.0$. Reported parameters correspond to the median of the marginalized posterior distributions with uncertainties indicating the central 68\% confidence interval.}
	
\end{deluxetable}

\begin{figure}[tbp]
	\centering
	\includegraphics[width=0.5\textwidth]{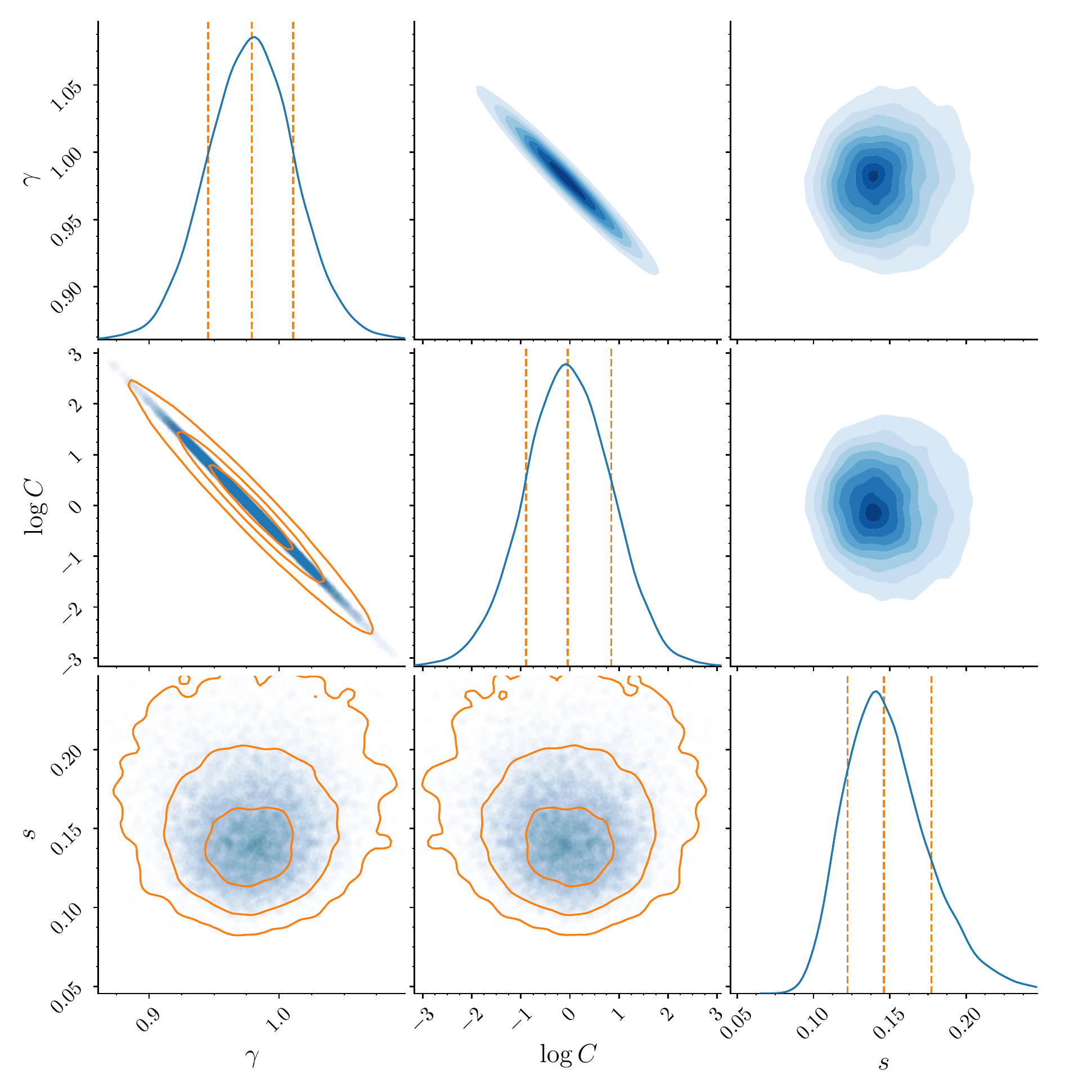} 
	\caption{The joint posterior distributions for our typical power-law fits to the line-line correlations (see Section~\ref{sec:lineline}) illustrate how samples for the power-law slope,$\gamma$, and constant, $\log C$, are highly correlated. The lines in the lower left corner show the 1$\sigma$, 2$\sigma$ and 3$\sigma$ contours on top of semi-transparent points from our MCMC sampling. The diagonal shows the marginalized distributions for each property with the distribution visualized from kernel density estimation. Along the diagonal the dashed lines indicate the median and middle 68\% confidence intervals. The upper right corner visualizes the detailed shape of the joint posteriors through kernel density estimation, showing well defined peaks.}
	\label{fig:linelinecorrPost}
\end{figure}

The probability distribution for each line luminosity pair were defined, in the case of FUMES targets, by their joint distribution created from the sampled line flux posteriors of the model fits (see Section~\ref{sec:linefits}) combined with the parallax measurement and its uncertainty. For literature data, we assumed Gaussian distributions for the line fluxes and combined with the parallax measurements to determine the joint luminosity distributions at each datum pair. The likelihood is defined by the product across the sample of stars with Equation~\ref{eq:linepow}. In $\log$-$\log$ space this is a linear model and we used a uniform prior on $\log C$, the offset, a Cauchy distribution for the power-law exponent, $\gamma$,\footnote{In practice we used a uniform distribution in the angle, $\theta$, which corresponds to the angle the line in $\log$-$\log$ space makes with the horizontal, and set $\gamma = \tan \theta$. } and a Half-Cauchy distribution with unity shape parameter for the scatter, $s$. These priors are minimally informative, and only marginally affect the posterior distributions in $\gamma$, $s$, and $C$, while improving Monte Carlo convergence efficiency. An example of the joint posterior distributions for these fits is shown in Figure~\ref{fig:linelinecorrPost}. The result of these fits are indicated by the dashed lines in Figures~\ref{fig:lineline_a}-\ref{fig:lineline_b}, and we tabulate the fit parameters in Table~\ref{tab:UVcorr}. 

The line pair with the smallest intrinsic scatter is C\textsc{iv}-N\textsc{v} ($\sim$0.1 dex). This is unsurprising as these two lines are optically thin and form at similar temperatures high in the transition region. We thus considered whether there may be a trend in intrinsic scatter relative to line formation temperature (see Table~\ref{tab:UVcorr}). The data do not reveal any suggestive trends although Mg\textsc{ii} and Ly$\alpha$ show the greatest scatter values, and correspond to the lowest formation temperature lines that we studied. On the contrary, C\textsc{ii} forms at similar temperatures to Mg\textsc{ii} and shows as small a scatter as the hotter Si\textsc{iv} transition region line. The larger scatter for Mg\textsc{ii} relation could be due to unaccounted for uncertainty associated with the ISM correction needed for the Mg\textsc{ii} flux estimates \citep{Youngblood2016}. This may also affect the Ly$\alpha$ correlation; however, the formation of Ly$\alpha$, its central profile, and its broad emission (Youngblood et al.\ \textit{accepted}) is likely more complicated \citep[][]{Peacock2019b}. Predicting that emission from the transition region C\textsc{iv} luminosity thus may not account for all of the processes impacting the Ly$\alpha$ flux. 

The rest of the line-line correlations show consistent scatters of 0.1-0.2 dex at fixed C\textsc{iv} luminosity. These measurements represent the intrinsic scatter of FUV line emission across the population, as all of the observations were either taken simultaneously or closely spaced in time \citep{France2016}. If this is indeed due to intrinsic variation, it may define the limit to which FUV features can be used to predict one another. Additionally, this intrinsic variation may differ between the `inactive' and `active' subsamples, however, our samples in each are not large enough to fully investigate. 

\section{Rotation-Activity Relation}\label{sec:rotact}

\subsection{Regression Model Fitting}\label{sec:rotact_fitting}

Canonical rotation-activity correlations with optical and X-ray emission \citep[e.g.,][]{Pizzolato2003, Wright2011, AstudilloDefru2017, Newton2017} have illustrated a strong rotational dependence typically characterized by a power-law distribution, and a saturated regime of activity for the fastest rotators. Since \cite{Noyes1984}, these relationships have been used to investigate the interplay between the internal convective motions and differential stellar rotation within an $\alpha$-$\Omega$ magnetic dynamo, thought to operate in stars with partly convective interiors \citep[e.g.,][]{Montesinos2001, Browning2006}. Mediated by the tachocline interface between the radiative core and convective envelope, the dynamo sustains the magnetic field through cyclic regeneration of toroidal and poloidal field. Theory suggests that the magnetic fields, and hence the resulting activity tracers should thus be mediated by the Rossby number, $Ro = P / \tau_{c}$, the ratio between the rotational period and the convective turn over time, which characterizes the timescales for bulk motions of internal convection. Recent results have further extended these studies across the boundary between fully convective and partly convective interiors \citep{Newton2017, Wright2018} with M dwarf samples, illustrating a continued rotational dependence for the activity tracers, despite the need for alternative dynamos for the lowest mass stars, which lack that tachocline interface.

With the FUMES sample, we can now probe these same mechanisms with new indicators of the magnetic activity in the FUV. To our emission line data, we fit a broken power-law as a function of $Ro$,

\begin{equation}
L_{y} / L_{\mathrm{bol}} =
\begin{cases}
(L_{y} / L_{\mathrm{bol}})_{\mathrm{sat}}, & Ro \le Ro_{c} \\
A (Ro)^{\eta} , &  Ro > Ro_{c} \; ,
\end{cases}
\label{eq:rotact}
\end{equation}

\noindent where $(L_{y} / L_{\mathrm{bol}})_{\mathrm{sat}} $ denotes the strength of emission line $y$, in the saturated regime relative to the star's bolometric luminosity, $\eta$ indicates the slope of the unsaturated regime, and $Ro_{c}$ is the critical Rossby number at which the activity transitions between regimes, with $A \equiv (L_{y} / L_{\mathrm{bol}})_{\mathrm{sat}} (Ro_{c})^{-\eta}$ to ensure continuity. To compute the Rossby number, we use the known rotation periods (see Table~\ref{tab:fumes}), and the empirical calibration for the convective turn overtime, $\tau_{c}$, as a function of mass from \cite{Wright2018}. We discuss systematic effects in this choice of calibration in Appendix~\ref{sec:ap_rossby}. Our segmented regression fit further includes a scatter term, $\sigma_{L}$, describing intrinsic dispersion in the observed luminosities at a fixed Rossby number, such that

\begin{equation}
 \log (L_{y} / L_{\mathrm{bol}})_{i} \, | \, \{ \beta \}  \sim\mathcal{N}(\log [L_{y} / L_{\mathrm{bol}} (Ro_{i})], \sigma_{L}) \; ,
\label{eq:rotact_prob}
\end{equation}

\noindent where the index $i$ indicates the data for each star, and $\{\beta\} = \{Ro_{c}, (L_{y} / L_{\mathrm{bol}})_{\mathrm{sat}}, \eta, \sigma_{L} \}$ is the set of four model parameters. Within this Bayesian model we employ as priors uniform distributions for $\log (L_{y} / L_{\mathrm{bol}})_{\mathrm{sat}}$ and $Ro_{c}$, a zero centered Cauchy distribution with unity shape parameter for the slope $\eta$ (see footnote 13), and Half-Cauchy distribution with unity shape parameter for $\sigma_{L}$.

Our fitting process using PyMC3 accounts for random uncertainty in both dimensions \citep{Kelly2007}, and possible error correlations between the assumed mass, and bolometric luminosity using the sampled posterior distributions from the work of Pineda et al.\ (\textit{in prep}). The error correlations between $M$ and $L_{\mathrm{bol}}$ are generally small for the field objects, but can be significant for the young stars (see Pineda et al.\ \textit{in prep}). We also included the scatter in the convective turnover time calibration from \cite{Wright2018} of 0.055 dex in $\log \tau_{c}$ at fixed mass. When using the literature periods, we further incorporated the quoted uncertainties in those measurements if available or used a 10\% Gaussian uncertainty if not, see Table~\ref{tab:fumes} (private communication, Newton, E.). Except for the Ly$\alpha$ measurements of some stars, the random uncertainties in the rotation-activity data are typically dominated by the error on the Rossby number driven by the scatter in the $\tau_{c}$ calibration. This careful accounting of the known sources of random error enabled us to carefully examine systematic effects with this fitting (see Appendix~\ref{sec:ap_rossby}).

\begin{deluxetable}{l c c c c c}
	\tablecaption{ Rotation-Activity Correlations\tablenotemark{a}
		\label{tab:RotActparam} }
	\tablehead{
		\colhead{Line} & \colhead{$N_{\mathrm{samp}}$}& \colhead{$\eta$} & \colhead{$Ro_{c}$} &  \colhead{$\log (L_{\mathrm{line}} / L_{\mathrm{bol}})_{\mathrm{sat}}$} & \colhead{$\sigma_{L}$} 
	}
	\startdata
	Ly$\alpha$ & 19 &  $-1.26 \pm ^{0.41}_{0.58} $& $0.21 \pm ^{0.16}_{0.11} $& $-3.58 \pm ^{0.16}_{0.15}$ & $0.32 \pm ^{0.10}_{0.07}$ \\
	Mg\textsc{ii} &12&  $-1.86 \pm ^{0.41}_{0.55} $& $0.20 \pm ^{0.11}_{0.08} $& $-3.99 \pm ^{0.14}_{0.15} $& $0.23 \pm ^{0.10}_{0.07}$\\		
	C\textsc{ii} & 15 & $-2.38 \pm ^{0.64}_{0.77} $& $0.24 \pm ^{0.12}_{0.11} $& $-5.25 \pm 0.16 $& $0.33 \pm ^{0.10}_{0.07}$ \\
	Si\textsc{iii} & 20& $-2.08 \pm ^{0.41}_{0.46} $& $0.20 \pm 0.07 $& $-5.37 \pm 0.13 $& $0.32 \pm ^{0.08}_{0.06}$  \\
	He\textsc{ii} &24&$-2.19 \pm ^{0.30}_{0.33} $& $0.19 \pm 0.05 $& $-4.92 \pm 0.11 $& $0.27 \pm ^{0.06}_{0.04}$ \\
	Si\textsc{iv} &21&  $-2.32 \pm ^{0.42}_{0.44} $& $0.24 \pm 0.07 $& $-5.38 \pm 0.14 $& $0.35 \pm ^{0.08}_{0.06}$ \\
	C\textsc{iv}	& 24&  $-2.02 \pm ^{0.39}_{0.41} $& $0.18 \pm ^{0.07}_{0.06} $& $-4.72 \pm 0.15 $& $0.37 \pm ^{0.08}_{0.06}$ \\	
	N\textsc{v}	 & 23& $-1.84 \pm ^{0.37}_{0.40} $& $0.19 \pm 0.07 $& $-5.37 \pm ^{0.14}_{0.13} $& $0.33 \pm ^{0.07}_{0.06}$ \\
	\enddata
	\tablenotetext{a}{Reported parameters correspond to the median of the marginalized posterior distribution with uncertainties indicating the central 68\% confidence interval.}
	
\end{deluxetable}

\begin{figure*}[tbp]
	\centering
	\includegraphics[width=\textwidth]{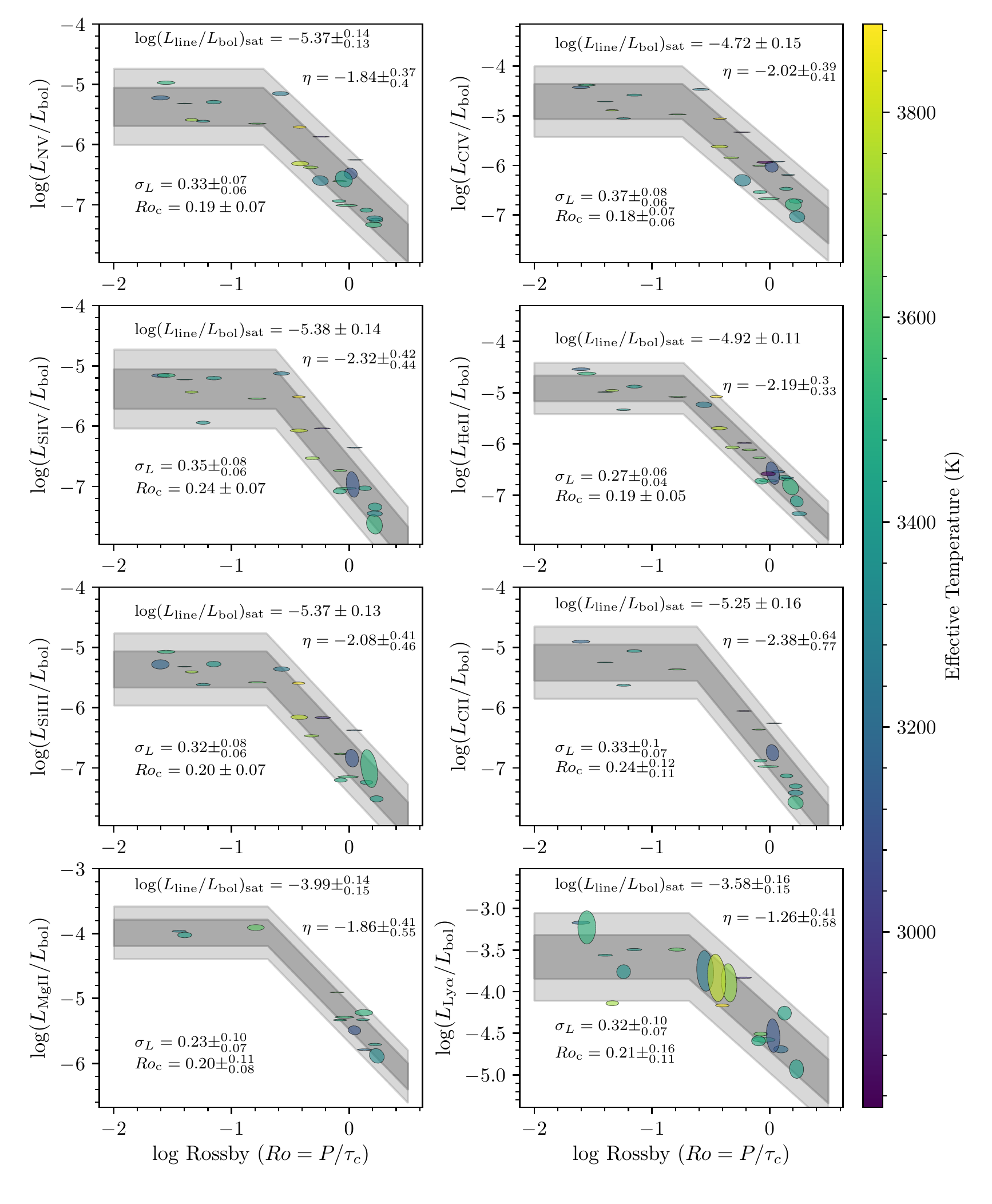} 
	\caption{Rotation-activity correlations are prominent across all of the UV emission features analyzed in this work. Individual data points are represented by 1$\sigma$ error ellipses, shaded to indicate each star's effective temperature. The best fit broken power law models (see Section~\ref{sec:rotact}) are shown in gray illustrating the 1$\sigma$ (dark gray) and 2$\sigma$ (light gray) scatters.}
	\label{fig:rotact_full}
\end{figure*}

\begin{figure}[htbp]
	\centering
	\includegraphics[width=0.5\textwidth]{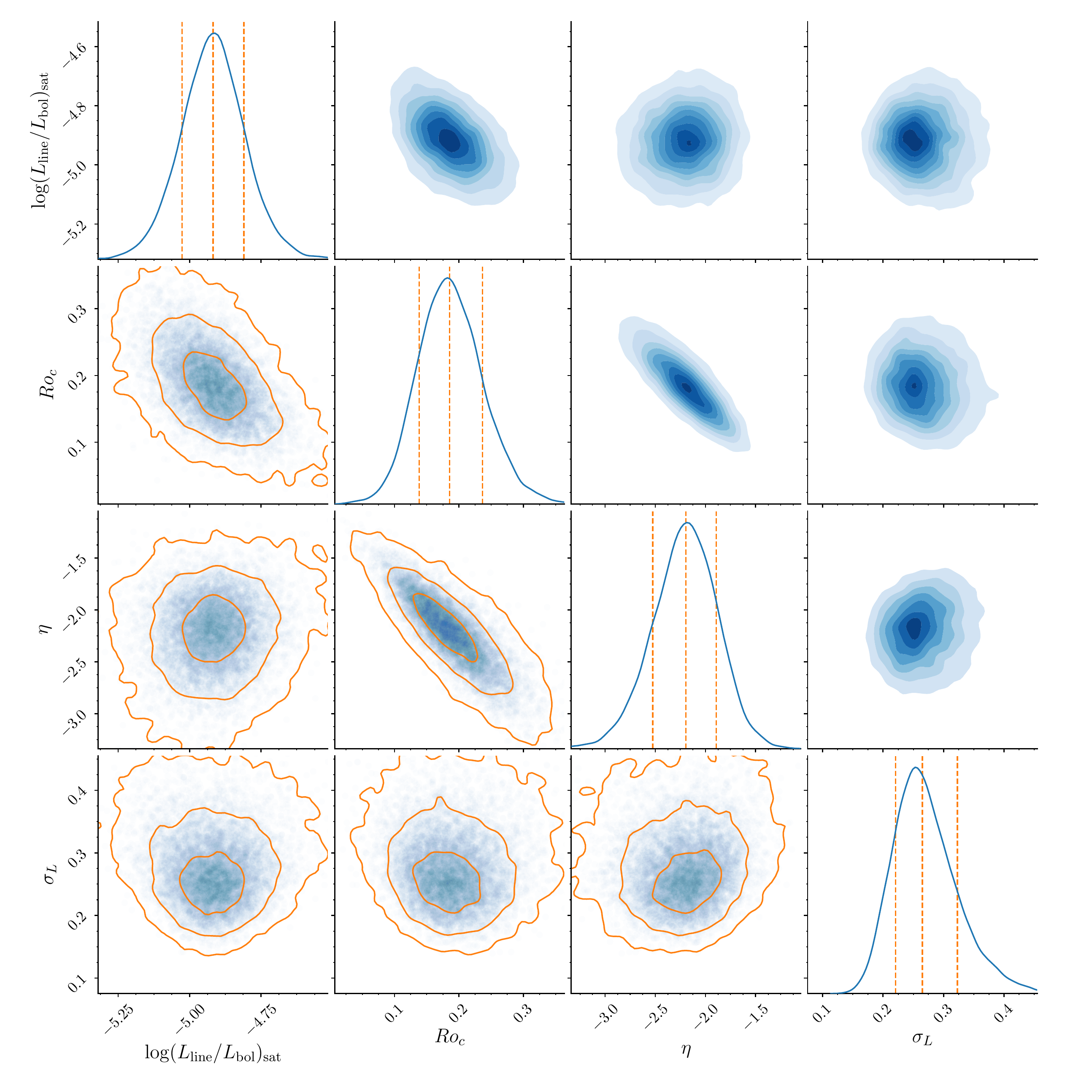} 
	\caption{The He\textsc{ii} joint posterior distributions (displayed as in Figure~\ref{fig:linelinecorrPost}) for the broken power-law fit (see Section~\ref{sec:rotact}) of the rotation-activity relation illustrate the correlation of the slope ($\eta$), and critical Rossby number ($Ro_{c}$). The posterior results for fits to the other features are qualitatively identical, with results shown in Table~\ref{tab:RotActparam}.}
	\label{fig:rotactPost}
\end{figure}

We applied these methods to eight UV emission lines, Ly$\alpha$, Mg\textsc{ii}, C\textsc{ii}, Si\textsc{iii}, He\textsc{ii}, Si\textsc{iv}, C\textsc{iv}, and N\textsc{v}, for which the samples permitted a detailed rotation-activity fit. We show the best parameter results in Table~\ref{tab:RotActparam} from the marginalized posteriors, with the fit solutions plotted in Figure~\ref{fig:rotact_full}. We also show an example of the joint posterior distributions for this fitting in Figure~\ref{fig:rotactPost}, illustrating how the critical Rossby number and unsaturated regime slope are generally correlated in our rotation-activity parameterization of Equation~\ref{eq:rotact}.

Because this regression analysis relies on stellar mass estimates for determining $Ro$, it may be potentially biased by our choice for the young stars to utilize the magnetic model based masses, which are larger than the non-magnetic model masses (see Pineda et al.\ \textit{in prep}), with a correspondingly larger implied $Ro$. This concern applies only to the five young stars in our sample (see Table~\ref{tab:fumes}). However, of the five, only AU Mic ($Ro = 0.11$-$0.17$) appears to be near the transition between saturated and unsaturated regimes. HIP~112312, HIP~17695, and CD-35 2722 have $Ro$ values well below 0.1 regardless of the model choice for the mass estimate, and HIP~23309 has a $Ro = 0.32$-$0.39$, larger than the literature values of $Ro_{c} \sim0.15$-$0.20$ \citep{Newton2017, Wright2018}. Because horizontal systematics within the saturated regime have no influence on the rotation-activity fits, and only one of the stars in a sample of $\gtrsim$20 targets per emission line might be affecting the analysis, we consider this systematic effect in young star masses to minimally impact our results. Furthermore, these young stars do not appear to be outliers in our rotation-activity plots (Figure~\ref{fig:rotact_full}). Although the difference in $Ro$ is small, only the result for Mg\textsc{ii} is likely to be impacted by the mass choice for AU~Mic, because of the limited target sample for that line. Correspondingly, the rotation-activity fit parameters for Mg\textsc{ii} have larger uncertainties.

\begin{figure}[tbp]
	\centering
	\includegraphics[width=0.5\textwidth]{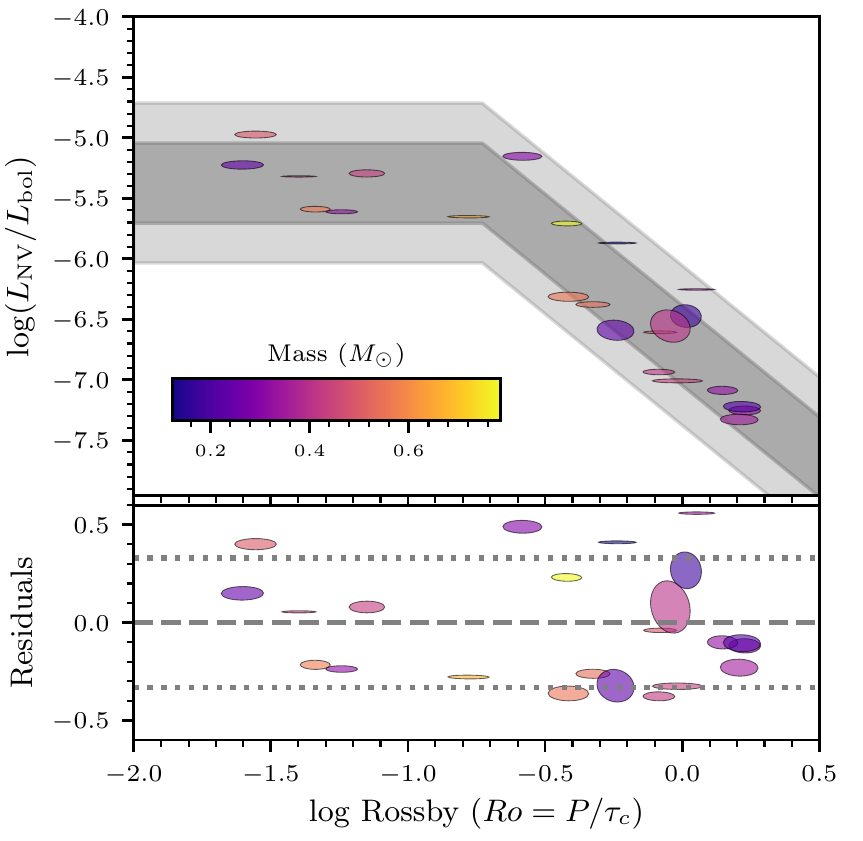} 
	\caption{Our rotation-activity model provides a good fit across our entire sample for each emission line, with the results for N\textsc{v} shown here again as an example, but now with error ellipses shaded according to stellar mass. The residuals are consistent with a normal distribution, and show no significant correlation with mass (Pearson coefficient of $-0.25$). The shaded region of top panel are as in Figure~\ref{fig:rotact_full}, with dotted-line in lower panel indicating that same scatter value, $\sigma_{L}$. }
	\label{fig:rotact_Nvmasses}
\end{figure}

We further consider the quality of our fits by examining the residuals, as illustrated in Figure~\ref{fig:rotact_Nvmasses}. The data residuals are consistent with being normally distributed about the best median rotation-activity relation, and show no apparent dependence on stellar mass. There may be some hints of an excess of data points below the median fit at large Rossby numbers ($\sim$1), but this sample is too small to be conclusive in this regard. This may be more evident in literature studies \citep[e.g.,][]{Newton2017}, which would indicate that at slow rotation periods the single power-law may be overestimating the magnetic activity.

With these considerations, for all of the emission lines, we deem the broken power-law to accurately describe the rotation dependence of the UV activity indicators across the entire sample, with each line showing slightly different emissions levels in the saturated regime, consistent critical Rossby values of $\sim$0.2, and a range of slopes in the unsaturated regime spanning $-1.3$ to $-2.4$ (see Table~\ref{tab:RotActparam}). For our M-dwarf sample the saturation levels of $\log (L_{\mathrm{line}} / L_{\mathrm{bol}})_{\mathrm{sat}}$ in the C\textsc{ii}, Si\textsc{iii}, N\textsc{v}, and Si\textsc{iv} lines all exceed those measured by \citet{France2018} for an ensemble of FGKM stars, which were all in the range of $-5.5$ to $-6$. In the saturated activity regime, M-dwarf FUV emissions exceed that of warmer stars relative to bolometric. For the power-law decay with decreasing rotation, all the lines except Ly$\alpha$ are consistent with a slope of $-2$, with Ly$\alpha$ corresponding to the shallowest slope in the rotation-activity analysis. While the error bar is large, because these data rely on model reconstructions of the emission flux, we consider if those assumptions may be systematically impacting this result.

\begin{figure}[tbp]
	\centering
	\includegraphics[width=0.5\textwidth]{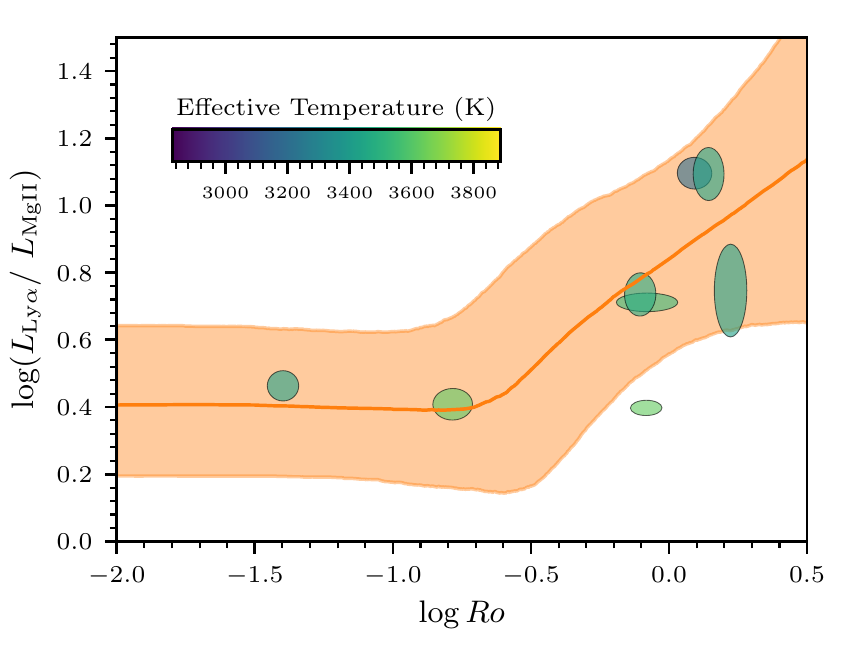} 
	\caption{The ratio of Ly$\alpha$ to Mg\textsc{ii} emission, which tracks the FUV/NUV stellar flux ratio, rises with increasing Rossby number over the course of stellar angular momentum evolution. The shaded area denotes the central 68\% confidence region defined by combining the rotation-activity fits to Ly$\alpha$ and Mg\textsc{ii}, see Section~\ref{sec:rotact_ratio}.}
	\label{fig:rotact_LyMg}
\end{figure}

\subsection{Ly$\alpha$ and Evolving Emission Line Ratios}\label{sec:rotact_ratio}

Our Ly$\alpha$ reconstructions (Youngblood et al.\ \textit{accepted}) and those from the literature sample \citep{Youngblood2016}, typically assume a Voigt-like profile for the emission, however, there is some evidence and theory suggesting that the ISM obscured Ly$\alpha$ profile peak may show an absorption self-reversal like that observed in other chromospheric lines \citep[e.g.,][]{Linsky1979, Redfield2002,Guinan2016,Youngblood2016, Peacock2019b}. Our Ly$\alpha$ reconstructions do not account for this effect, and we may thus be overestimating the Ly$\alpha$ flux. However, for this effect to impact the fitted slope in the rotation-activity unsaturated regime, the strength of self-reversals would also need to depend on the Rossby number. Theoretical Ly$\alpha$ line profiles from \cite{Peacock2019b} for GJ~832, GJ~176, and GJ~436 also show strong core absorption due to non-LTE effects, such that the reconstructed fluxes are greater by a factor of $\sim$2. A systematic effect at this level or stronger between active and inactive M-dwarfs is needed to explain the shallow Ly$\alpha$ slope. However, more data is required to validate whether these theoretical profiles match those produced by nature. There is currently no evidence for such a systematic difference in Ly$\alpha$ core profiles between active and inactive M-dwarfs.

If typical Ly$\alpha$ lines only show weak self-reversals in M-dwarfs as suggested empirically by \cite{Guinan2016} and \cite{Bourrier2017c}, then the shallow slope could indicate the persistence of Ly$\alpha$ emission at slow rotation rates even as other high-energy features decay more rapidly. As the most prominent emission lines in the FUV and NUV respectively, we consider the evolution of the ratio of Ly$\alpha$ to Mg\textsc{ii} emissions. We illustrate this in Figure~\ref{fig:rotact_LyMg}, showing the emission ratio as a function of Rossy number. The shaded region denotes the ratio of the power-law fits in the two lines from Section~\ref{sec:rotact_fitting} consistent with the uncertainties in the parameter estimates to the 1$\sigma$ level. More data are needed to better refine rotation-activity relationships in both features, but the rise in ratio with angular momentum evolution indicates that the relative strengths of different portions of the high-energy spectrum changes over time. Evolutionary changes in the spectral illumination has implications for the prevalent photochemistry and atmospheric history of any exoplanetary systems orbiting low-mass hosts. For example, stronger FUV relative to NUV emissions may drive a build-up of photochemical ozone in planetary atmospheres \citep[e.g.][]{Gao2015,Harman2015}. Although a full account of the relative contribution of Mg\textsc{ii} to the total NUV luminosity is necessary for a complete description, our data suggest that this effect may increases as stars spin-down over time. Furthermore, such an evolutionary effect may be more prominent for early M-dwarfs, and not as significant for late M-dwarfs \citep{Schneider2018}.

\begin{figure}[tbp]
	\centering
	\includegraphics[width=0.5\textwidth]{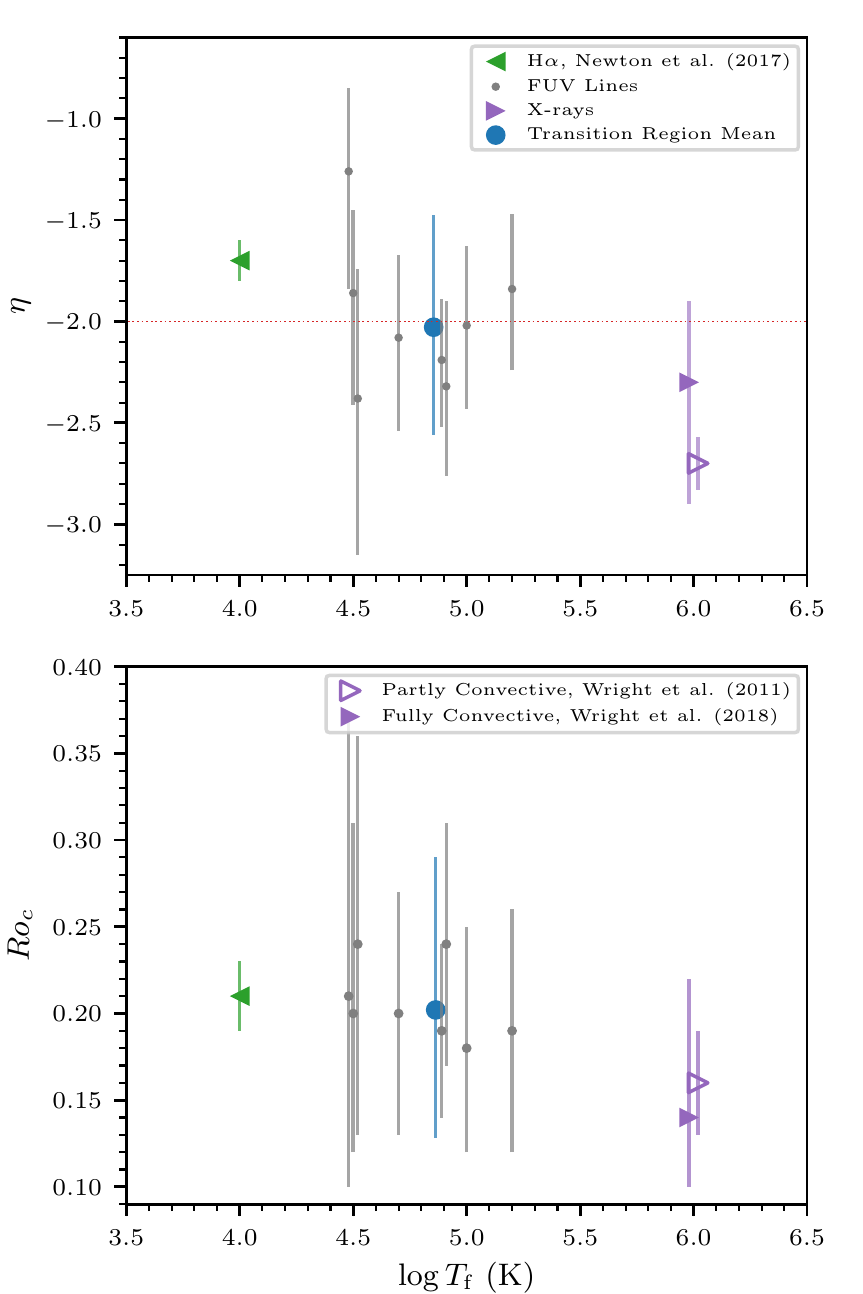} 
	\caption{Our new results for the transition region rotation-activity relationships, unsaturated slope, $\eta$ (\textit{Top}), and critical Rossby number, $Ro_{c}$ (\textit{bottom}), enable a comparison across features probing distinct regions of the stellar upper atmosphere. As activity declines with slower rotation, the layers of the atmosphere from the chromosphere to the corona may respond differently, see Section~\ref{sec:rotact_temp}.}
	\label{fig:rotact_tf}
\end{figure}

\subsection{Trends with Formation Temperature}\label{sec:rotact_temp}

Our results across all of the lines (Table~\ref{tab:RotActparam}) reveal that most of the best fit slopes are around the canonical value of $-2$ implied by a distributed dynamo model \citep{Noyes1984} in which dynamo amplification occurs throughout the convective region. Whereas in an interface dynamo like that based on a tachocline there are additional dependencies which lead to deviations away from this canonical slope of $-2$ \citep{Parker1993,Charbonneau1997,Wright2011}. To compare these results from the different lines, we plot our fit parameters for $\eta$ and $Ro_{c}$ as a function of line formation temperature in Figure~\ref{fig:rotact_tf}. The UV lines are plotted at the temperatures listed in Table~\ref{tab:UVcorr}, to which we also add the H$\alpha$ results of \cite{Newton2017} as representative of the chromosphere, and the X-ray results of \cite{Wright2011,Wright2018} corresponding to the corona using a nominal temperature of $10^{6}$ K, although X-ray flux contributions extend to higher temperatures. In Figure~\ref{fig:rotact_tf}, we show the mean value representative of the transition region, taking the combined posteriors across all the lines in each parameter, yielding medians and central 68\% confidence intervals of $\eta  = -2.02 \pm ^{0.55}_{0.53}$ and $Ro_{c} = 0.20\pm^{0.09}_{0.07}$. Our mean result is consistent to within the uncertainties with both the chromospheric and coronal results in the literature. 

However, there may be a trend in the unsaturated rotation-activity slope as a function of formation temperature. While this is largely a consequence of different values for the H$\alpha$ and X-ray results, our UV data fall directly in between. As stars spin down in the unsaturated regime, the coronal X-ray emissions appear to decline rapidly with deeper atmospheric layers showing a slower decay in their magnetic activity. The onset of activity decline with slower rotation begins first in X-rays, and only after some angular momentum evolution the transition region and chromospheric features also begin to decline, since the best fit $Ro_{c}$ are greater for the deeper layers. 

If these trends hold from analyses of larger UV samples, they will help reveal the role of magnetic heating processes in mediating the relationship between the internal magnetic dynamo and the emission features. While the rotation-activity relationships have been used to constrain stellar dynamos, this connection is indirect. The emission is necessarily a consequence of the non-thermal magnetic heating, and if the different layers exhibit distinct power-law slopes, then this effect of the heating processes needs to be accounted for when making dynamo inferences from rotation-activity relationships. The trend evident in the top panel of Figure~\ref{fig:rotact_tf} for the unsaturated slope suggests that whether the dominant process is Alfv\'{e}n wave heating or nano-flare reconnection, the decline in non-thermal heating with weaker average field strengths \citep[e.g.,][]{Shulyak2017} as stars spin down takes place from the outer most layers inward. In other words, the magnetic heating processes persist more strongly in deeper atmospheric layers across angular momentum evolution affecting the relative strengths of emission features formed in different layers of the atmosphere.

There are however some important caveats to the comparisons implicit in Figure~\ref{fig:rotact_tf}. Between this work and the studies of \cite{Newton2017} and \cite{Wright2011,Wright2018}, the samples are not identical. Our work and \cite{Newton2017} use similar mixed age samples of M-dwarfs, whereas the X-ray data of \cite{Wright2018} also includes more massive stars, although normalizing by the bolometric luminosity and examining the rotation relationship in $Ro$ is designed to account for these possible differences. More significantly, as we discuss in Appendix~\ref{sec:ap_rossby}, differences in the $Ro$ calibration can impart systematic effects on the rotation-activity analyses. \cite{Newton2017} used the calibration of \cite{Wright2011}, whereas we used that of \cite{Wright2018}. Relative to our work the \cite{Newton2017} results for critical Rossby number and unsaturated slope could be systematically greater than if they used the same calibration. The magnitude of this effect however can be small and depends on the stellar sample (see Appendix~\ref{sec:ap_rossby}). Moreover, there may have been issues with the analyses of \cite{Wright2011} leading to a much steeper power-law slope \citep{Reiners2014}. Nevertheless, it appears that in the chromosphere $\eta$ is likely shallower than $-2$, whereas in the corona it is steeper than $-2$, with the transition region value from our work right in the middle. Consistent methods and calibrations need to be applied across the different wavebands to discern whether there are indeed rotation-activity trends as a function of where in the stellar atmosphere the emission originates.

\begin{figure}[tbp]
	\centering
	\includegraphics[width=0.5\textwidth]{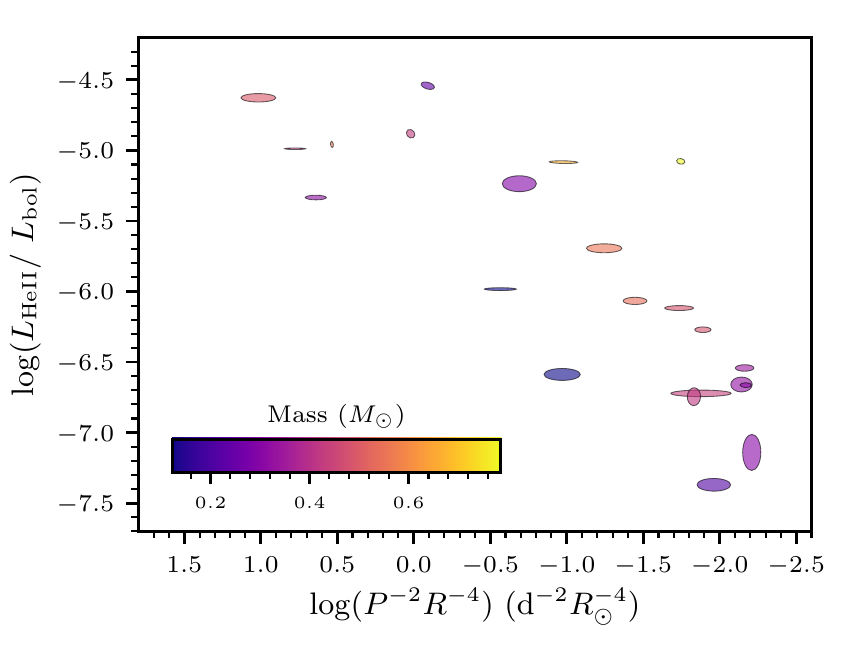} 
	\caption{The UV emission of our sample, traced by He\textsc{ii}, plotted as representative 1$\sigma$ ellipses against the rotation period and radius provides an alternative description for the rotation-activity relationship. These data show a possible mass dependence in the activity, see Section~\ref{sec:rotact_RP}.}
	\label{fig:rotact_rein}
\end{figure}

\subsection{Alternatives to Rossby Number?}\label{sec:rotact_RP}

Potential issues with a choice of calibration can be avoided by circumventing the Rossby number entirely. \cite{Reiners2014} argue for a more general approach in analyzing the rotation-activity relationships of low-mass stars, and conclude that the activity dependence can be well described by a combination of rotation period and radius, specifically $L_{X}/ L_{\mathrm{bol}} = k P^{-2} R^{-4}$, where $k$ is a scaling constant. We consider how this relationship may apply to our UV data. In Figure~\ref{fig:rotact_rein}, we show the He\textsc{ii} luminosity normalized by stellar bolometric luminosity against the stellar $P^{-2}R^{-4}$. Like the Rossby scaling, Figure~\ref{fig:rotact_rein} illustrates a potentially power-law decay in emission with a scatter of activity at a given abscissa. Although our sample is much smaller, there are hints of the mass dependence in activity using this scaling that was illustrated by \cite{Newton2017} with H$\alpha$ data by which more massive stars preferentially lie to the right of the locus of points in Figure~\ref{fig:rotact_rein}. The mass dependence appears to increase the intrinsic scatter at fixed rotation in the activity beyond the $\sim$0.27 dex scatter evident in our fits to $L_{\mathrm{HeII}} / L_{\mathrm{bol}}$ vs.\ $Ro$.

While it is evident that rotation still plays a key role in the magnetic emission of fully convective stars, an empirical Rossby number may not be the most appropriate scaling to uncover the underlying relationships governing the dynamo action, and possible differences with partly convective stars. It appears to work sufficiently well across the chromosphere, transition region and corona, however, by defining $Ro$ specifically to minimize scatter in the rotation-activity diagrams across this full range of stars, this procedure may be masking real differences in the evolution of magnetic activity between partly and fully convective objects \citep[e.g.,][]{Magaudda2020}. If these two stellar mass regimes did indeed show distinct rotation-activity correlations, i.e., statistically distinguishable rotational dependencies of their activity decay, the process of defining an empirical Rossby number across both samples would be averaging together these possible differences, since it presupposes that both samples follow the same power-law relation.

\section{Temporal Evolution}\label{sec:age_evo}

In Section~\ref{sec:rotact}, we used the stellar rotation period as a proxy for examining magnetic activity throughout the lifetimes of low-mass stars. However, using gyrochronology \citep[e.g.,][]{Barnes2003}, we can also compare our FUV activity indicators directly to stellar ages using literature relations connecting rotation period and age. While this practice is well established for solar-like stars \citep[e.g.,][]{Angus2015, Gallet2015}, the process is much more difficult for M-dwarfs \citep[e.g.,][]{Guinan2016}, which lack a multitude of targets with well determined ages. Nevertheless, \cite{Engle2018RNAAS} have developed empirical relations for M0-M1 and M2.5-M6 dwarfs to estimate ages from measured rotation periods. Although such gyrochronology relations for M-dwarfs remain ongoing work, we use those results in this paper to provide an approximate indication of the expected behavior across the low-mass star regime.

\subsection{UV Emissions Across Time}\label{sec:uvtime}

\begin{figure}[tbp]
	\centering
	\includegraphics[width=0.5\textwidth]{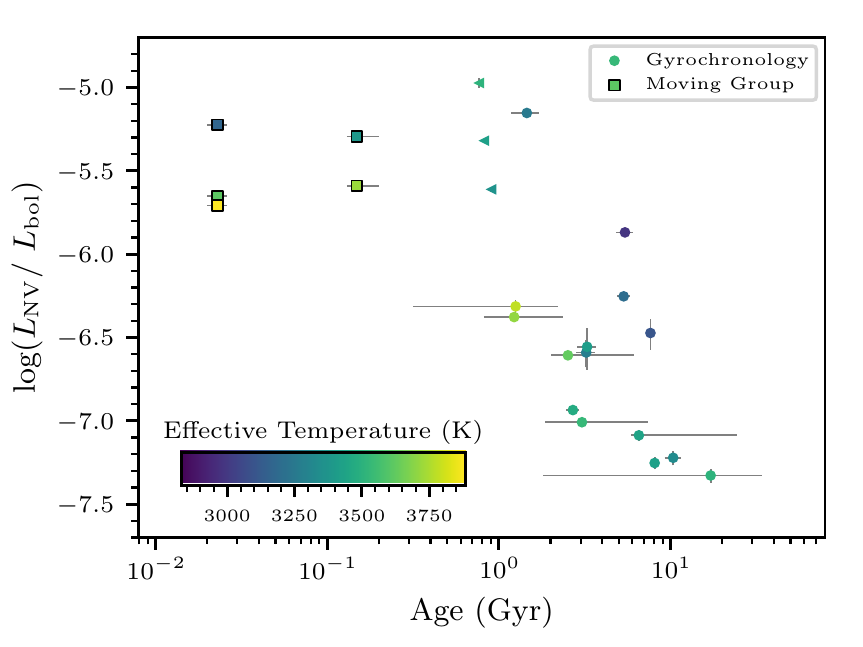} 
	\caption{The N\textsc{v} luminosity of our sample as a function of age illustrates how young M-stars exhibit saturated FUV emissions which last for $\sim$1 Gyr, before declining. The cooler M-dwarfs also show strong UV emissions for longer into their lifetimes than warmer objects. Age uncertainties (1$\sigma$) are determined from the group age determination in the literature, or from the uncertainties in the gyrochronology relations, with upper limits for the rapid rotation star ages indicated as triangles, see Section~\ref{sec:age_evo}.}
	\label{fig:evoNV}
\end{figure}

We thus show the evolution of FUV activity with age in Figure~\ref{fig:evoNV}, focusing on the N\textsc{v} emission feature as representative of all of the other FUV lines (Sections~\ref{sec:lineline} and~\ref{sec:rotact}). We chose the N\textsc{v} line for this analysis as it is tightly correlated with C\textsc{iv}, and existing literature relations allow us to also estimate EUV emission from their measurements, see below. 

For the five objects with moving group membership (see Table~\ref{tab:fumes}), we used the mean group age with error, and the gyrochronology relations for the rest of the sample. To assess the uncertainties on the ages from the gyrochronology calibration, we used the reported uncertainties of the parameters of the best-fit relations \citep{Engle2018RNAAS} and the period uncertainties as in Section~\ref{sec:rotact_fitting}, sampling each appropriately, assuming independently Gaussian distributed variables in computing the distribution for the age estimate.\footnote{\cite{Engle2018RNAAS} did not report a scatter about their best fit rotation-age relations, so we could not incorporate that uncertainty in these age estimates. For the M0-M1 bin, we ignored the uncertainty on their power-law exponent, as the results are too sensitive to its value and we cannot properly account for its correlation with the other parameters that would mitigate this effect.} Because the \cite{Engle2018RNAAS} rotation-age relations are defined by two bins of spectral type ranges, for our sample we used the appropriate relation corresponding to the optical spectral types listed in Table~\ref{tab:fumes}. However, some of our objects have spectral types, M1.5-M2, between the defined ranges for the rotation-age relations. For these objects we combined the estimates from both the M0-M1 and M2.5-M6 relations to determine the age estimate and its uncertainty. Three of these stars were rapid rotators without known ages: AD~Leo, EV Lac, and LP~247-13. Given the large uncertainty in the gyrochronology ages, especially early on where the rotational evolution has yet to converge, we plot these age estimates as 3$\sigma$ upper limits (triangles in Figure~\ref{fig:evoNV}).

In Figure~\ref{fig:evoNV}, we also shade the points according to their effective temperature. Based on these age estimates, Figure~\ref{fig:evoNV} illustrates how cooler, later M-dwarfs persist with strong activity levels for a longer duration of their early lifetimes. Considering the mean level of $\log(L_{\mathrm{NV}} / L_{\mathrm{bol}})_{\mathrm{sat}} \sim -5.37$ and 0.33 dex scatter, some of the cooler objects may display near saturation level activity beyond an age of 1 Gyr. In contrast, the warmer M-dwarfs, by this age, appear to have declined in their magnetic emissions by an order of magnitude relative to the saturated regime. Using \textit{GALEX} photometry, \cite{Schneider2018} also found this divergence in UV evolution between early and mid-M dwarfs.

\begin{figure}[tbp]
	\centering
	\includegraphics[width=0.5\textwidth]{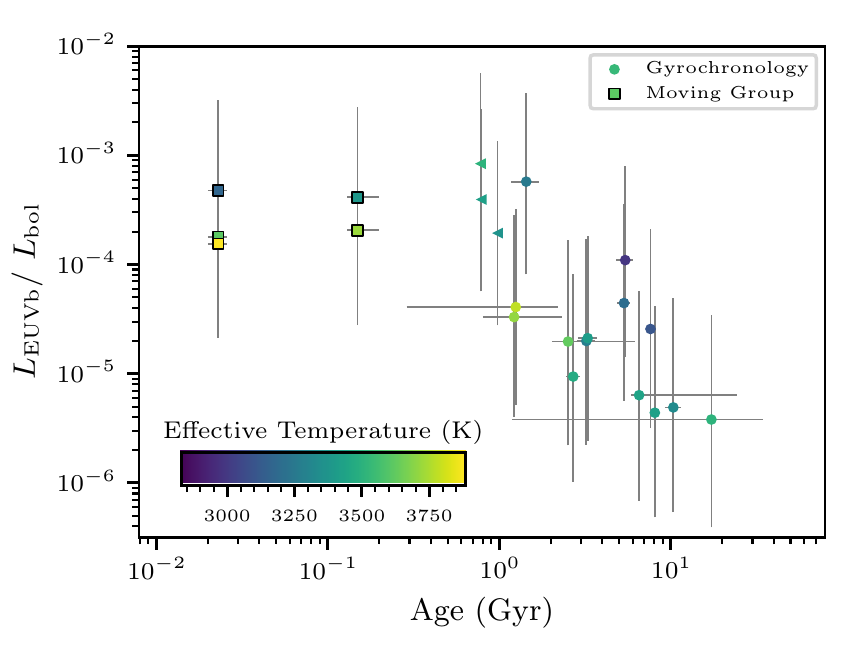} 
	\caption{Based on available scaling relations from the FUV \citep{France2018}, we can estimate the EUVb (90-360 \AA) luminosity of low-mass stars, similar to Figure~\ref{fig:evoNV}. The temporal evolution of EUV emission drops $\sim$2 orders of magnitude from youth to field ages, see Section~\ref{sec:age_evo}.}
	\label{fig:evoEUV}
\end{figure}

Because of its importance to exoplanetary atmospheric heating, we also considered what our FUV evolutionary data imply for the extreme ultraviolet portion of the high-energy spectrum. To estimate the EUV from the FUV data, we used the empirical relation of \cite{France2018} between N\textsc{v} emission and the blue portion of the EUV passband, EUVb (90-360 \AA). While not the full EUV spectrum, this is the portion for which available data and relatively low ISM attenuation have permitted any empirical estimates in the low-mass star regime. This EUVb band corresponds to about half of the total EUV energy from M-dwarf stars \citep{Fontenla2016,France2018}. In Figure~\ref{fig:evoEUV}, we plot this EUV evolution with the same ages as discussed above, and EUVb luminosities determined from the N\textsc{v} emission, incorporating the parameter uncertainties and 0.24 dex scatter in the \cite{France2018} relation. For young active M-dwarf stars the EUV luminosity is $\sim$$10^{-3.5}$ relative to bolometric, with these emissions lasting for the first several hundreds of Myrs of their lives and perhaps longer.

\begin{figure}[tbp]
	\centering
	\includegraphics[width=0.5\textwidth]{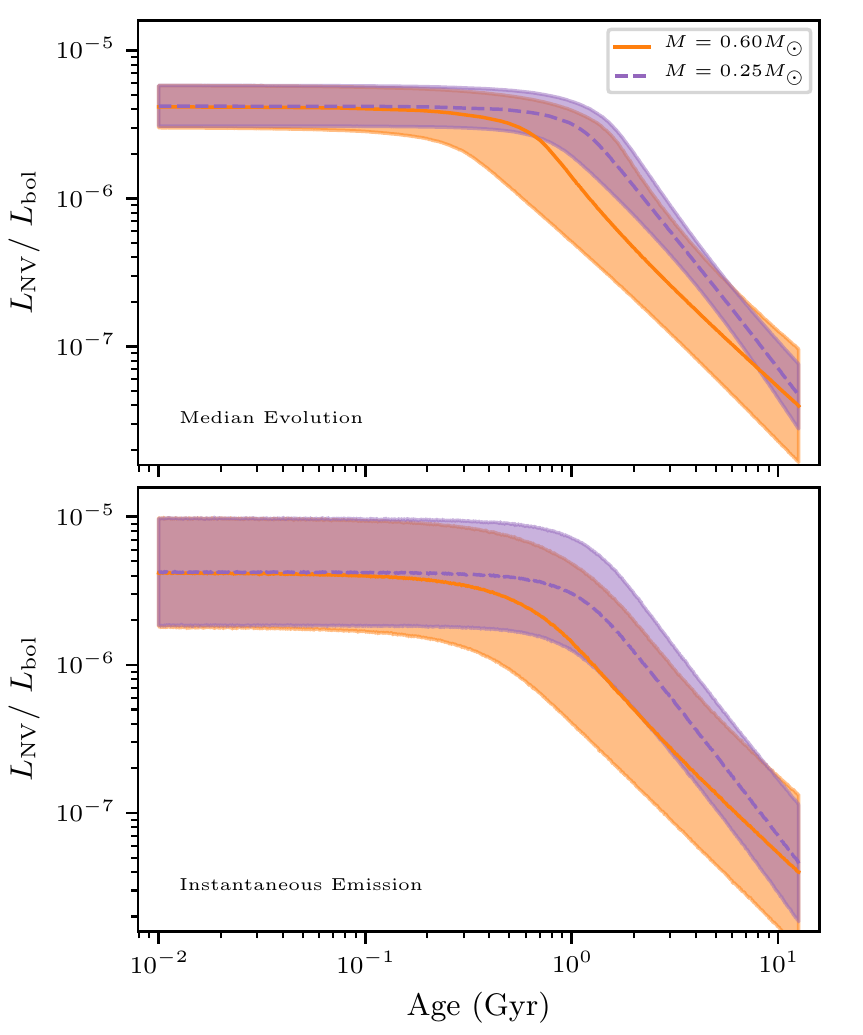} 
	\caption{ The early and mid M-dwarf FUV temporal evolution implied by our rotation-activity correlation in N\textsc{v}, and the empirical rotation-age relations, show possible differences across time, with ranges indicating the central 68\% confidence interval about the median. The top panel focuses on just the average evolution, and the bottom panel includes the likely range for emission accounting for the observed scatter across the populations ($\sim$0.33 dex).}
	\label{fig:evoModel}
\end{figure}

While these empirical relations for age, and difficult to detect portions of the spectrum, like the EUV, provide general estimates for the evolution of activity and high-energy emission, Figure~\ref{fig:evoEUV} further illustrates the broad uncertainty inherent in using these estimates as inputs to exoplanetary evolution modeling. Refined rotation-age relations in the M-dwarf regime, and expanded samples of stellar EUV detections, from missions like ESCAPE \citep{France2019}, will greatly expand the utility of these methods for exoplanetary applications. 

Assuming the Rossby parameterization removes a significant mass dependence, we use the rotation-activity correlation regression fits together with the assumed rotation evolution to calculate the activity evolution of early and mid M-dwarfs. In Figure~\ref{fig:evoModel}, we plot separate curves for the evolution of N\textsc{v} emission with stellar age, for a 0.6 $M_{\odot}$ M-dwarf, corresponding to the M0-M1 rotational evolution, and a 0.25 $M_{\odot}$ M-dwarf, corresponding to the M2.5-M6 rotational evolution. The top panel of Figure~\ref{fig:evoModel}, shows the expected median evolution for each mass star with its central 68\% confidence level. The uncertainties are determined at a given age by sampling the uncertainty distributions of the rotation-age relation, the calibration between mass and convective turn over time, and the joint posterior of the rotation-activity fit. The bottom panel of Figure~\ref{fig:rotact_Nvmasses} additionally includes the scatter at fixed Rossby number (see Section~\ref{sec:rotact}), to illustrate the range of likely emission values accounting for intrinsic scatter in the observed population. More data are required to assess; however, we expect this scatter to include the effects of rotational variability, activity cycles, and possible metallicity variations. While the nominal N\textsc{v} emission levels relative to bolometric are higher for lower mass objects at the same age, there is little difference in the evolution within the uncertainties.

\subsection{Accumulated UV Evolutionary Histories}\label{sec:accumUV}

\begin{figure}[tbp]
	\centering
	\includegraphics[width=0.5\textwidth]{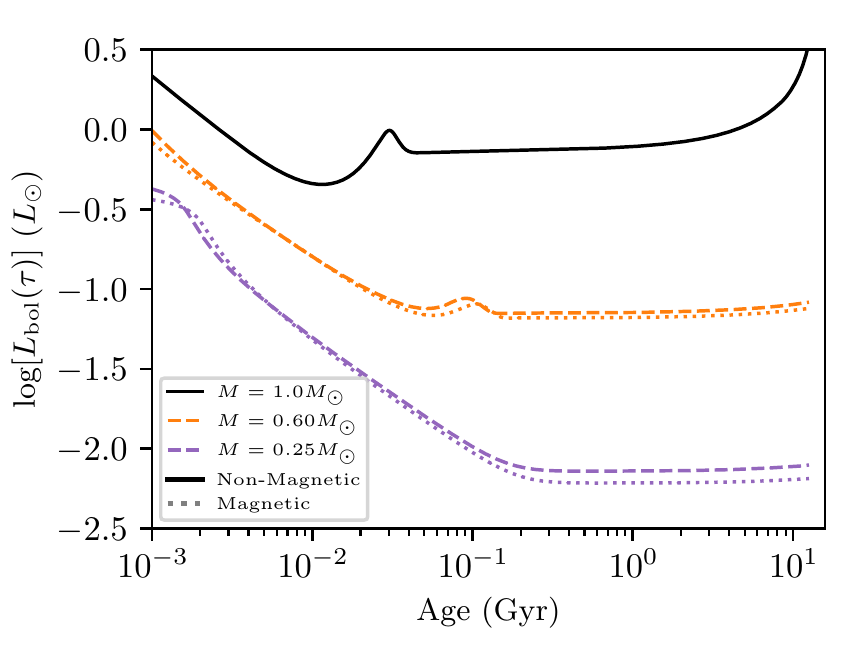} 
	\caption{The bolometric luminosity evolution in the stellar models used in this work to calculate the history of low-mass star UV illumination.}
	\label{fig:evoLum}
\end{figure}

With possibly distinct UV luminosity evolution between early and mid M-dwarfs, as shown in Figure~\ref{fig:evoModel}, we calculate the typical evolution of the UV illumination for exoplanetary systems using the median N\textsc{v} rotation-activity relation (Section~\ref{sec:rotact}), and the age-rotation evolution from \citep{Engle2018RNAAS}. The average evolution for individual systems may vary by up to a factor of two. As a measure of the illumination histories, we assess how much quiescent UV energy a given exoplanet accumulates over time as

\begin{equation}
\mathcal{E}(t) = \int_{0}^{t} F_{\mathrm{UV}}(\tau) d\tau = \int_{0}^{t} \frac{ \mathcal{R}(\tau) L_{\mathrm{bol}}(\tau)}{4\pi d^{2}} d\tau  \; ,
\label{eq:Eaccum}
\end{equation}

\noindent where $F$ is the UV flux impinging on the planet at an average distance $d$ from the host, and we integrate from 0 to an age $t$. The UV flux can be rewritten as a combination of the emission ratio evolution, $\mathcal{R}(\tau) = L_{y} / L_{\mathrm{bol}}$, determined from Section~\ref{sec:rotact}, and $L_{\mathrm{bol}}(\tau)$, which accounts for changes in bolometric luminosity with stellar evolution. In evaluating Equation~\ref{eq:Eaccum}, we use evolutionary models \citep{Dotter2008,Feiden2016} for the bolometric luminosity, which can evolve substantially at young ages (see Figure~\ref{fig:evoLum}), and allows a self-consistent metric across several Gyrs. We compared the accumulated UV energies for planets around a 0.25 $M_{\odot}$ host and a 0.60 $M_{\odot}$ host with that of a planet around a Sun-like star.

In order to compare with the lower-mass objects, the UV emission history for Sun-like stars required a different treatment from our approach with M-dwarfs. Using a sample of G-dwarfs, \cite{Ribas2005} determined the EUV evolutionary history of Sun-like stars, as a simple power-lay decay with time across 0.1-6.7 Gyr. Since results from \cite{France2018} suggest that the N\textsc{v} emission and EUV of solar and lower-mass stars are related by a simple multiplicative factor, we consider the Sun-like N\textsc{v} emission to follow the same power-law defined by \cite{Ribas2005}. We thus scaled their result to the Sun's current N\textsc{v} luminosity from Duvvuri et al. (\textit{accepted}) using the solar minimum quiescent Solar irradiance spectrum of \cite{Woods2009}:

\begin{equation}
F_{\mathrm{N\textsc{v}},\odot} =0.105 \, \tau^{-1.23}, \mathrm{erg} \; \mathrm{s}^{-1} \; \mathrm{cm}^{-2}, \; 0.1 < \tau < 6.7 \; \mathrm{Gyr} \; , 
\label{eq:solNv}
\end{equation}

\noindent where $\tau$ is in units of Gyrs. We take the current age of the Sun to be 4.6 Gyr, and this flux is evaluated at a distance of 1~AU. Although the EUV bands from \cite{Ribas2005} and \cite{France2018} are different, this conversion allows a representative evolutionary history for the Sun-like N\textsc{v} emissions, which we can compare with the activity-age evolution determined for M-dwarfs in Section~\ref{sec:uvtime}. Since Equation~\ref{eq:solNv} is only applicable to a particular age interval, to extrapolate to younger ages we consider the UV emissions at $\tau$ = 0.1 Gyr to correspond to a saturated level relative to bolometric that is constant for $\tau <$0.1 Gyr, $\log L_{\mathrm{N\textsc{v}}}/L_{\mathrm{bol}} = -5.75$, set by the model bolometric luminosity at $\tau = 0.1$ Gyr. We then let the bolometric luminosity change according to the non-magnetic Dartmouth stellar models for $\tau <$0.1 Gyr \citep[e.g.,][]{Dotter2008}.

\begin{figure}[tbp]
	\centering
	\includegraphics[width=0.5\textwidth]{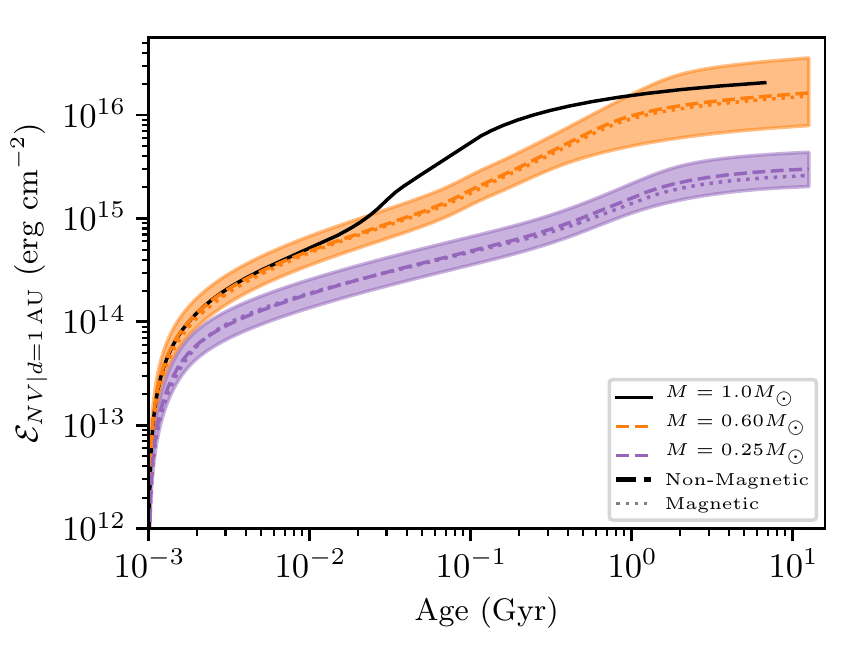} 
	\caption{The cumulative N\textsc{v} energy experienced by a planetary system at a distance of 1~AU from its host star, according to Equation~\ref{eq:Eaccum}, for each of 3 host masses, accounting for rotation-activity and bolometric luminosity evolution. For the M-dwarfs we include two lines each to illustrate the evolution with magnetic and non-magnetic models. The absolute high-energy fluxes at other bands can be scaled from the N\textsc{v} emissions. The shaded bands indicate the central 68\% confidence interval for the median evolution of the UV emissions using non-magnetic models, the corresponding magnetic model interval (not shown) is similar.}
	\label{fig:evoAbsCum}
\end{figure}

We show our results in Figure~\ref{fig:evoAbsCum}, as the accumulated N\textsc{v} emission experienced by a planet at a fixed distance of 1~AU from its host, for the 3 planet host cases, 0.25 $M_{\odot}$, 0.60 $M_{\odot}$, and 1.0 $M_{\odot}$, starting at 1~Myr, the lower limit of the evolutionary models. Around the higher-mass objects the absolute N\textsc{v} illumination is typically higher than around the lower-mass stars. The shaded bands in Figure~\ref{fig:evoAbsCum} propagate the uncertainty in the median evolution (as in Figure~\ref{fig:evoModel}) through the integral of Equation~\ref{eq:Eaccum}. We do not include the intrinsic scatter because we focus here on the cumulative average history from the rotation-activity analysis. Our treatment, however, excludes uncertainty from unknown systematics with the bolometric luminosity evolution, although at least with regards to the role of magnetism, its contribution may be minor (see Figure~\ref{fig:evoLum}). \citet{Ribas2005} did not report uncertainties for their EUV evolution in Sun-like stars, and we thus do not include it either.

\begin{figure}[tbp]
	\centering
	\includegraphics[width=0.5\textwidth]{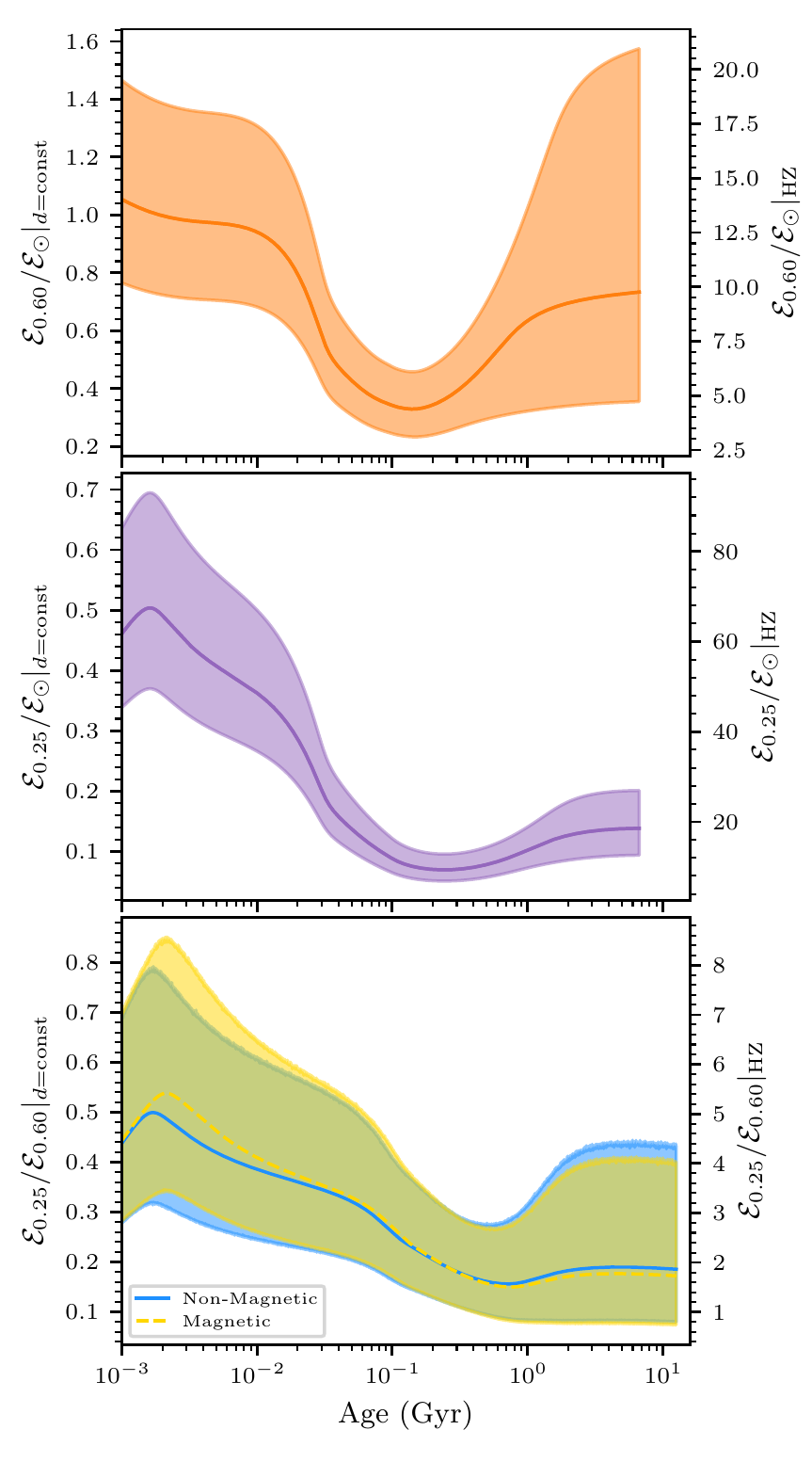} 
	\caption{A ratio comparison of the accumulated UV energies over evolutionary history impacting planets orbiting stars of 0.25, 0.60 and 1.0 $M_{\odot}$, following Equation~\ref{eq:Eaccum}. The left-side axes correspond to planets at equal distances around the respective hosts, and the right-side axes correspond to planet distances at which the 5~Gyr bolometric instellations are equivalent, i.e., comparing the field-age habitable zones around each host. \textit{Top} - Early-M dwarf relative to Sun-like star, \textit{middle} - mid-M dwarf relative to Sun-like star, and \textit{Bottom} - mid-M dwarf relative to early-M dwarf, using both magnetic (dashed) and non-magnetic (solid) models for the stellar luminosity evolution, see Section~\ref{sec:accumUV}. In each case, the more massive hosts deliver more UV energy to planets at the same distance, but within their respective habitable zones, planets orbiting the lower-mass star experience greater levels of UV emission. The shaded regions of each panel indicate the central 68\% confidence region accounting for the known uncertainty in the median rotational evolution of FUV emissions.}
	\label{fig:evoRatios}
\end{figure}

In Figure~\ref{fig:evoRatios}, we also show the ratios ($\mathcal{E}_{0.60}/\mathcal{E}_{\odot}$, $\mathcal{E}_{0.25}/\mathcal{E}_{\odot}$, and $\mathcal{E}_{0.25}/\mathcal{E}_{0.60}$) of the accumulated UV energies, propagating the uncertainty in the M-dwarf median rotational evolution. The left-side axes give the result for planets at equal distances around each star, and the right-side axes indicate the same ratio, but for the respective habitable zones. This further assumes planets at that distance have remained there throughout history. The right-axis curves are thus a constant multiple of the left-axis, as in

\begin{equation}
	\frac{	\mathcal{E}_{a}}{ \mathcal{E}_{b}} \biggr\rvert_{\mathrm{HZ}} = \frac{  \int_{0}^{t} \mathcal{R}_{a} \, L_{\mathrm{bol},a}  \, d\tau  }{ \int_{0}^{t} \mathcal{R}_{b} \, L_{\mathrm{bol}, b} \,  d\tau} \times \frac{d^{2}_{\mathrm{HZ},b}}{ d^{2}_{\mathrm{HZ},a}}  \; ,
	\label{eq:Eaccum_ratio_left}
\end{equation}

\noindent where the habitable zone distances are determined by the 5 Gyr bolometric instellations. The right-side axes in Figure~\ref{fig:evoRatios} thus indicate the relative accumulated UV energies for planets in the respective field age habitable zones of each star.

In Figure~\ref{fig:evoRatios}, we illustrate in the low-mass star regime that at a constant distance exoplanets orbiting higher-mass objects experience a greater absolute accumulation of UV energy. However, within the respective habitable zones around each host the planets orbiting the lower-mass star intercept much more UV energy. For example, relative to the Earth around the Sun, the same planet in the habitable zone around the 0.25 $M_{\odot}$ star will accumulate $\sim$20 times more UV energy by the time it has reached field age, which is $\sim$2 times more than the same planet in the habitable zone around the 0.60 $M_{\odot}$ star. Within the known uncertainty of the median rotational evolution these figures can vary by factors of a couple. The curves of Figure~\ref{fig:evoRatios} reflect at early times differential bolometric luminosity evolution modulated by activity-age evolution at late times, with modulations of the evolution dictated by the age at which each star begins their power-law decay of UV activity. The bottom panel of Figure~\ref{fig:evoRatios} also shows the results using both magnetic (dashed line) and non-magnetic (solid line) models to illustrate that similarity of the luminosity evolution, and the robustness of our results to this potential systematic effect.

These results were derived specifically using N\textsc{v} emissions; however, the line is broadly representative of the entire FUV band given the strong correlations between emission lines (Section~\ref{sec:lineline}). Thus, although the absolute energy levels determined from Equation~\ref{eq:Eaccum}, and shown in Figure~\ref{fig:evoAbsCum}, will vary between choice of UV feature, the ratios of Figure~\ref{fig:evoRatios} are broadly representative of total FUV emissions. Moreover, to the extent that the N\textsc{v} emission is directly proportional to EUV emissions \citep{France2018}, the ratios of Figure~\ref{fig:evoRatios} translate exactly to the relative exposure of planets to EUV emissions around each stellar host. This is a key result given the importance of EUV fluxes for exoplanetary atmospheric heating and escape.

\begin{figure}[tbp]
	\centering
	\includegraphics[width=0.5\textwidth]{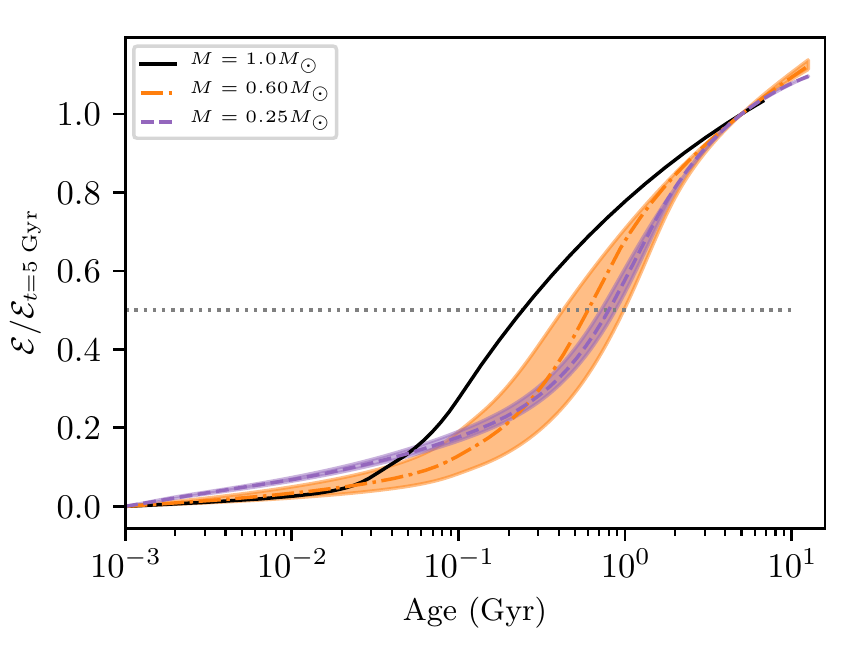} 
	\caption{The UV energy accumulation (see Equation~\ref{eq:Eaccum}) as a function of time (as in Figure~\ref{fig:evoAbsCum}) relative to the total at $t= 5$ Gyrs, using non-magnetic stellar evolutionary models. During the first Gyr of their lifetimes, planets around each low-mass stellar host receive at least 50\% of the total quiescent UV energy exposure that will have accumulated by an age of 5 Gyr.}
	\label{fig:evoRelAccum}
\end{figure}

In Figure~\ref{fig:evoRelAccum}, we further show the importance of different epochs in the total accumulated UV exposure for exoplanetary systems. Relative to the cumulative quiescent UV energy experienced at an age of 5~Gyrs, for planets around low-mass hosts, they reach total energetic exposures exceeding $\sim$50\% by ages of 800, 600, and 250~Myr respectively for hosts of mass 0.25, 0.60, and 1.0 $M_{\odot}$. These results quantify the importance of the first Gyr of stellar lifetimes in their total energetic input to exoplanetary systems. These early ages clearly need to be taken into account when considering the evolutionary history of planetary systems, and their response to high-energy emissions.

\section{Conclusions}\label{sec:conc}

The UV data of the FUMES sample presented in this paper combined with the literature data has allowed us to examine how the far ultraviolet spectroscopic emissions of low-mass stars change with angular momentum evolution over time. At fast rotation rates (Rossby number, $Ro\lesssim0.2 $), corresponding to young stellar ages ($\lesssim$1) Gyr, the FUV lines exhibit saturated emissions at levels of $10^{-3.5}$-$10^{-5.4}$ relative to the stellar bolometric luminosity, depending on the specific line, with Ly$\alpha$ being the strongest feature. As the stars spin-down with age, these emission levels drop by $\sim$2 orders of magnitude for typical field ages. However, this decline appears to be weakest in Ly$\alpha$, which will change the spectroscopic balance of energy output (i.e., FUV/NUV ratio) between young and old M-dwarfs. This evolutionary behavior is evident throughout the M-dwarf sample for both early and mid-to-late M-dwarfs. Because these stars show similar FUV luminosities relative to bolometric, the early M-dwarfs are more UV luminous in absolute terms, but with potentially different spin-down behaviors, the cooler stars may emit at the saturation levels for a greater duration of their early lifetimes. This evolutionary behavior may have several implications across stellar astrophysics and exoplanetary science. 

\subsection{Stellar Atmospheres and Dynamos}

The FUV emission lines directly probe the transition region of the stellar coronal atmosphere. Our spectroscopic data showed how the different features change with rotation, directly implying how the atmospheric structure changes across time, between active and inactive low-mass stars. These data can thus be used to directly assess those changes through stellar chromospheric and coronal models \citep[e.g.,][]{Fontenla2016,Peacock2019b}. The comparison of the rotation-activity relationships across H$\alpha$, the FUV, and X-rays, however, suggests more significant changes in the corona with spin-down relative to the changes in the deeper layers of the atmosphere. These differences must be a direct consequence of how the non-thermal heating processes change with rotation rate. Models of the magnetic heating process itself in M-dwarfs must be able to account for these evolutionary effects, and their significance at different layers of the atmosphere. 

This aspect of non-thermal chromospheric/coronal heating is an important consideration in making dynamo inferences, because the observed rotation-activity relationships in different wavebands, often used for this endeavor, are mediated by the magnetic heating, and are not directly defined by the dynamo processes. The multi-wavelength rotation-activity relationships need to be reconciled with more homogeneous methodologies, including the possibly different behaviors in different upper atmospheric layers in order to provide definitive conclusions with respect to dynamo theory. Crucially, while our analysis shows that a Rossby scaling works well for normalizing the rotation-activity relationships in both partly and fully convective M-dwarfs, the empirical scaling may be masking real differences in these populations. This effect is potentially evident in alternative scalings, however, larger samples are required to examine these differences in the UV, to compare with other wavebands \citep[e.g.,][]{Magaudda2020}.

\subsection{Exoplanets}

The stellar high-energy spectrum largely defines the prevalent photochemistry and mass-loss history of exoplanetary systems. While these effects have often been investigated in the context of individual nearby planets around low-mass stars, the present day observations of the planetary atmosphere have been shaped by the cumulative history of these stellar emissions. Our spectroscopic FUV data provide its rotational evolution for M-dwarfs directly, which can be transformed using rotation-age gyrochronology relationships. Although the latter remain uncertain for M-dwarfs, their improvement will greatly improve our assessments of this evolutionary history. Employing literature scaling relations using FUV emission features then enables estimates across the high-energy spectrum, including the EUV \citep[e.g.,][]{France2016,France2018}. To understand exoplanetary atmospheres this stellar emission history needs to be taken into account. 

Of particular importance is the likely difference between early and mid-to-late M-dwarfs, with regard to how long they persist in exhibiting near saturation level activity. By old field ages, planets in similar orbits around these two different kinds of hosts will have experienced likely different histories in high-energy radiative environments \citep[e.g.,][]{Luger2015}. Moreover, changes in the relative significance of FUV or NUV emissions over time will influence the prevalent exoplanetary atmospheric molecules that are observable today. Our results enable a way to account for these effects across different emission features and wavebands when considering new exoplanetary systems.

Using our rotation-activity correlation fits (Section~\ref{sec:rotact}), assuming no residual mass dependence, we can predict the most prominent FUV emission features in quiescence from a known rotation period and mass to generally within 0.3 dex of intrinsic scatter. This scatter pertains to the sample population and is likely a consequence of the combined effects of activity cycles, rotation variations in visible active regions, metallicity differences, and/or the stochastic nature of magnetic heating. As an example, we imagine a 0.4 $M_{\odot}$ star with 60 d rotation period, with 3\% uncertainty on the mass and 5\% in the rotation period. Accounting for scatter in the Rossby number calibration and our best fit parameters, our rotation-activity correlations would predict a mean value of $\log ( L_{\mathrm{CIV}} / L_{\mathrm{bol}} ) = -6.21 \pm 0.16$. With improved rotation-age relations, our data will enable a more comprehensive assessment of the high-energy radiative input to exoplanetary systems across time.

\section{Summary}\label{sec:summary}

In this paper, we have examined the far ultraviolet emission of M-dwarf stars, as probes of the stellar upper atmosphere and non-thermal magnetic heating, their rotational evolution, and possible implications for planetary systems orbiting these kinds of hosts. Additional FUMES papers will discuss the Ly$\alpha$ reconstructions (Youngblood et. al.\ \textit{accepted}), and time variability in the UV/optical emissions (Duvvuri et al.\ \textit{in prep}). For this work, our primary findings are summarized below.

\begin{itemize}
	\item We reported emission line-emission line correlations across $\sim$3 orders of magnitude for the FUV features with respect to C\textsc{iv} emissions, revealing $\sim$0.1-0.2 dex of intrinsic scatter between FUV features defining the extent to which such features can be used to predict one another across the M-dwarf population, see Section~\ref{sec:lineline}. 
	\item We provided rotation-activity correlations as a function of Rossby number across 8 UV features, including Ly$\alpha$, with typical power-law slopes of $-2$ and critical Rossby numbers of 0.2, see Section~\ref{sec:rotact}.
	\item The decay of Ly$\alpha$ emission with rotation is likely weaker than it is for other FUV features, implying evolutionary changes in the relative balance of UV spectroscopic emissions, see Section~\ref{sec:rotact_ratio}.
	\item A possible trend in the rotation-activity correlations as a function of atmospheric layer points to the importance of disentangling magnetic heating effects through the stellar atmosphere when investigating the dynamo dependence on rotation rate, see Section~\ref{sec:rotact_temp}. 
	\item We demonstrated systematic effects in the resulting fit parameters for rotation-activity correlations (power-law slope, critical Rossby number) when utilizing different empirical calibrations for the convective turn over time as a function of mass, see Appendix~\ref{sec:ap_rossby}. 
	\item Mid-to-late M-dwarfs may exhibit saturation level FUV activity for a longer duration of their early lifetimes relative to early M-dwarfs, with correspondingly distinct histories of high-energy emission impacting exoplanetary systems around these hosts, see Section~\ref{sec:age_evo}. 
	\item Planets in the habitable zones around mid-to-late M-dwarfs, at field ages, will have accumulated $\sim$2$\times$ more EUV exposure than planets around early M-dwarfs, and 20$\times$ more exposure than planets in the habitable zones around Sun-like stars, see Section~\ref{sec:accumUV}.
	\item For planets orbiting low-mass stars, the majority of energetic UV exposure accumulated by the age of 5 Gyrs was experienced during the saturated phase of activity evolution, lasting $\sim$1~Gyr, see Section~\ref{sec:accumUV}.
\end{itemize}

\section*{Acknowledgments}

The authors would like to thank Zachary Berta-Thompson, Elisabeth Newton, and Girish Duvvuri for useful discussions and commentary in the preparation of this work. The authors also acknowledge R.\ O.\ Parke Loyd and  Alejandro Nu\~{n}ez for providing additional information used in this paper. We further thank the anonymous referee for their comments in the preparation of this publication.

Support for Program numbers HST-GO 14640, 14633 and 15071 were provided by NASA
through a grant from the Space Telescope Science Institute, which is
operated by the Association of Universities for Research in Astronomy,
Incorporated, under NASA contract NAS5-26555.

A.Y. acknowledges support by an appointment to the NASA Postdoctoral Program
at Goddard Space Flight Center, administered by USRA through a contract with NASA.

Based on observations at Cerro Tololo Inter-American Observatory, National Optical Astronomy Observatory (NOAO Prop. ID 2017B-0316 ; PI: J. Pineda), which is operated by the Association of Universities for Research in Astronomy (AURA) under a cooperative agreement with the National Science Foundation. 

Based on observations obtained with the Apache Point Observatory 3.5-meter telescope, which is owned and operated by the Astrophysical Research Consortium.

\facility{HST (STIS), Blanco (ARCoIRIS), APO (TSPEC)}


\section*{Appendix}
\renewcommand{\thesubsection}{\Alph{subsection}}


\subsection{Concerning the Use of Surface Fluxes}\label{sec:ap_surfaceF}

\begin{figure}[tbp]
	\centering
	\includegraphics[width=0.5\textwidth]{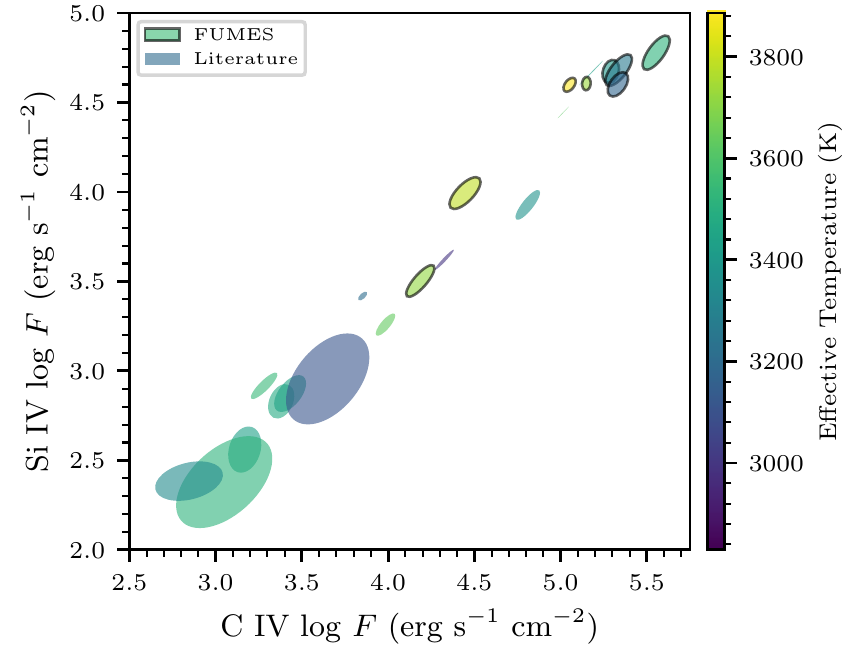} 
	\caption{The measured emission strengths of C\textsc{iv} and Si\textsc{iv} as in Figures~\ref{fig:lineline_a} and~\ref{fig:lineline_a}, but shown as surface flux instead of luminosity utilizing our radius determinations from Table~\ref{tab:fumes}. Line-line regression analyzes not accounting for correlated uncertainties, as shown here by slanted ellipses, will yield biased results, see Appendix~\ref{sec:ap_surfaceF}.} 
	\label{fig:ap_surfaceF}
\end{figure}

In Section~\ref{sec:lineline}, we correlated FUV line flux measurements, illustrating tight relationships between different emission features probing distinct temperatures of the transition region. In the literature these kinds of correlations have been expressed similarly, but instead of luminosity, they have been expressed using the surface flux, normalizing the luminosities by the stellar surface area \citep[e.g.,][]{Wood2005,Youngblood2017}. While this attempts to normalize the emissions accounting for the area of the emitting region to enable comparisons across different kinds of stars, the inclusion of the radii introduces additional uncertainty and increases the error correlations, as the radius uncertainty will dominate the error budget relative to the parallax and flux measurements.\footnote{This effect was less significant in the past when parallaxes were not known as precisely (pre-\textit{Gaia}).} In Figure~\ref{fig:ap_surfaceF}, we illustrate this effect using the same data for C\textsc{iv} and Si\textsc{iv} that comprise the upper right most panel of Figure~\ref{fig:lineline_a}, but transforming the emission measurements to surface flux with the radius determinations from Table~\ref{tab:fumes}. The representative 2$\sigma$ error ellipses now are all highly inclined, revealing the strong correlations between the uncertainties in each quantity. 

The power-law slope estimated in such line-line correlations using the surface flux should be identical to the luminosity approach used in this work, however, the uncertainties on such estimates do not accurately reflect the nature of the underlying data if they do not account for this correlated error, and are less precisely constrained when accounting for the actual radii uncertainty in the analysis. This additional uncertainty may obscure intrinsic scatter in the analyzed correlations. We therefore recommend luminosity as a currently more robust choice when defining predictive relations between stellar emission features. However, whenever surface fluxes are necessary, a careful accounting of possibly correlations should be included.

\subsection{Systematics with the Convective Turnover Time }\label{sec:ap_rossby}

In Section~\ref{sec:rotact}, we analyzed the rotation-activity relation of M-dwarfs in FUV emission lines with the FUMES and literature samples. We used the empirical calibration of \cite{Wright2018} to estimate the convective turn over time from the stellar mass in computing the Rossby number, $Ro = P / \tau_{c} $, for each star. This empirical calibration is based on minimizing the scatter in the X-ray rotation-activity correlation of low-mass stars. \cite{Wright2011} details the typical procedures used in developing this kind of empirical calibration. Since the $Ro$ is generally closely related to the internal dynamo action \citep[although see][]{Reiners2014}, the use of this kind of calibration enables a dynamo comparison amongst stars with convective interiors, from F to M stars. However, changes in the kind of magnetic dynamo that generates field in fully convective stars, for example $\alpha^2$ instead of $\alpha$-$\Omega$ \citep[e.g.,][]{Browning2008}, suggest that there is no physical reason a single such calibration should work across that full range. Moreover, with relatively deeper convective zones, a single representative value for the time scale of convective motions is likely an increasingly poor approximation with decreasing stellar mass. Accordingly, although the empirical calibration reduces the X-ray rotation-activity scatter, it may not be representative of the dynamo behavior in the fully convective regime.

\begin{figure}[tbp]
	\centering
	\includegraphics[width=0.5\textwidth]{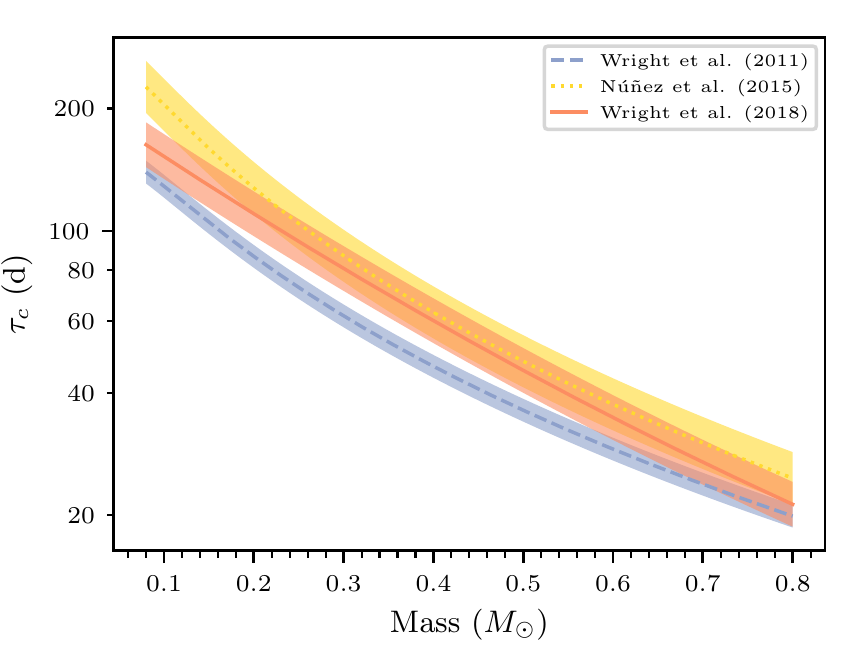} 
	\caption{Different calibrations for the convective turn over time as a function of mass yield systematic differences in rotation-activity analyses when used to compute a characteristic Rossby number, $Ro$, see Appendix~\ref{sec:ap_rossby}.} 
	\label{fig:ap_convTcomp}
\end{figure}

Nevertheless, using these relations provides a means to compare to literature results and test how the X-ray calibrated relation applies to the activity at other wavelengths, as we have done in Section~\ref{sec:rotact}. As detailed in this Appendix, we further investigated how those results are impacted by the choice of empirical calibration for the convective turnover time as a function of stellar mass. In Figure~\ref{fig:ap_convTcomp}, we plot three such literature calibrations from \cite{Wright2011}, \cite{Nunez2015}, and \cite{Wright2018}, including the scatter about those relationships, reported in those works, or obtained via private communication (0.064 dex, Nu\~{n}ez, A.). The \cite{Nunez2015} calibration uses the same data as \cite{Wright2011}, but assumes the canonical value for the best fit slope ($-$2) instead of the best fit value from \cite{Wright2011}, and the \cite{Wright2018} calibration, which sits in between the other two, updates the \cite{Wright2011} result with more fully convective stars and a slightly different functional form. 

We re-fit the rotation-activity data presented in Section~\ref{sec:rotact}, using the same methods, but using the two additional literature calibrations to determine their impact on the best fit parameters. These results are shown in Table~\ref{tab:ap_sysConvT_ro} for the critical Rossby number, $Ro_{c}$, and in Table~\ref{tab:ap_sysConvT_eta} for the slope of the unsaturated regime, $\eta$. The effect of the different calibrations is most readily illustrated by comparing the results using \cite{Wright2011} vs. \cite{Nunez2015}, as they have a greater separation in $\tau_{c}$-$M$ space (see Figure~\ref{fig:ap_convTcomp}). The \cite{Nunez2015} calibration gives higher convective turnover times at a given mass than \cite{Wright2011}. Consequently, the best fit $Ro_{c}$ is systematically smaller using \cite{Nunez2015} than it is when using the \cite{Wright2011} calibration --- higher $\tau_{c}$ corresponds to smaller $Ro$. Between the two calibrations as applied to our data, this yielded a systematic offset of $\sim$0.04-0.05 in $Ro_{c}$, see Table~\ref{tab:ap_sysConvT_ro}.

\begin{deluxetable}{l c c c c}
	\tablecaption{ Critical Rossby Systematics: $Ro_{c}$ \tablenotemark{a}
		\label{tab:ap_sysConvT_ro} }
	\tablehead{
		\colhead{Line} & \colhead{\cite{Wright2011}} & \colhead{\cite{Nunez2015}} & \colhead{\cite{Wright2018}} 
	}
	\startdata
	Ly$\alpha$ & $0.25 \pm ^{0.20}_{0.12} $& $0.19 \pm ^{0.14}_{0.10} $& $  \mathbf{  0.21 \pm ^{0.16}_{0.11}   }$  \\
	Mg\textsc{ii} & $0.23 \pm ^{0.14}_{0.09} $& $0.19 \pm ^{0.11}_{0.08} $& $  \mathbf{  0.20 \pm ^{0.11}_{0.08}   }$ \\		
	C\textsc{ii} & $0.28 \pm ^{0.15}_{0.14} $& $0.20 \pm ^{0.10}_{0.09} $& $  \mathbf{  0.24 \pm ^{0.12}_{0.11}   }$   \\
	Si\textsc{iii} &  $0.23 \pm  0.08 $& $0.18 \pm 0.06 $& $  \mathbf{  0.20 \pm  0.07    }$   \\
	He\textsc{ii} &$0.22 \pm 0.06 $& $0.16 \pm ^{0.05}_{0.04} $& $  \mathbf{  0.19 \pm 0.05   }$   \\
	Si\textsc{iv} & $0.28 \pm 0.08 $& $0.22 \pm 0.06 $& $  \mathbf{  0.24 \pm 0.07   }$   \\
	C\textsc{iv}	 & $0.22 \pm 0.08 $& $0.17 \pm 0.06$& $  \mathbf{  0.18 \pm ^{0.07}_{0.06}   }$ \\	
	N\textsc{v}	 & $0.22 \pm 0.09 $& $0.18 \pm 0.06 $& $  \mathbf{  0.19 \pm 0.07   }$  \\
	\enddata
	\tablenotetext{a}{Reported parameters correspond to the median of the marginalized posterior distribution with uncertainties indicating the central 68\% confidence interval. The reference for each column indicates the source used for the calibration of the mass dependent convective turn over time.}
	
\end{deluxetable}

\begin{deluxetable}{l c c c c}
	\tablecaption{ Unsaturated Slope Systematics: $\eta$ \tablenotemark{a}
		\label{tab:ap_sysConvT_eta} }
	\tablehead{
		\colhead{Line} & \colhead{\cite{Wright2011}} & \colhead{\cite{Nunez2015}} & \colhead{\cite{Wright2018}} 
	}
	\startdata
	Ly$\alpha$ & $-1.21 \pm ^{0.39}_{0.54} $& $-1.25 \pm ^{0.42}_{0.56} $& $  \mathbf{  -1.26 \pm ^{0.41}_{0.58}   }$  \\
	Mg\textsc{ii} & $-1.72 \pm ^{0.37}_{0.51} $& $-1.86 \pm ^{0.44}_{0.58} $& $  \mathbf{  -1.86 \pm ^{0.41}_{0.55}   }$  \\		
	C\textsc{ii} & $-2.25 \pm ^{0.62}_{0.72} $& $-2.25 \pm ^{0.57}_{0.70} $& $  \mathbf{  -2.38 \pm ^{0.64}_{0.77}   }$  \\
	Si\textsc{iii} & $-1.94 \pm ^{0.39}_{0.41} $& $-2.10 \pm ^{0.41}_{0.42} $& $  \mathbf{  -2.08 \pm ^{0.41}_{0.46}   }$  \\
	He\textsc{ii} &$-2.11 \pm ^{0.27}_{0.29} $& $-2.15 \pm ^{0.30}_{0.32} $& $  \mathbf{  -2.19 \pm ^{0.30}_{0.33}   }$ \\
	Si\textsc{iv} & $-2.21 \pm ^{0.40}_{0.41} $& $-2.31 \pm ^{0.39}_{0.42} $& $  \mathbf{  -2.32 \pm ^{0.42}_{0.44}   }$ \\
	C\textsc{iv}	 &$-1.92 \pm ^{0.37}_{0.39} $& $-2.04 \pm ^{0.37}_{0.39} $& $  \mathbf{  -2.02 \pm ^{0.39}_{0.41}   }$  \\	
	N\textsc{v}	 &$-1.76 \pm ^{0.37}_{0.38} $& $-1.88 \pm ^{0.36}_{0.37} $& $  \mathbf{  -1.84 \pm ^{0.37}_{0.40}   }$  \\
	\enddata
	\tablenotetext{a}{Reported parameters correspond to the median of the marginalized posterior distribution with uncertainties indicating the central 68\% confidence interval. The reference for each column indicates the source used for the calibration of the mass dependent convective turn over time.}
	
\end{deluxetable}

\begin{figure}[tbp]
	\centering
	\includegraphics[width=0.5\textwidth]{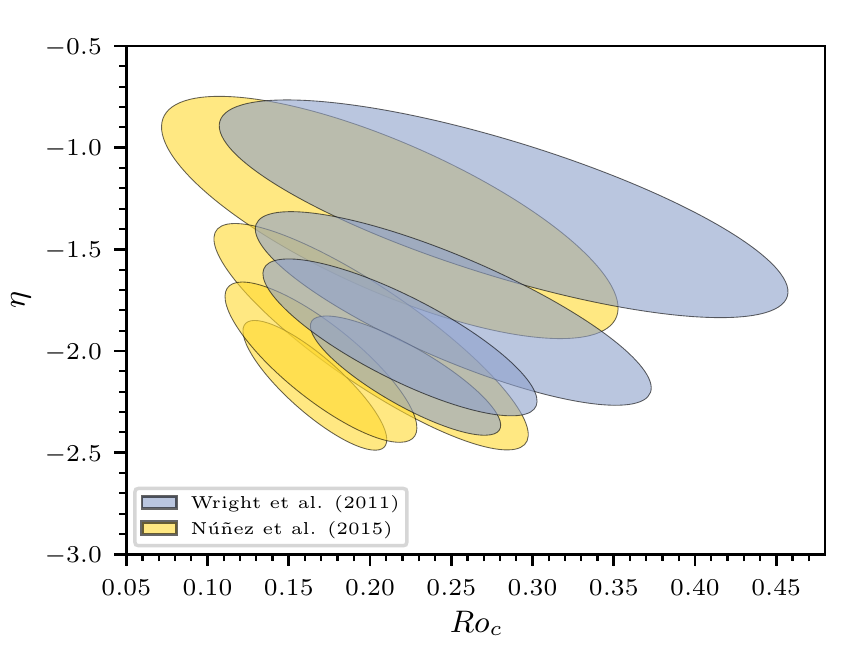} 
	\caption{Changing the assumed calibration for convective turnover time as a function of stellar mass when fitting the canonical rotation-activity relationships systematically affects the best fit parameters for the unsaturated slope, $\eta$, and critical Rossby number, $Ro_{c}$. The representative error ellipses shown here for four of the FUV line rotation-activity fits collectively shift, yellow \citep{Nunez2015} relative to blue \citep{Wright2011}, when using distinct calibrations, see Appendix~\ref{sec:ap_rossby}. } 
	\label{fig:ap_etaRoss}
\end{figure}

This effect on the best fit critical Rossby number is relatively intuitive given the direct impact on the convective turnover time between calibration choices. However, we also observe a systematic difference in the best fit slope from the rotation-activity analysis when changing between empirical calibrations. With our data, larger assumed convective time scales (smaller $Ro$) yielded systematically steeper slopes (more negative) for the unsaturated regime, see Table~\ref{tab:ap_sysConvT_eta}. Although small, this effect is generally evident across the eight different lines we analyzed. To illustrate this systematic effect, we show representative ellipses for the joint posterior distributions of $Ro_{c}$-$\eta$ across 4 FUV lines in Figure~\ref{fig:ap_etaRoss}. The error ellipses for each individual line shift to the left (smaller $Ro_{c}$), and down (steeper $\eta$) when changing calibrations from \cite{Wright2011} to \cite{Nunez2015}. Because these fit parameters are correlated, it is perhaps unsurprising that systematic effects would appear in both $Ro_{c}$ and $\eta$, however, the systematic shift is not in the same direction, as the critical Rossby number and slope posteriors are anti-correlated, not correlated.

We attribute this systematic effect to the non-linearity of the calibrations as applied to individual samples. If the choice of empirical calibration scaled the assumed Rossby numbers of all of the stars in the same way, we could expect the best fit slope of the unsaturated regime to remain constant. A comparison of the calibrations for $\tau_{c} (M)$ (see Figure~\ref{fig:ap_convTcomp}), shows that this is generally not the case. Thus, depending on the sample of stars, some objects shift in $Ro$ space more than others. A large number of fully convective stars in the sample would likely increase the magnitude of this systematic effect on the best-fit slopes between the calibrations of \cite{Wright2011} and \cite{Nunez2015}, as that is where those functions largely diverge. It is therefore difficult to estimate the extent of this systematic effect on the slopes without doing the entirety of the analysis with multiple calibrations for $\tau_{c}$ for each sample of stars. This makes comparisons somewhat more difficult across the literature as methods have been updated and evolved over time, with different stellar samples. Future comparisons across wavebands and samples will greatly benefit from homogeneous analysis methodologies. The framework presented in this paper for the rotation-activity work (Section~\ref{sec:rotact}), accounts for known uncertainties across all available data, possible correlations, and includes a measure of the intrinsic scatter within the regression fit. The presence of these systematics effects, especially when using a quantity as uncertain as the convective turnover time scale, also supports the argument in favor of finding simpler descriptions that capture the relevant physics for characterizing the dependence of activity on stellar physical and rotational properties, as discussed in \cite{Reiners2014}. We tested some of those methods in Section~\ref{sec:rotact_RP}.

\bibliographystyle{aasjournal}
\bibliography{fumes1_paper1sub_arXiv}

\begin{thebibliography}{}
\expandafter\ifx\csname natexlab\endcsname\relax\def\natexlab#1{#1}\fi
\providecommand{\url}[1]{\href{#1}{#1}}
\providecommand{\dodoi}[1]{doi:~\href{http://doi.org/#1}{\nolinkurl{#1}}}
\providecommand{\doeprint}[1]{\href{http://ascl.net/#1}{\nolinkurl{http://ascl.net/#1}}}
\providecommand{\doarXiv}[1]{\href{https://arxiv.org/abs/#1}{\nolinkurl{https://arxiv.org/abs/#1}}}

\bibitem[{{Alonso-Floriano} {et~al.}(2015){Alonso-Floriano}, {Morales},
  {Caballero}, {Montes}, {Klutsch}, {Mundt}, {Cort{\'e}s-Contreras}, {Ribas},
  {Reiners}, {Amado}, {Quirrenbach}, \& {Jeffers}}]{AlonsoFloriano2015}
{Alonso-Floriano}, F.~J., {Morales}, J.~C., {Caballero}, J.~A., {et~al.} 2015,
  \aap, 577, A128, \dodoi{10.1051/0004-6361/201525803}

\bibitem[{{Angus} {et~al.}(2015){Angus}, {Aigrain}, {Foreman-Mackey}, \&
  {McQuillan}}]{Angus2015}
{Angus}, R., {Aigrain}, S., {Foreman-Mackey}, D., \& {McQuillan}, A. 2015,
  \mnras, 450, 1787, \dodoi{10.1093/mnras/stv423}

\bibitem[{{Astudillo-Defru} {et~al.}(2017){Astudillo-Defru}, {Delfosse},
  {Bonfils}, {Forveille}, {Lovis}, \& {Rameau}}]{AstudilloDefru2017}
{Astudillo-Defru}, N., {Delfosse}, X., {Bonfils}, X., {et~al.} 2017, \aap, 600,
  A13, \dodoi{10.1051/0004-6361/201527078}

\bibitem[{{Ayres} {et~al.}(2003){Ayres}, {Brown}, {Harper}, {Osten}, {Linsky},
  {Wood}, \& {Redfield}}]{Ayres2003}
{Ayres}, T.~R., {Brown}, A., {Harper}, G.~M., {et~al.} 2003, \apj, 583, 963,
  \dodoi{10.1086/345409}

\bibitem[{{Barnes}(2003)}]{Barnes2003}
{Barnes}, S.~A. 2003, \apj, 586, 464, \dodoi{10.1086/367639}

\bibitem[{{Barnes}(2010)}]{Barnes2010}
---. 2010, \apj, 722, 222, \dodoi{10.1088/0004-637X/722/1/222}

\bibitem[{{Bell} {et~al.}(2015){Bell}, {Mamajek}, \& {Naylor}}]{Bell2015}
{Bell}, C. P.~M., {Mamajek}, E.~E., \& {Naylor}, T. 2015, \mnras, 454, 593,
  \dodoi{10.1093/mnras/stv1981}

\bibitem[{{Bourrier} {et~al.}(2017){Bourrier}, {Ehrenreich}, {Allart},
  {Wyttenbach}, {Semaan}, {Astudillo-Defru}, {Gracia-Bern{\'a}}, {Lovis},
  {Pepe}, {Thomas}, \& {Udry}}]{Bourrier2017c}
{Bourrier}, V., {Ehrenreich}, D., {Allart}, R., {et~al.} 2017, \aap, 602, A106,
  \dodoi{10.1051/0004-6361/201730542}

\bibitem[{{Bourrier} {et~al.}(2018){Bourrier}, {Lovis}, {Beust}, {Ehrenreich},
  {Henry}, {Astudillo-Defru}, {Allart}, {Bonfils}, {S{\'e}gransan}, {Delfosse},
  {Cegla}, {Wyttenbach}, {Heng}, {Lavie}, \& {Pepe}}]{Bourrier2018}
{Bourrier}, V., {Lovis}, C., {Beust}, H., {et~al.} 2018, \nat, 553, 477,
  \dodoi{10.1038/nature24677}

\bibitem[{{Boyajian} {et~al.}(2012){Boyajian}, {von Braun}, {van Belle},
  {McAlister}, {ten Brummelaar}, {Kane}, {Muirhead}, {Jones}, {White},
  {Schaefer}, {Ciardi}, {Henry}, {L{\'o}pez-Morales}, {Ridgway}, {Gies}, {Jao},
  {Rojas-Ayala}, {Parks}, {Sturmann}, {Sturmann}, {Turner}, {Farrington},
  {Goldfinger}, \& {Berger}}]{Boyajian2012}
{Boyajian}, T.~S., {von Braun}, K., {van Belle}, G., {et~al.} 2012, \apj, 757,
  112, \dodoi{10.1088/0004-637X/757/2/112}

\bibitem[{{Browning}(2008)}]{Browning2008}
{Browning}, M.~K. 2008, \apj, 676, 1262, \dodoi{10.1086/527432}

\bibitem[{{Browning} {et~al.}(2006){Browning}, {Miesch}, {Brun}, \&
  {Toomre}}]{Browning2006}
{Browning}, M.~K., {Miesch}, M.~S., {Brun}, A.~S., \& {Toomre}, J. 2006, \apjl,
  648, L157, \dodoi{10.1086/507869}

\bibitem[{{Charbonneau} \& {MacGregor}(1997)}]{Charbonneau1997}
{Charbonneau}, P., \& {MacGregor}, K.~B. 1997, \apj, 486, 502,
  \dodoi{10.1086/304485}

\bibitem[{{Cushing} {et~al.}(2004){Cushing}, {Vacca}, \&
  {Rayner}}]{Cushing2004}
{Cushing}, M.~C., {Vacca}, W.~D., \& {Rayner}, J.~T. 2004, \pasp, 116, 362,
  \dodoi{10.1086/382907}

\bibitem[{{Donati} {et~al.}(2008){Donati}, {Morin}, {Petit}, {Delfosse},
  {Forveille}, {Auri{\`e}re}, {Cabanac}, {Dintrans}, {Fares}, {Gastine},
  {Jardine}, {Ligni{\`e}res}, {Paletou}, {Ramirez Velez}, \&
  {Th{\'e}ado}}]{Donati2008}
{Donati}, J.-F., {Morin}, J., {Petit}, P., {et~al.} 2008, \mnras, 390, 545,
  \dodoi{10.1111/j.1365-2966.2008.13799.x}

\bibitem[{{Dotter} {et~al.}(2008){Dotter}, {Chaboyer}, {Jevremovi{\'c}},
  {Kostov}, {Baron}, \& {Ferguson}}]{Dotter2008}
{Dotter}, A., {Chaboyer}, B., {Jevremovi{\'c}}, D., {et~al.} 2008, The
  Astrophysical Journal Supplement Series, 178, 89, \dodoi{10.1086/589654}

\bibitem[{{Dressing} \& {Charbonneau}(2015)}]{Dressing2015}
{Dressing}, C.~D., \& {Charbonneau}, D. 2015, \apj, 807, 45,
  \dodoi{10.1088/0004-637X/807/1/45}

\bibitem[{{Engle} \& {Guinan}(2018)}]{Engle2018RNAAS}
{Engle}, S.~G., \& {Guinan}, E.~F. 2018, Research Notes of the American
  Astronomical Society, 2, 34, \dodoi{10.3847/2515-5172/aab1f8}

\bibitem[{{Feiden}(2016)}]{Feiden2016}
{Feiden}, G.~A. 2016, Astronomy and Astrophysics, 593, A99,
  \dodoi{10.1051/0004-6361/201527613}

\bibitem[{{Feiden} \& {Chaboyer}(2013)}]{Feiden2013}
{Feiden}, G.~A., \& {Chaboyer}, B. 2013, The Astrophysical Journal, 779, 183,
  \dodoi{10.1088/0004-637X/779/2/183}

\bibitem[{{Feiden} \& {Chaboyer}(2014)}]{Feiden2014}
---. 2014, The Astrophysical Journal, 789, 53,
  \dodoi{10.1088/0004-637X/789/1/53}

\bibitem[{{Fontenla} {et~al.}(2016){Fontenla}, {Linsky}, {Witbrod}, {France},
  {Buccino}, {Mauas}, {Vieytes}, \& {Walkowicz}}]{Fontenla2016}
{Fontenla}, J.~M., {Linsky}, J.~L., {Witbrod}, J., {et~al.} 2016, \apj, 830,
  154, \dodoi{10.3847/0004-637X/830/2/154}

\bibitem[{{France} {et~al.}(2018){France}, {Arulanantham}, {Fossati}, {Lanza},
  {Loyd}, {Redfield}, \& {Schneider}}]{France2018}
{France}, K., {Arulanantham}, N., {Fossati}, L., {et~al.} 2018, \apjs, 239, 16,
  \dodoi{10.3847/1538-4365/aae1a3}

\bibitem[{{France} {et~al.}(2013){France}, {Froning}, {Linsky}, {Roberge},
  {Stocke}, {Tian}, {Bushinsky}, {D{\'e}sert}, {Mauas}, {Vieytes}, \&
  {Walkowicz}}]{France2013}
{France}, K., {Froning}, C.~S., {Linsky}, J.~L., {et~al.} 2013, \apj, 763, 149,
  \dodoi{10.1088/0004-637X/763/2/149}

\bibitem[{{France} {et~al.}(2016){France}, {Parke Loyd}, {Youngblood}, {Brown},
  {Schneider}, {Hawley}, {Froning}, {Linsky}, {Roberge}, {Buccino},
  {Davenport}, {Fontenla}, {Kaltenegger}, {Kowalski}, {Mauas}, {Miguel},
  {Redfield}, {Rugheimer}, {Tian}, {Vieytes}, {Walkowicz}, \&
  {Weisenburger}}]{France2016}
{France}, K., {Parke Loyd}, R.~O., {Youngblood}, A., {et~al.} 2016, \apj, 820,
  89, \dodoi{10.3847/0004-637X/820/2/89}

\bibitem[{{France} {et~al.}(2019){France}, {Fleming}, {Drake}, {Mason},
  {Youngblood}, {Bourrier}, {Fossati}, {Froning}, {Koskinen}, {Kruczek},
  {Lipscy}, {McEntaffer}, {Romaine}, {Siegmund}, \& {Wilkinson}}]{France2019}
{France}, K., {Fleming}, B.~T., {Drake}, J.~J., {et~al.} 2019, in Society of
  Photo-Optical Instrumentation Engineers (SPIE) Conference Series, Vol. 11118,
  \procspie, 1111808, \dodoi{10.1117/12.2526859}

\bibitem[{{Gallet} \& {Bouvier}(2015)}]{Gallet2015}
{Gallet}, F., \& {Bouvier}, J. 2015, \aap, 577, A98,
  \dodoi{10.1051/0004-6361/201525660}

\bibitem[{{Gao} {et~al.}(2015){Gao}, {Hu}, {Robinson}, {Li}, \&
  {Yung}}]{Gao2015}
{Gao}, P., {Hu}, R., {Robinson}, T.~D., {Li}, C., \& {Yung}, Y.~L. 2015, \apj,
  806, 249, \dodoi{10.1088/0004-637X/806/2/249}

\bibitem[{{Garraffo} {et~al.}(2015){Garraffo}, {Drake}, \&
  {Cohen}}]{Garraffo2015}
{Garraffo}, C., {Drake}, J.~J., \& {Cohen}, O. 2015, \apj, 807, L6,
  \dodoi{10.1088/2041-8205/807/1/L6}

\bibitem[{{Garraffo} {et~al.}(2018){Garraffo}, {Drake}, {Dotter}, {Choi},
  {Burke}, {Moschou}, {Alvarado-G{\'o}mez}, {Kashyap}, \&
  {Cohen}}]{Garraffo2018}
{Garraffo}, C., {Drake}, J.~J., {Dotter}, A., {et~al.} 2018, \apj, 862, 90,
  \dodoi{10.3847/1538-4357/aace5d}

\bibitem[{{Guinan} {et~al.}(2016){Guinan}, {Engle}, \& {Durbin}}]{Guinan2016}
{Guinan}, E.~F., {Engle}, S.~G., \& {Durbin}, A. 2016, \apj, 821, 81,
  \dodoi{10.3847/0004-637X/821/2/81}

\bibitem[{{Hall}(2008)}]{Hall2008}
{Hall}, J.~C. 2008, Living Reviews in Solar Physics, 5, 2,
  \dodoi{10.12942/lrsp-2008-2}

\bibitem[{{Harman} {et~al.}(2015){Harman}, {Schwieterman}, {Schottelkotte}, \&
  {Kasting}}]{Harman2015}
{Harman}, C.~E., {Schwieterman}, E.~W., {Schottelkotte}, J.~C., \& {Kasting},
  J.~F. 2015, \apj, 812, 137, \dodoi{10.1088/0004-637X/812/2/137}

\bibitem[{{Hartman} {et~al.}(2011){Hartman}, {Bakos}, {Noyes}, {Sip{\H{o}}cz},
  {Kov{\'a}cs}, {Mazeh}, {Shporer}, \& {P{\'a}l}}]{Hartman2011}
{Hartman}, J.~D., {Bakos}, G.~{\'A}., {Noyes}, R.~W., {et~al.} 2011, \aj, 141,
  166, \dodoi{10.1088/0004-6256/141/5/166}

\bibitem[{{Hawley} {et~al.}(1996){Hawley}, {Gizis}, \& {Reid}}]{Hawley1996}
{Hawley}, S.~L., {Gizis}, J.~E., \& {Reid}, I.~N. 1996, \aj, 112, 2799,
  \dodoi{10.1086/118222}

\bibitem[{{Hawley} \& {Pettersen}(1991)}]{Hawley1991}
{Hawley}, S.~L., \& {Pettersen}, B.~R. 1991, \apj, 378, 725,
  \dodoi{10.1086/170474}

\bibitem[{{Hawley} {et~al.}(2007){Hawley}, {Walkowicz}, {Allred}, \&
  {Valenti}}]{Hawley2007}
{Hawley}, S.~L., {Walkowicz}, L.~M., {Allred}, J.~C., \& {Valenti}, J.~A. 2007,
  \pasp, 119, 67, \dodoi{10.1086/510561}

\bibitem[{{Hawley} {et~al.}(2003){Hawley}, {Allred}, {Johns-Krull}, {Fisher},
  {Abbett}, {Alekseev}, {Avgoloupis}, {Deustua}, {Gunn}, {Seiradakis}, {Sirk},
  \& {Valenti}}]{Hawley2003b}
{Hawley}, S.~L., {Allred}, J.~C., {Johns-Krull}, C.~M., {et~al.} 2003, \apj,
  597, 535, \dodoi{10.1086/378351}

\bibitem[{{Houdebine} {et~al.}(2017){Houdebine}, {Mullan}, {Bercu}, {Paletou},
  \& {Gebran}}]{Houdebine2017}
{Houdebine}, E.~R., {Mullan}, D.~J., {Bercu}, B., {Paletou}, F., \& {Gebran},
  M. 2017, \apj, 837, 96, \dodoi{10.3847/1538-4357/aa5cad}

\bibitem[{{Kane} {et~al.}(2017){Kane}, {von Braun}, {Henry}, {Waters},
  {Boyajian}, \& {Mann}}]{Kane2017}
{Kane}, S.~R., {von Braun}, K., {Henry}, G.~W., {et~al.} 2017, \apj, 835, 200,
  \dodoi{10.3847/1538-4357/835/2/200}

\bibitem[{{Kelly}(2007)}]{Kelly2007}
{Kelly}, B.~C. 2007, \apj, 665, 1489, \dodoi{10.1086/519947}

\bibitem[{{Kirkpatrick}(2005)}]{Kirkpatrick2005}
{Kirkpatrick}, J.~D. 2005, \araa, 43, 195,
  \dodoi{10.1146/annurev.astro.42.053102.134017}

\bibitem[{{Kirkpatrick} {et~al.}(1991){Kirkpatrick}, {Henry}, \&
  {McCarthy}}]{Kirkpatrick1991}
{Kirkpatrick}, J.~D., {Henry}, T.~J., \& {McCarthy}, Donald~W., J. 1991, \apjs,
  77, 417, \dodoi{10.1086/191611}

\bibitem[{{Kirkpatrick} {et~al.}(1995){Kirkpatrick}, {Henry}, \&
  {Simons}}]{Kirkpatrick1995}
{Kirkpatrick}, J.~D., {Henry}, T.~J., \& {Simons}, D.~A. 1995, \aj, 109, 797,
  \dodoi{10.1086/117323}

\bibitem[{{Kirkpatrick} {et~al.}(1999){Kirkpatrick}, {Reid}, {Liebert},
  {Cutri}, {Nelson}, {Beichman}, {Dahn}, {Monet}, {Gizis}, \&
  {Skrutskie}}]{Kirkpatrick1999}
{Kirkpatrick}, J.~D., {Reid}, I.~N., {Liebert}, J., {et~al.} 1999, \apj, 519,
  802, \dodoi{10.1086/307414}

\bibitem[{Klimchuk(2006)}]{Klimchuk2006}
Klimchuk, J.~A. 2006, Solar Physics, 234, 41, \dodoi{10.1007/s11207-006-0055-z}

\bibitem[{{K{\"u}ker} {et~al.}(2019){K{\"u}ker}, {R{\"u}diger}, {Olah}, \&
  {Strassmeier}}]{Kuker2019}
{K{\"u}ker}, M., {R{\"u}diger}, G., {Olah}, K., \& {Strassmeier}, K.~G. 2019,
  \aap, 622, A40, \dodoi{10.1051/0004-6361/201833173}

\bibitem[{{Linsky}(1980)}]{Linsky1980}
{Linsky}, J.~L. 1980, \araa, 18, 439,
  \dodoi{10.1146/annurev.aa.18.090180.002255}

\bibitem[{{Linsky}(2017)}]{Linsky2017}
---. 2017, \araa, 55, 159, \dodoi{10.1146/annurev-astro-091916-055327}

\bibitem[{{Linsky} {et~al.}(1979){Linsky}, {Worden}, {McClintock}, \&
  {Robertson}}]{Linsky1979}
{Linsky}, J.~L., {Worden}, S.~P., {McClintock}, W., \& {Robertson}, R.~M. 1979,
  \apjs, 41, 47, \dodoi{10.1086/190607}

\bibitem[{{Loyd} \& {France}(2014)}]{Loyd2014}
{Loyd}, R.~O.~P., \& {France}, K. 2014, The Astrophysical Journal Supplement
  Series, 211, 9, \dodoi{10.1088/0067-0049/211/1/9}

\bibitem[{{Loyd} {et~al.}(2016){Loyd}, {France}, {Youngblood}, {Schneider},
  {Brown}, {Hu}, {Linsky}, {Froning}, {Redfield}, {Rugheimer}, \&
  {Tian}}]{Loyd2016}
{Loyd}, R.~O.~P., {France}, K., {Youngblood}, A., {et~al.} 2016, \apj, 824,
  102, \dodoi{10.3847/0004-637X/824/2/102}

\bibitem[{{Loyd} {et~al.}(2018){Loyd}, {France}, {Youngblood}, {Schneider},
  {Brown}, {Hu}, {Segura}, {Linsky}, {Redfield}, {Tian}, {Rugheimer}, {Miguel},
  \& {Froning}}]{Loyd2018}
---. 2018, \apj, 867, 71, \dodoi{10.3847/1538-4357/aae2bd}

\bibitem[{{Luger} \& {Barnes}(2015)}]{Luger2015}
{Luger}, R., \& {Barnes}, R. 2015, Astrobiology, 15, 119,
  \dodoi{10.1089/ast.2014.1231}

\bibitem[{{Magaudda} {et~al.}(2020){Magaudda}, {Stelzer}, {Covey}, {Raetz},
  {Matt}, \& {Scholz}}]{Magaudda2020}
{Magaudda}, E., {Stelzer}, B., {Covey}, K.~R., {et~al.} 2020, arXiv e-prints,
  arXiv:2004.02904.
\newblock \doarXiv{2004.02904}

\bibitem[{{Mallonn} {et~al.}(2018){Mallonn}, {Herrero}, {Juvan}, {von Essen},
  {Rosich}, {Ribas}, {Granzer}, {Alexoudi}, \& {Strassmeier}}]{Mallonn2018}
{Mallonn}, M., {Herrero}, E., {Juvan}, I.~G., {et~al.} 2018, \aap, 614, A35,
  \dodoi{10.1051/0004-6361/201732300}

\bibitem[{{Mann} {et~al.}(2015){Mann}, {Feiden}, {Gaidos}, {Boyajian}, \& {von
  Braun}}]{Mann2015}
{Mann}, A.~W., {Feiden}, G.~A., {Gaidos}, E., {Boyajian}, T., \& {von Braun},
  K. 2015, \apj, 804, 64, \dodoi{10.1088/0004-637X/804/1/64}

\bibitem[{{Mann} {et~al.}(2019){Mann}, {Dupuy}, {Kraus}, {Gaidos}, {Ansdell},
  {Ireland}, {Rizzuto}, {Hung}, {Dittmann}, \& {Factor}}]{Mann2019}
{Mann}, A.~W., {Dupuy}, T., {Kraus}, A.~L., {et~al.} 2019, \apj, 871, 63,
  \dodoi{10.3847/1538-4357/aaf3bc}

\bibitem[{{Meadows} {et~al.}(2018){Meadows}, {Reinhard}, {Arney}, {Parenteau},
  {Schwieterman}, {Domagal-Goldman}, {Lincowski}, {Stapelfeldt}, {Rauer},
  {DasSarma}, {Hegde}, {Narita}, {Deitrick}, {Lustig-Yaeger}, {Lyons},
  {Siegler}, \& {Grenfell}}]{Meadows2018}
{Meadows}, V.~S., {Reinhard}, C.~T., {Arney}, G.~N., {et~al.} 2018,
  Astrobiology, 18, 630, \dodoi{10.1089/ast.2017.1727}

\bibitem[{{Meibom} {et~al.}(2015){Meibom}, {Barnes}, {Platais}, {Gilliland},
  {Latham}, \& {Mathieu}}]{Meibom2015}
{Meibom}, S., {Barnes}, S.~A., {Platais}, I., {et~al.} 2015, \nat, 517, 589,
  \dodoi{10.1038/nature14118}

\bibitem[{{Messina} {et~al.}(2010){Messina}, {Desidera}, {Turatto},
  {Lanzafame}, \& {Guinan}}]{Messina2010}
{Messina}, S., {Desidera}, S., {Turatto}, M., {Lanzafame}, A.~C., \& {Guinan},
  E.~F. 2010, \aap, 520, A15, \dodoi{10.1051/0004-6361/200913644}

\bibitem[{{Montesinos} {et~al.}(2001){Montesinos}, {Thomas}, {Ventura}, \&
  {Mazzitelli}}]{Montesinos2001}
{Montesinos}, B., {Thomas}, J.~H., {Ventura}, P., \& {Mazzitelli}, I. 2001,
  \mnras, 326, 877, \dodoi{10.1046/j.1365-8711.2001.04476.x}

\bibitem[{{Morin} {et~al.}(2008){Morin}, {Donati}, {Petit}, {Delfosse},
  {Forveille}, {Albert}, {Auri{\`e}re}, {Cabanac}, {Dintrans}, {Fares},
  {Gastine}, {Jardine}, {Ligni{\`e}res}, {Paletou}, {Ramirez Velez}, \&
  {Th{\'e}ado}}]{Morin2008}
{Morin}, J., {Donati}, J.~F., {Petit}, P., {et~al.} 2008, \mnras, 390, 567,
  \dodoi{10.1111/j.1365-2966.2008.13809.x}

\bibitem[{{Morley} {et~al.}(2017){Morley}, {Kreidberg}, {Rustamkulov},
  {Robinson}, \& {Fortney}}]{Morley2017}
{Morley}, C.~V., {Kreidberg}, L., {Rustamkulov}, Z., {Robinson}, T., \&
  {Fortney}, J.~J. 2017, \apj, 850, 121, \dodoi{10.3847/1538-4357/aa927b}

\bibitem[{{Narain} \& {Ulmschneider}(1996)}]{Narain1996}
{Narain}, U., \& {Ulmschneider}, P. 1996, \ssr, 75, 453,
  \dodoi{10.1007/BF00833341}

\bibitem[{{Newton} {et~al.}(2014){Newton}, {Charbonneau}, {Irwin},
  {Berta-Thompson}, {Rojas-Ayala}, {Covey}, \& {Lloyd}}]{Newton2014}
{Newton}, E.~R., {Charbonneau}, D., {Irwin}, J., {et~al.} 2014, \aj, 147, 20,
  \dodoi{10.1088/0004-6256/147/1/20}

\bibitem[{{Newton} {et~al.}(2015){Newton}, {Charbonneau}, {Irwin}, \&
  {Mann}}]{Newton2015}
{Newton}, E.~R., {Charbonneau}, D., {Irwin}, J., \& {Mann}, A.~W. 2015, \apj,
  800, 85, \dodoi{10.1088/0004-637X/800/2/85}

\bibitem[{{Newton} {et~al.}(2017){Newton}, {Irwin}, {Charbonneau}, {Berlind},
  {Calkins}, \& {Mink}}]{Newton2017}
{Newton}, E.~R., {Irwin}, J., {Charbonneau}, D., {et~al.} 2017, \apj, 834, 85,
  \dodoi{10.3847/1538-4357/834/1/85}

\bibitem[{{Newton} {et~al.}(2016){Newton}, {Irwin}, {Charbonneau},
  {Berta-Thompson}, {Dittmann}, \& {West}}]{Newton2016}
---. 2016, \apj, 821, 93, \dodoi{10.3847/0004-637X/821/2/93}

\bibitem[{{Newton} {et~al.}(2018){Newton}, {Mondrik}, {Irwin}, {Winters}, \&
  {Charbonneau}}]{Newton2018}
{Newton}, E.~R., {Mondrik}, N., {Irwin}, J., {Winters}, J.~G., \&
  {Charbonneau}, D. 2018, \aj, 156, 217, \dodoi{10.3847/1538-3881/aad73b}

\bibitem[{{Noyes} {et~al.}(1984){Noyes}, {Hartmann}, {Baliunas}, {Duncan}, \&
  {Vaughan}}]{Noyes1984}
{Noyes}, R.~W., {Hartmann}, L.~W., {Baliunas}, S.~L., {Duncan}, D.~K., \&
  {Vaughan}, A.~H. 1984, \apj, 279, 763, \dodoi{10.1086/161945}

\bibitem[{{N{\'u}{\~n}ez} {et~al.}(2015){N{\'u}{\~n}ez}, {Ag{\"u}eros},
  {Covey}, {Hartman}, {Kraus}, {Bowsher}, {Douglas}, {L{\'o}pez-Morales},
  {Pooley}, {Posselt}, {Saar}, \& {West}}]{Nunez2015}
{N{\'u}{\~n}ez}, A., {Ag{\"u}eros}, M.~A., {Covey}, K.~R., {et~al.} 2015, \apj,
  809, 161, \dodoi{10.1088/0004-637X/809/2/161}

\bibitem[{{Owen} \& {Jackson}(2012)}]{Owen2012}
{Owen}, J.~E., \& {Jackson}, A.~P. 2012, \mnras, 425, 2931,
  \dodoi{10.1111/j.1365-2966.2012.21481.x}

\bibitem[{{Parker}(1993)}]{Parker1993}
{Parker}, E.~N. 1993, \apj, 408, 707, \dodoi{10.1086/172631}

\bibitem[{{Peacock} {et~al.}(2019){Peacock}, {Barman}, {Shkolnik},
  {Hauschildt}, {Baron}, \& {Fuhrmeister}}]{Peacock2019b}
{Peacock}, S., {Barman}, T., {Shkolnik}, E.~L., {et~al.} 2019, \apj, 886, 77,
  \dodoi{10.3847/1538-4357/ab4f6f}

\bibitem[{{Pizzolato} {et~al.}(2003){Pizzolato}, {Maggio}, {Micela},
  {Sciortino}, \& {Ventura}}]{Pizzolato2003}
{Pizzolato}, N., {Maggio}, A., {Micela}, G., {Sciortino}, S., \& {Ventura}, P.
  2003, \aap, 397, 147, \dodoi{10.1051/0004-6361:20021560}

\bibitem[{{Ranjan} {et~al.}(2017){Ranjan}, {Wordsworth}, \&
  {Sasselov}}]{Ranjan2017}
{Ranjan}, S., {Wordsworth}, R., \& {Sasselov}, D.~D. 2017, \apj, 843, 110,
  \dodoi{10.3847/1538-4357/aa773e}

\bibitem[{{Rayner} {et~al.}(2009){Rayner}, {Cushing}, \& {Vacca}}]{Rayner2009}
{Rayner}, J.~T., {Cushing}, M.~C., \& {Vacca}, W.~D. 2009, \apjs, 185, 289,
  \dodoi{10.1088/0067-0049/185/2/289}

\bibitem[{{Redfield} \& {Linsky}(2002)}]{Redfield2002}
{Redfield}, S., \& {Linsky}, J.~L. 2002, \apjs, 139, 439,
  \dodoi{10.1086/338650}

\bibitem[{{Redfield} \& {Linsky}(2004)}]{Redfield2004}
---. 2004, \apj, 613, 1004, \dodoi{10.1086/423311}

\bibitem[{{Reid} {et~al.}(1995){Reid}, {Hawley}, \& {Gizis}}]{Reid1995}
{Reid}, I.~N., {Hawley}, S.~L., \& {Gizis}, J.~E. 1995, \aj, 110, 1838,
  \dodoi{10.1086/117655}

\bibitem[{{Reiners} \& {Mohanty}(2012)}]{Reiners2012b}
{Reiners}, A., \& {Mohanty}, S. 2012, \apj, 746, 43,
  \dodoi{10.1088/0004-637X/746/1/43}

\bibitem[{{Reiners} {et~al.}(2014){Reiners}, {Sch{\"u}ssler}, \&
  {Passegger}}]{Reiners2014}
{Reiners}, A., {Sch{\"u}ssler}, M., \& {Passegger}, V.~M. 2014, \apj, 794, 144,
  \dodoi{10.1088/0004-637X/794/2/144}

\bibitem[{{Ribas} {et~al.}(2005){Ribas}, {Guinan}, {G{\"u}del}, \&
  {Audard}}]{Ribas2005}
{Ribas}, I., {Guinan}, E.~F., {G{\"u}del}, M., \& {Audard}, M. 2005, \apj, 622,
  680, \dodoi{10.1086/427977}

\bibitem[{{Rojas-Ayala} {et~al.}(2012){Rojas-Ayala}, {Covey}, {Muirhead}, \&
  {Lloyd}}]{RojasAyala2012}
{Rojas-Ayala}, B., {Covey}, K.~R., {Muirhead}, P.~S., \& {Lloyd}, J.~P. 2012,
  \apj, 748, 93, \dodoi{10.1088/0004-637X/748/2/93}

\bibitem[{{Rutten} {et~al.}(1989){Rutten}, {Schrijver}, {Zwaan}, {Duncan}, \&
  {Mewe}}]{Rutten1989}
{Rutten}, R.~G.~M., {Schrijver}, C.~J., {Zwaan}, C., {Duncan}, D.~K., \&
  {Mewe}, R. 1989, \aap, 219, 239

\bibitem[{{Scalo} {et~al.}(2007){Scalo}, {Kaltenegger}, {Segura}, {Fridlund},
  {Ribas}, {Kulikov}, {Grenfell}, {Rauer}, {Odert}, {Leitzinger}, {Selsis},
  {Khodachenko}, {Eiroa}, {Kasting}, \& {Lammer}}]{Scalo2007}
{Scalo}, J., {Kaltenegger}, L., {Segura}, A.~G., {et~al.} 2007, Astrobiology,
  7, 85, \dodoi{10.1089/ast.2006.0125}

\bibitem[{{Schneider} \& {Shkolnik}(2018)}]{Schneider2018}
{Schneider}, A.~C., \& {Shkolnik}, E.~L. 2018, \aj, 155, 122,
  \dodoi{10.3847/1538-3881/aaaa24}

\bibitem[{{Shkolnik} {et~al.}(2009){Shkolnik}, {Liu}, \& {Reid}}]{Shkolnik2009}
{Shkolnik}, E., {Liu}, M.~C., \& {Reid}, I.~N. 2009, \apj, 699, 649,
  \dodoi{10.1088/0004-637X/699/1/649}

\bibitem[{{Shkolnik} \& {Barman}(2014)}]{Shkolnik2014a}
{Shkolnik}, E.~L., \& {Barman}, T.~S. 2014, \aj, 148, 64,
  \dodoi{10.1088/0004-6256/148/4/64}

\bibitem[{{Shulyak} {et~al.}(2017){Shulyak}, {Reiners}, {Engeln}, {Malo},
  {Yadav}, {Morin}, \& {Kochukhov}}]{Shulyak2017}
{Shulyak}, D., {Reiners}, A., {Engeln}, A., {et~al.} 2017, Nature Astronomy, 1,
  0184, \dodoi{10.1038/s41550-017-0184}

\bibitem[{{Skumanich}(1972)}]{Skumanich1972}
{Skumanich}, A. 1972, \apj, 171, 565, \dodoi{10.1086/151310}

\bibitem[{{Stelzer} {et~al.}(2013){Stelzer}, {Marino}, {Micela},
  {L{\'o}pez-Santiago}, \& {Liefke}}]{Stelzer2013}
{Stelzer}, B., {Marino}, A., {Micela}, G., {L{\'o}pez-Santiago}, J., \&
  {Liefke}, C. 2013, \mnras, 431, 2063, \dodoi{10.1093/mnras/stt225}

\bibitem[{{Su{\'a}rez Mascare{\~n}o} {et~al.}(2016){Su{\'a}rez Mascare{\~n}o},
  {Rebolo}, \& {Gonz{\'a}lez Hern{\'a}ndez}}]{SuarezMascareno2016}
{Su{\'a}rez Mascare{\~n}o}, A., {Rebolo}, R., \& {Gonz{\'a}lez Hern{\'a}ndez},
  J.~I. 2016, \aap, 595, A12, \dodoi{10.1051/0004-6361/201628586}

\bibitem[{{Su{\'a}rez Mascare{\~n}o} {et~al.}(2015){Su{\'a}rez Mascare{\~n}o},
  {Rebolo}, {Gonz{\'a}lez Hern{\'a}ndez}, \& {Esposito}}]{SuarezMascareno2015}
{Su{\'a}rez Mascare{\~n}o}, A., {Rebolo}, R., {Gonz{\'a}lez Hern{\'a}ndez},
  J.~I., \& {Esposito}, M. 2015, \mnras, 452, 2745,
  \dodoi{10.1093/mnras/stv1441}

\bibitem[{{Terrien} {et~al.}(2015){Terrien}, {Mahadevan}, {Deshpande}, \&
  {Bender}}]{Terrien2015}
{Terrien}, R.~C., {Mahadevan}, S., {Deshpande}, R., \& {Bender}, C.~F. 2015,
  \apjs, 220, 16, \dodoi{10.1088/0067-0049/220/1/16}

\bibitem[{{Tian} \& {Ida}(2015)}]{Tian2015}
{Tian}, F., \& {Ida}, S. 2015, Nature Geoscience, 8, 177,
  \dodoi{10.1038/ngeo2372}

\bibitem[{{Torres} {et~al.}(2006){Torres}, {Quast}, {da Silva}, {de La Reza},
  {Melo}, \& {Sterzik}}]{Torres2006}
{Torres}, C.~A.~O., {Quast}, G.~R., {da Silva}, L., {et~al.} 2006, \aap, 460,
  695, \dodoi{10.1051/0004-6361:20065602}

\bibitem[{{Vacca} {et~al.}(2003){Vacca}, {Cushing}, \& {Rayner}}]{Vacca2003}
{Vacca}, W.~D., {Cushing}, M.~C., \& {Rayner}, J.~T. 2003, \pasp, 115, 389,
  \dodoi{10.1086/346193}

\bibitem[{{van Saders} {et~al.}(2016){van Saders}, {Ceillier}, {Metcalfe},
  {Silva Aguirre}, {Pinsonneault}, {Garc{\'\i}a}, {Mathur}, \&
  {Davies}}]{vanSaders2016}
{van Saders}, J.~L., {Ceillier}, T., {Metcalfe}, T.~S., {et~al.} 2016, \nat,
  529, 181, \dodoi{10.1038/nature16168}

\bibitem[{{Vanderburg} {et~al.}(2020){Vanderburg}, {Rowden}, {Bryson},
  {Coughlin}, {Batalha}, {Collins}, {Latham}, {Mullally}, {Col{\'o}n}, {Henze},
  {Huang}, \& {Quinn}}]{Vanderburg2020}
{Vanderburg}, A., {Rowden}, P., {Bryson}, S., {et~al.} 2020, \apjl, 893, L27,
  \dodoi{10.3847/2041-8213/ab84e5}

\bibitem[{{Vidotto} {et~al.}(2014){Vidotto}, {Gregory}, {Jardine}, {Donati},
  {Petit}, {Morin}, {Folsom}, {Bouvier}, {Cameron}, {Hussain}, {Marsden},
  {Waite}, {Fares}, {Jeffers}, \& {do Nascimento}}]{Vidotto2014b}
{Vidotto}, A.~A., {Gregory}, S.~G., {Jardine}, M., {et~al.} 2014, \mnras, 441,
  2361, \dodoi{10.1093/mnras/stu728}

\bibitem[{{von Braun} {et~al.}(2011){von Braun}, {Boyajian}, {Kane}, {van
  Belle}, {Ciardi}, {L{\'o}pez-Morales}, {McAlister}, {Henry}, {Jao}, {Riedel},
  {Subasavage}, {Schaefer}, {ten Brummelaar}, {Ridgway}, {Sturmann},
  {Sturmann}, {Mazingue}, {Turner}, {Farrington}, {Goldfinger}, \&
  {Boden}}]{vonBraun2011}
{von Braun}, K., {Boyajian}, T.~S., {Kane}, S.~R., {et~al.} 2011, \apjl, 729,
  L26, \dodoi{10.1088/2041-8205/729/2/L26}

\bibitem[{{von Braun} {et~al.}(2014){von Braun}, {Boyajian}, {van Belle},
  {Kane}, {Jones}, {Farrington}, {Schaefer}, {Vargas}, {Scott}, \& {ten
  Brummelaar}}]{vonBraun2014}
{von Braun}, K., {Boyajian}, T.~S., {van Belle}, G.~T., {et~al.} 2014, \mnras,
  438, 2413, \dodoi{10.1093/mnras/stt2360}

\bibitem[{{Wahhaj} {et~al.}(2011){Wahhaj}, {Liu}, {Biller}, {Clarke},
  {Nielsen}, {Close}, {Hayward}, {Mamajek}, {Cushing}, {Dupuy}, {Tecza},
  {Thatte}, {Chun}, {Ftaclas}, {Hartung}, {Reid}, {Shkolnik}, {Alencar},
  {Artymowicz}, {Boss}, {de Gouveia Dal Pino}, {Gregorio-Hetem}, {Ida},
  {Kuchner}, {Lin}, \& {Toomey}}]{Wahhaj2011}
{Wahhaj}, Z., {Liu}, M.~C., {Biller}, B.~A., {et~al.} 2011, \apj, 729, 139,
  \dodoi{10.1088/0004-637X/729/2/139}

\bibitem[{{Wilson} {et~al.}(2004){Wilson}, {Henderson}, {Herter}, {Matthews},
  {Skrutskie}, {Adams}, {Moon}, {Smith}, {Gautier}, {Ressler}, {Soifer}, {Lin},
  {Howard}, {LaMarr}, {Stolberg}, \& {Zink}}]{Wilson2004}
{Wilson}, J.~C., {Henderson}, C.~P., {Herter}, T.~L., {et~al.} 2004, in
  \procspie, Vol. 5492, Ground-based Instrumentation for Astronomy, ed.
  A.~F.~M. {Moorwood} \& M.~{Iye}, 1295--1305, \dodoi{10.1117/12.550925}

\bibitem[{{Wood} {et~al.}(2005){Wood}, {Redfield}, {Linsky}, {M{\"u}ller}, \&
  {Zank}}]{Wood2005}
{Wood}, B.~E., {Redfield}, S., {Linsky}, J.~L., {M{\"u}ller}, H.-R., \& {Zank},
  G.~P. 2005, The Astrophysical Journal Supplement Series, 159, 118,
  \dodoi{10.1086/430523}

\bibitem[{{Woods} {et~al.}(2009){Woods}, {Chamberlin}, {Harder}, {Hock},
  {Snow}, {Eparvier}, {Fontenla}, {McClintock}, \& {Richard}}]{Woods2009}
{Woods}, T.~N., {Chamberlin}, P.~C., {Harder}, J.~W., {et~al.} 2009, \grl, 36,
  L01101, \dodoi{10.1029/2008GL036373}

\bibitem[{{Wright} \& {Drake}(2016)}]{Wright2016}
{Wright}, N.~J., \& {Drake}, J.~J. 2016, \nat, 535, 526,
  \dodoi{10.1038/nature18638}

\bibitem[{{Wright} {et~al.}(2011){Wright}, {Drake}, {Mamajek}, \&
  {Henry}}]{Wright2011}
{Wright}, N.~J., {Drake}, J.~J., {Mamajek}, E.~E., \& {Henry}, G.~W. 2011,
  \apj, 743, 48, \dodoi{10.1088/0004-637X/743/1/48}

\bibitem[{{Wright} {et~al.}(2018){Wright}, {Newton}, {Williams}, {Drake}, \&
  {Yadav}}]{Wright2018}
{Wright}, N.~J., {Newton}, E.~R., {Williams}, P. K.~G., {Drake}, J.~J., \&
  {Yadav}, R.~K. 2018, \mnras, 479, 2351, \dodoi{10.1093/mnras/sty1670}

\bibitem[{{Youngblood} {et~al.}(2016){Youngblood}, {France}, {Loyd}, {Linsky},
  {Redfield}, {Schneider}, {Wood}, {Brown}, {Froning}, {Miguel}, {Rugheimer},
  \& {Walkowicz}}]{Youngblood2016}
{Youngblood}, A., {France}, K., {Loyd}, R.~O.~P., {et~al.} 2016, \apj, 824,
  101, \dodoi{10.3847/0004-637X/824/2/101}

\bibitem[{{Youngblood} {et~al.}(2017){Youngblood}, {France}, {Loyd}, {Brown},
  {Mason}, {Schneider}, {Tilley}, {Berta-Thompson}, {Buccino}, {Froning},
  {Hawley}, {Linsky}, {Mauas}, {Redfield}, {Kowalski}, {Miguel}, {Newton},
  {Rugheimer}, {Segura}, {Roberge}, \& {Vieytes}}]{Youngblood2017}
---. 2017, \apj, 843, 31, \dodoi{10.3847/1538-4357/aa76dd}

\bibitem[{{Zuckerman} \& {Song}(2004)}]{Zuckerman2004}
{Zuckerman}, B., \& {Song}, I. 2004, \araa, 42, 685,
  \dodoi{10.1146/annurev.astro.42.053102.134111}

\end{thebibliography}

\end{document}